\documentclass[sigconf]{acmart}

\usepackage{soul}
\usepackage{subcaption}
\usepackage[inline]{enumitem}
\usepackage{graphicx}
\usepackage{xcolor}
\usepackage{anonymous-acm}
\usepackage{cleveref}[2012/02/15]% v0.18.4; 
\usepackage{multirow}
\usepackage{acmart-taps}

\newif\ifproofread%
\newcommand{\changemarker}[1]{%
    \ifproofread%
    \else
    \fi
}
\proofreadtrue%

\AtBeginDocument{%
  \providecommand\COTwo{CO$_2$}}

\copyrightyear{2023} 
\acmYear{2023} 
\setcopyright{rightsretained} 
\acmConference[CHI '23]{Proceedings of the 2023 CHI Conference on Human Factors in Computing Systems}{April 23--28, 2023}{Hamburg, Germany}
\acmBooktitle{Proceedings of the 2023 CHI Conference on Human Factors in Computing Systems (CHI '23), April 23--28, 2023, Hamburg, Germany}\acmDOI{10.1145/3544548.3580675}
\acmISBN{978-1-4503-9421-5/23/04} % ChkTeX 8

\begin{document}

\newtheorem{study}{Study}
\renewcommand{\thetheorem}{\arabic{study}}
\newcommand\note[2] 
 {\todo[fancyline]{\textbf{#1}: #2}}
 
%% The "title'' command has an optional parameter,
%% allowing the author to define a "short title'' to be used in page headers.
\title[Save A Tree or 6 kg of CO2?]{Save A Tree or 6 kg of CO2? Understanding Effective Carbon Footprint Interventions for Eco-Friendly Vehicular Choices}

%%
%% The "author'' command and its associated commands are used to define
%% the authors and their affiliations.
%% Of note is the shared affiliation of the first two authors, and the
%% "authornote'' and "authornotemark'' commands
%% used to denote shared contribution to the research.

\author{Vikram Mohanty}
\authornote{Author is currently at Virginia Tech.}
\email{vikrammohanty@vt.edu}
\orcid{0000-0001-6296-3134} % ChkTeX 8
\affiliation{%
  \institution{Toyota Research Institute}
  \streetaddress{4400 El Camino Real}
  \city{Los Altos}
  \state{California}
  \country{USA}
  \postcode{94022}
}

\author{Alexandre Filipowicz}
\email{alex.filipowicz@tri.global}
\orcid{0000-0002-1311-386X} % ChkTeX 8
\affiliation{%
  \institution{Toyota Research Institute}
  \streetaddress{4400 El Camino Real}
  \city{Los Altos}
  \state{California}
  \country{USA}
  \postcode{94022}
}

\author{Nayeli Bravo}
\email{nayeli.bravo@tri.global}
\orcid{0000-0001-9238-9831} % ChkTeX 8
\affiliation{%
  \institution{Toyota Research Institute}
  \streetaddress{4400 El Camino Real}
  \city{Los Altos}
  \state{California}
  \country{USA}
  \postcode{94022}
}

\author{Scott Carter}
\email{scott.carter@tri.global}
\affiliation{%
  \institution{Toyota Research Institute}
  \streetaddress{4400 El Camino Real}
  \city{Los Altos}
  \state{California}
  \country{USA}
  \postcode{94022}
}

\author{David A. Shamma}
\email{aymans@acm.org}
% \email{ayman.shamma@tri.global}
\orcid{0000-0003-2399-9374} % ChkTeX 8
\affiliation{%
  \institution{Toyota Research Institute}
  \streetaddress{4400 El Camino Real}
  \city{Los Altos}
  \state{California}
  \country{USA}
  \postcode{94022}
}

%%
%% By default, the full list of authors will be used in the page
%% headers. Often, this list is too long, and will overlap
%% other information printed in the page headers. This command allows
%% the author to define a more concise list
%% of authors' names for this purpose.
\renewcommand{\shortauthors}{Mohanty, Filipowicz, et al.}

%%
%% The abstract is a short summary of the work to be presented in the
%% article.
\begin{abstract}
  From ride-hailing to car rentals, consumers are often presented with eco-friendly options. 
  Beyond highlighting a ``green'' vehicle and \COTwo{} emissions, \COTwo{} equivalencies have been designed to provide understandable amounts; we ask which equivalencies will lead to eco-friendly decisions.
  We conducted five ride-hailing scenario surveys where participants picked between regular and eco-friendly options, testing equivalencies, social features, and valence-based interventions.
  Further, we tested a car-rental embodiment to gauge how an individual (needing a car for several days) might behave versus the immediate ride-hailing context.
  We find that participants are more likely to choose green rides when presented with additional information about emissions; \COTwo{} by weight was found to be the most effective.
  Further, we found that information framing---be it individual or collective footprint, positive or negative valence---had an impact on participants’ choices.
  Finally, we discuss how our findings inform the design of effective interventions for reducing car-based carbon-emissions.
\end{abstract}




\maketitle

%%%% 1.introduction.tex starts here %%%%

\section{Introduction}
\begin{figure}
  \centering
  \begin{subfigure}[b]{0.45\columnwidth}
    \centering
    \includegraphics[height=7pc]{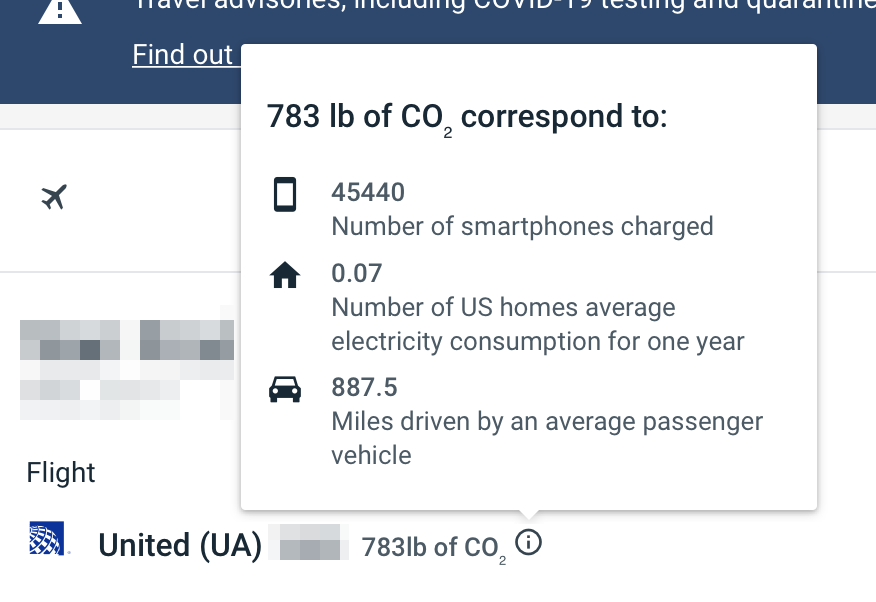}
    \caption{\label{fig:united}}
    \Description{An image of a flight booking in progress that shows both raw CO2 emissions and equivalencies for the flight. The flight would emit 783 lbs of CO2, which is equivalent to the emissions from charging 45440 smartphones, or the annual energy usage of 0.07 homes, or 887.5 miles driven by an average passenger vehicle.}
  \end{subfigure}  
  \hspace{1pc}
  \begin{subfigure}[b]{0.45\columnwidth}
    \centering
    \includegraphics[height=7pc]{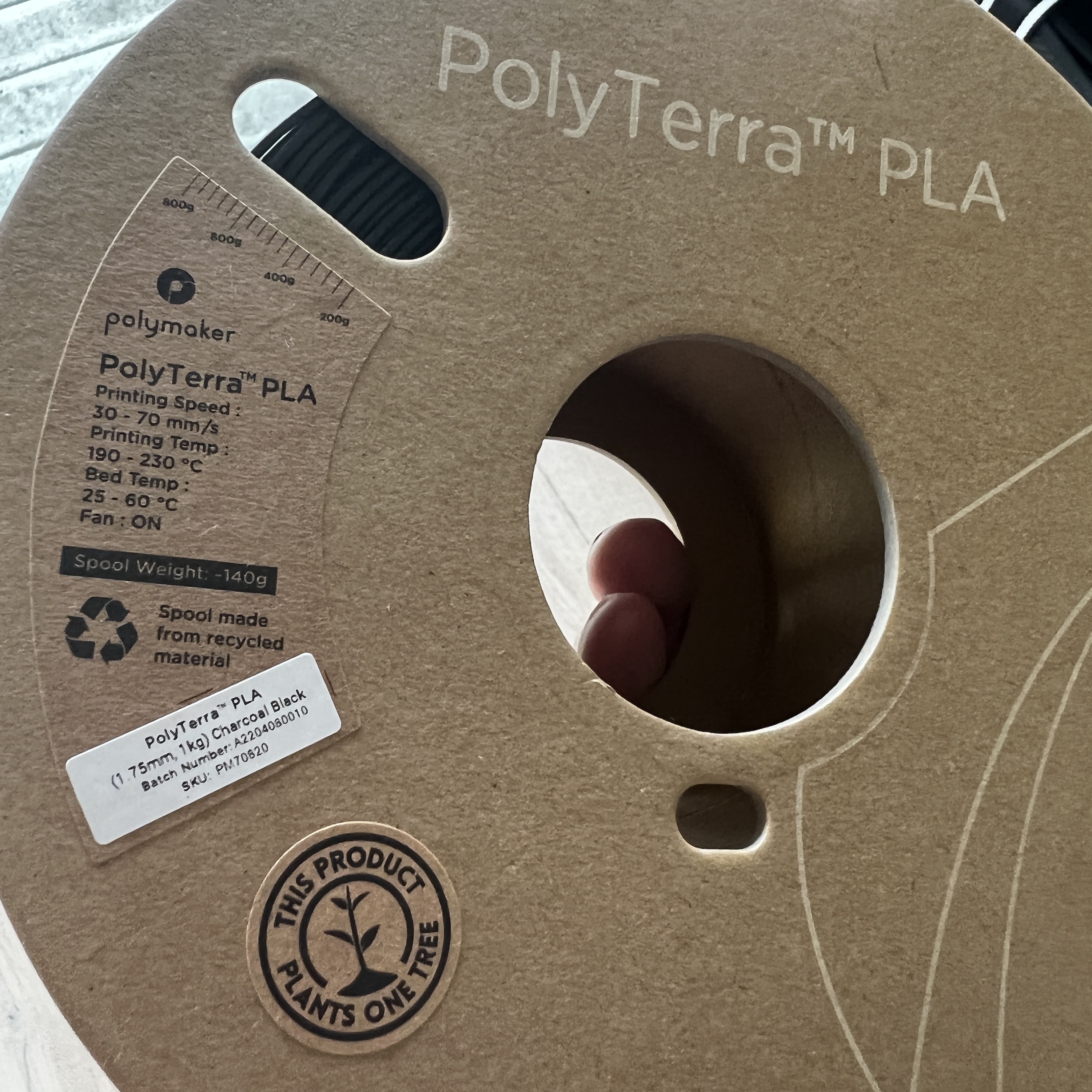}
    \caption{\label{fig:polyterra}}
    \Description{An image of a Polyterra 3D printing PLA filament that
      displays environmental impact in terms of an equivalency: this product plants one tree.}
  \end{subfigure}
  \caption{\label{fig:others} Equivalencies in the wild. (a) While booking a flight, carbon equivalencies are shown to the buyer. (b) A spool of 3D printer filament with a sticker denoting ``This product plants one tree.''}
\end{figure}

There is a growing interest in \emph{eco-feedback interfaces}, interfaces designed to communicate the environmental impact of different products and services, across a wide range of domains, including energy consumption~\cite{yun2013sustainability,holmes2007eco}, waste disposal~\cite{holstius2004infotropism}, and transportation~\cite{park2017ecotrips,sanguinetti2017greenfly,froehlich2009ubigreen}. These interfaces display eco-feedback information that can be simple (e.g., showing a green leaf designating eco-friendly options), or more complex, taking the form of direct information about carbon emissions (e.g., pounds of \COTwo{}; Figure~\ref{fig:united}), information about carbon-equivalent activities (e.g., miles driven; Figure~\ref{fig:united}), or information about incentives or offsets associated with a product (e.g., tree planting; Figure~\ref{fig:polyterra}). However, it is unclear how consumers process this eco-feedback and information about carbon emissions. In this article, we explore the effectiveness of different carbon-based eco-feedback interventions ---  whether consumers prefer direct \COTwo\ emissions, simpler heuristic interventions (e.g., green logos), or more relatable \COTwo{}-equivalent activities. We examine this topic in the context of personal transportation: ride-sharing apps (sometimes called ride-hailing) and vehicle rentals. 

Eco-friendly options have steadily grown in the car-sharing space. Options like hybrid and battery electric vehicles are steadily becoming more common in both ride-sharing (e.g., through services like Uber Green and JustGrab Green) and vehicle rental-companies~\cite{green_2022,grab_2022,lyft_2019,woods_2022}. Additionally, consumers generally choose ride-sharing and vehicle rental options using online platforms.
% With ride-sharing, choices are made quickly and have an impact over a short period of time.
% In contrast, vehicle rentals choice can factor longer temporal horizons (e.g., days, weeks or months), which has implications for factors like fueling costs and/or where to charge a battery electric vehicle.
This paper explores the influence of emission information on car-sharing choices. We examine how displaying \COTwo\ emissions information, \COTwo\ equivalencies, and other information might help people choose eco-friendly ride-sharing options.

A number of tools exist to help people calculate their carbon footprint~\cite{west2016evaluating,froehlich2009ubigreen, andersson2020novel}.
These tools aim to provide context for emissions by linking them directly to everyday activities (similar to the success that adding calorie information to food has had on eating habits~\cite{bollinger2011calorie}).
To help provide relatable information, the Environmental Protection
Agency (EPA) created a Greenhouse Gas Equivalency
Calculator\footnote{\url{https://www.epa.gov/energy/greenhouse-gas-equivalencies-calculator}
  (Accessed February 2023).} that aims to ``translate abstract measurements into concrete terms you can understand''~\cite{epa_calc_2022}. This tool converts \COTwo\ emissions into \COTwo{}-equivalent activities (e.g., pounds of \COTwo\ converted to miles driven; as of writing, the calculator computes 22 different equivalencies).
Tools such as these can have an important role to play in the transportation sector. Transportation is a leading cause of US-based \COTwo\ emissions~\cite{ivanova2016environmental} and a number of interventions are being applied to reduce transportation-based emissions. Carbon calculators have been created to help users measure and track the emission due to their transportation habits~\cite{ajufo2021automated,park2017ecotrips}. \COTwo{}-related information has also been provided alongside transportation options, with the intention of informing people's choices (e.g., \COTwo\ from flying, Monroney window stickers). In addition to information-based interventions, numerous interventions have tried to motivated consumers towards more sustainable options through social interventions (e.g., on social media~\cite{gabrielli2014design,gabrielli2013designing}) and reward/incentive programs~\cite{mahmoodi2021using}.
However, it is unclear whether \COTwo\ and emission equivalencies resonates with consumers, how these interventions compare to non-\COTwo\ related interventions, and whether any of these  interventions influence consumer choices.

\begin{figure*}
    \centering
    \includegraphics[width=0.75\textwidth]{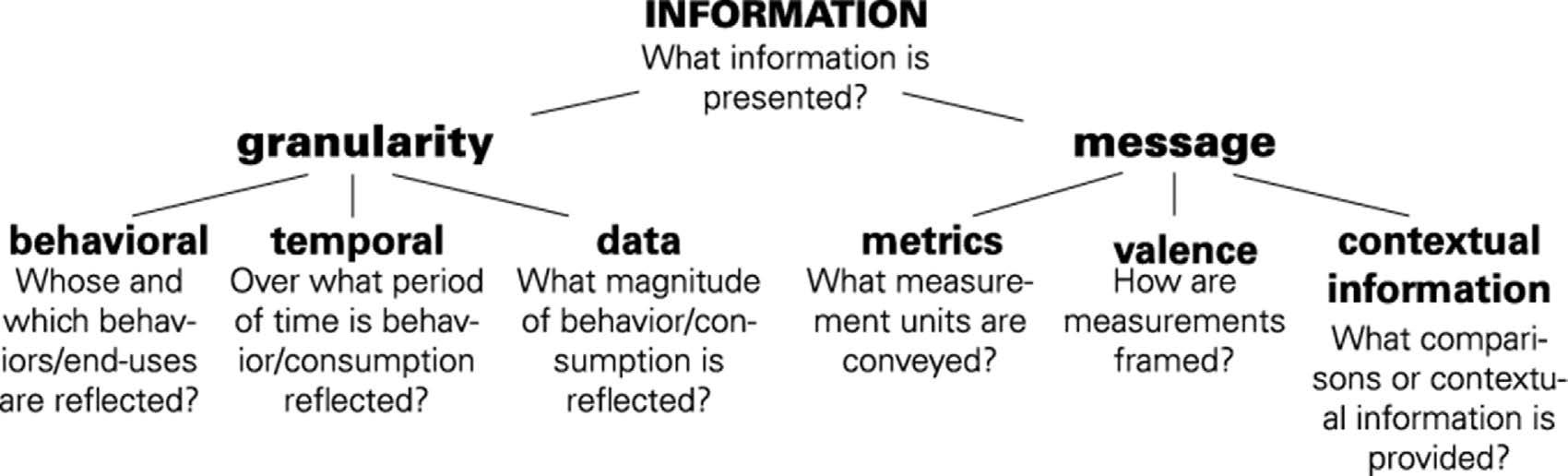}
    \caption{Sub-categories and dimensions of eco-feedback information (reprinted from Sanguinetti et al.~\cite{sanguinetti2018information} with permission from Elsevier).}
    \Description{A tree diagram of eco-feedback information, broken into two dimensions --- granularity and message. Granularity is further broken into three sub-categories: behavioral, temporal, and data. Message is broken into three sub-categories: metrics, valence, and contextual information.}
\label{fig:ecofeedbackframework}
\end{figure*}

In this paper, we examine whether information about \COTwo\ influences people's decisions in ride-sharing and rental-car interfaces through multiple studies. We guide these studies by exploring different facets of eco-feedback information as proposed in a framework by Sanguinetti et al.~\cite{sanguinetti2018information} (see Figure~\ref{fig:ecofeedbackframework}). We first explored whether information about \COTwo\ emissions and \COTwo\ emission equivalencies influence people's ride-sharing choices and how these interventions compare with other non-\COTwo\ related intervention types. We then followed up with a second set of studies to examine how contextual information, valence, and additional explanations play a role in communicating carbon emissions. Then, we conducted a third set of ride-sharing studies that more closely examined how people perceive and use raw \COTwo\ emission values to make their decision. Finally, we examined how people process emission-related information in a vehicle rental context and explore how these \COTwo\ interventions influence choices with longer temporal consideration.

%%%% 1.introduction.tex ends here %%%%

%%%% 2.relatedwork.tex starts here %%%%

\section{Related Work}
Many human activities emit \COTwo{} and other damaging greenhouse gas (GHG) emissions including transportation, power generation, and household energy usage~\cite{epa_report_2022}.
To tackle this ongoing climate crisis, a wide range of entities are taking action---be it businesses pledging ``carbon neutrality''~\cite{pledge_2022}, governments introducing climate policies~\cite{house_2022,spain_2022}, or non-profit organizations drawing public attention towards environmental issues~\cite{catf_2022}.
In line with this trend, eco-feedback technologies---\emph{technologies that provide feedback about resource consumption and environmental impact}---are also becoming a prevalent method to encourage individual consumers to adopt energy-saving practices~\cite{sanguinetti2018information,walz2014gamifying,froehlich2010design}.

\subsection{Eco-Feedback Systems}
Over the years, HCI research has explored the design and deployment of different eco-feedback interfaces spanning across different contexts: e.g\ food consumption~\cite{zapico2016eco,shakeri2021envirofy}, household energy usage~\cite{yun2013sustainability,holmes2007eco}, green transit~\cite{froehlich2009ubigreen,park2017ecotrips}, waste disposal~\cite{holstius2004infotropism}.
Multiple eco-feedback systems have shown promising results in transforming user behavior towards being environmentally conscious, be it smart thermostats engaging users to actively save energy~\cite{stopps2021residential} or carbon footprint calculators encouraging users to reflect and reduce their emissions~\cite{west2016evaluating}.

Personal transportation is the largest contributor to carbon emissions both in the US~\cite{epa_report_2022, ivanova2016environmental} and Europe~\cite{statista_2022}.
To tackle this particular problem, prior work has explored the use of personal eco-feedback interfaces to nudge users towards greener commute options (e.g., carpooling, biking) through feedback on tracked transportation habits~\cite{froehlich2009ubigreen,park2017ecotrips}.
Multiple studies have also shown promising results with in-car dashboards displaying eco-feedback on fuel efficiency and driving behavior~\cite{walz2014gamifying}.
Recently, navigation apps like Google Maps have introduced eco-friendly routing aimed at lower fuel consumption~\cite{lozano-aguilera_2022}.

These eco-feedback interfaces have largely focused on either alternative transportation solutions or optimization during driving. However, less focus has been placed on how information is processed within car-sharing interfaces like ride-sharing or vehicle rentals. As ride-sharing and other ``mobility as a service'' platforms gain in popularity~\cite{goodall2017rise,jittrapirom2017mobility}, understanding how people use and process carbon information in these contexts can play an important role in helping people make greener transportation choices.
% \todo[author=vikram,fancyline]{Gotta think of a line that highlights why these contexts are equally important compared to the alternative solutions.}
Ridesharing companies such as Uber, Lyft, and Grab have started providing eco-friendly rides (a hybrid or battery electric vehicle) in multiple cities around the world~\cite{lyft_2019,grab_2022,green_2022}.
Similarly, rental companies have also added electric vehicles to their fleets~\cite{woods_2022}.
% Car manufacturers all around the world are increasingly adding electric vehicles to their lineup~\cite{bartlett_preston_2022}.
However, it is not clear what kind of eco-feedback information should be provided (and how) to nudge users to make these eco-friendly vehicular choices. 

% Some evidence suggests that simply providing GHG information when users select different options can nudge them to make greener flight choices~\cite{sanguinetti2017greenfly}. 

% In this paper, we investigate the influence if different \COTwo\ and non-\COTwo\ related information interventions influence eco-friendly choices in the context of ridesharing and car rentals. 

\subsection{Communicating Carbon Emissions}

Seen as an extension of persuasive technology~\cite{fogg2002persuasive}, prior work on eco-feedback interfaces has largely focused on tool design, especially on sensory and display embodiments, rather than behavioral theory~\cite{sanguinetti2018information,froehlich2010design}.
Studies have also shown that these eco-feedback systems may not always translate into desired energy-saving behaviors~\cite{hansson2021decade}.
Froehlich et al.\ called attention to engaging with environmental psychology theories to inform the design of eco-feedback interfaces~\cite{froehlich2010design}. Sanguinetti et al.\ proposed a framework mapping different aspects of eco-feedback design to behavioral change~\cite{sanguinetti2018information}, suggesting three high-level designs of eco-feedback design---\emph{information}, \emph{timing}, and \emph{display}.
In this paper, we focus on the \emph{information} aspect of eco-feedback design, specifically on understanding effective ways to communicate carbon emissions to consumers making vehicular choices.

While terms like \emph{carbon footprint} have been part of our lexicon for decades~\cite{safire_2008}, it is still unclear what the most effective currency for carbon emissions might be.
While most carbon footprint calculators or eco-feedback interfaces communicate emissions using \COTwo\ by weight~\cite{ajufo2021automated,west2016evaluating}, there are still be gaps in our understanding of the levels of “carbon literacy”  amongst the general public.
Targeted goals related to calorie levels for food consumption~\cite{link_gunnars_2021} or footsteps per day~\cite{rieck_2020} have become a common part of our daily lives with the ubiquity of tracking devices~\cite{peng2021habit}. However, the same cannot be said for carbon emissions even though calculators are becoming more accessible.
Through this work, we aim to improve our understanding of how to represent carbon emissions as a starting point for bridging gaps in carbon literacy levels.

Different eco-feedback interfaces have used different metrics to represent carbon emissions, ranging from direct \COTwo\ by weight~\cite{sanguinetti2017greenfly} to different types of carbon emissions equivalencies such as how many trees or polar bears would be saved, how many seconds of toxic gases in a volcanic eruption would be emitted, or how many tons of fossilized material is embodied in fuel consumed~\cite{park2017ecotrips,froehlich2009ubigreen}.
The equivalencies provided by the United State EPA's carbon calculator are meant to be “concrete understandable terms”, such as gallons of gasoline, pounds of coal, bags of waste, or number of trees~\cite{epa_calc_2022}.
However, little is known about how end-users respond to these metrics or if these metrics to communicate carbon emissions effectively.

In this paper, we conduct multiple studies to better understand how to communicate carbon emissions in the context of vehicular choices.
To guide our studies, we use the design framework proposed by Sanguinetti et al.~\cite{sanguinetti2018information} to explore different dimensions of eco-feedback information.

\subsubsection{Design Framework for Eco-Feedback Information}

Sanguinetti et al.~\cite{sanguinetti2018information} divide eco-feedback information into two high-level categories: (a) \emph{message}---the content itself and (b) \emph{granularity}---how fine or coarse the information is.
The authors describe three dimensions of \emph{message}:
\begin{itemize}
\item \textbf{Message:} The framework discusses the trade-offs between direct measurement units such as \COTwo{} and equivalencies that may seem more familiar to the user (e.g., trees).
  Although direct measurement units enable objective analysis, consumers may not understand these scientific units nor find them actionable.
  Although less precise, metaphorical units may seem more tangible or familiar to people, and motivate behavioral change.
  All studies in the current article investigate the effectiveness of both scientific units and different equivalencies (listed on the EPA website) across multiple studies.
\item \textbf{Valence:} The framing given to eco-feedback, whether positive or negative, can have an impact on how it is perceived and used by users.
  Schwartz's Norm Activation Model~\cite{schwartz1977normative} suggests that negative valence might appeal to users more for calling attention to the negative consequences of resource consumption, and thus result in eco-friendly behavior.
  Study~\aptLtoX[graphic=no,type=env]{2b}{\ref{study:2:valence}} investigates the influence of different valence framing on the how people use eco-feedback information.
\item \textbf{Contextual Information:} Feedback Intervention Theory suggests that feedback, when compared to a standard or a target, can impact behavior~\cite{kluger1996effects}.
  Prior work has shown promising results of energy consumption reduction in multiple settings upon exposure to different types of goal-setting, be it historical comparisons (i.e., one's own consumption from the past) or social comparisons (i.e., consumption levels of other members in the community)~\cite{gabrielli2014design}.
  Froehlich et al.\ also discuss studies where comparisons are sought after but do not necessarily impact behavior~\cite{froehlich2010design}.
  Study~\aptLtoX[graphic=no,type=env]{3b}{\ref{study:3:target}} explores whether goal-setting provides a context in which people can better understand carbon.
\end{itemize}

Similarly, the authors describe three different types of \emph{information granularity}:
\begin{itemize}
\item \textbf{Behavioral Granularity:} This describes who (i.e., target consumers) and which type of behavior (i.e., specific or generic) is being targeted.
  Studies have shown promising results with both high granularity (i.e., when individual consumption is targeted) and low granularity (i.e., when collective feedback is provided)~\cite{fischer2008feedback,gabrielli2014design}.
  Study~\aptLtoX[graphic=no,type=env]{2a}{\ref{study:2:social}} investigates how individual and collective feedback can be leveraged to communicate carbon emissions.
\item \textbf{Temporal Granularity:} User behavior can be impacted by whether eco-feedback information is provided at small intervals or accumulated over longer durations~\cite{van2001prototype}.
  Study~\ref{study:4:rental} investigates how temporal granularity plays a role in communicating carbon emissions in the context of vehicle rentals.
\item \textbf{Data Granularity:} User behavior can be affected by the resolution of eco-feedback information, whether information is provided in fine-grained numeric data or coarser representations (e.g., light that changes between two colors)~\cite{dahlinger2018impact}.
  Our studies investigate data granularity by testing fine-grained emissions information (e.g., \COTwo\ emission values) and coarse representations (e.g., green leaf icon) across multiple studies.
\end{itemize}

\section{Phase 1: Carbon Equivalencies in Ride-Hailing Choices}

Our goal is to understand both whether carbon equivalencies lead ride-share passengers towards greener choices and to identify which equivalencies passengers consider relatable and understandable. 
\begin{study}\label{study:1}
We conducted a survey with a mock-up of a ride-haling app to establish a base understanding of \COTwo\ interventions when picking a short ride in which 
a participant is shown a ride with two options: a standard ride and a
green ride. 
Beyond ride choice, MaxDiff questions further examined what equivalencies are relatable and useful.
\end{study}
This study has two main hypotheses.
\aptLtoX[graphic=no,type=env]{% Coding for XML/HTML generation
\begin{enumerate*}%[label=\textbf{H\ref{study:1}.\arabic*}]
\item[\textbf{H1.1}]\label{hypo:1:core} Interventions that provide information about carbon increase the likelihood that passengers choose carbon-friendly options.
\item[\textbf{H1.2}]\label{hypo:1:equiv} Interventions that convey relatable carbon equivalents increase the likelihood of choosing carbon-friendly ride options over detailed \COTwo\ information.
\end{enumerate*}}{\begin{enumerate*}[label=\textbf{H\ref{study:1}.\arabic*}]
\item\label{hypo:1:core} Interventions that provide information about carbon increase the likelihood that passengers choose carbon-friendly options.
\item\label{hypo:1:equiv} Interventions that convey relatable carbon equivalents increase the likelihood of choosing carbon-friendly ride options over detailed \COTwo\ information.
\end{enumerate*}}

\begin{figure*}
  \includegraphics[width=\textwidth]{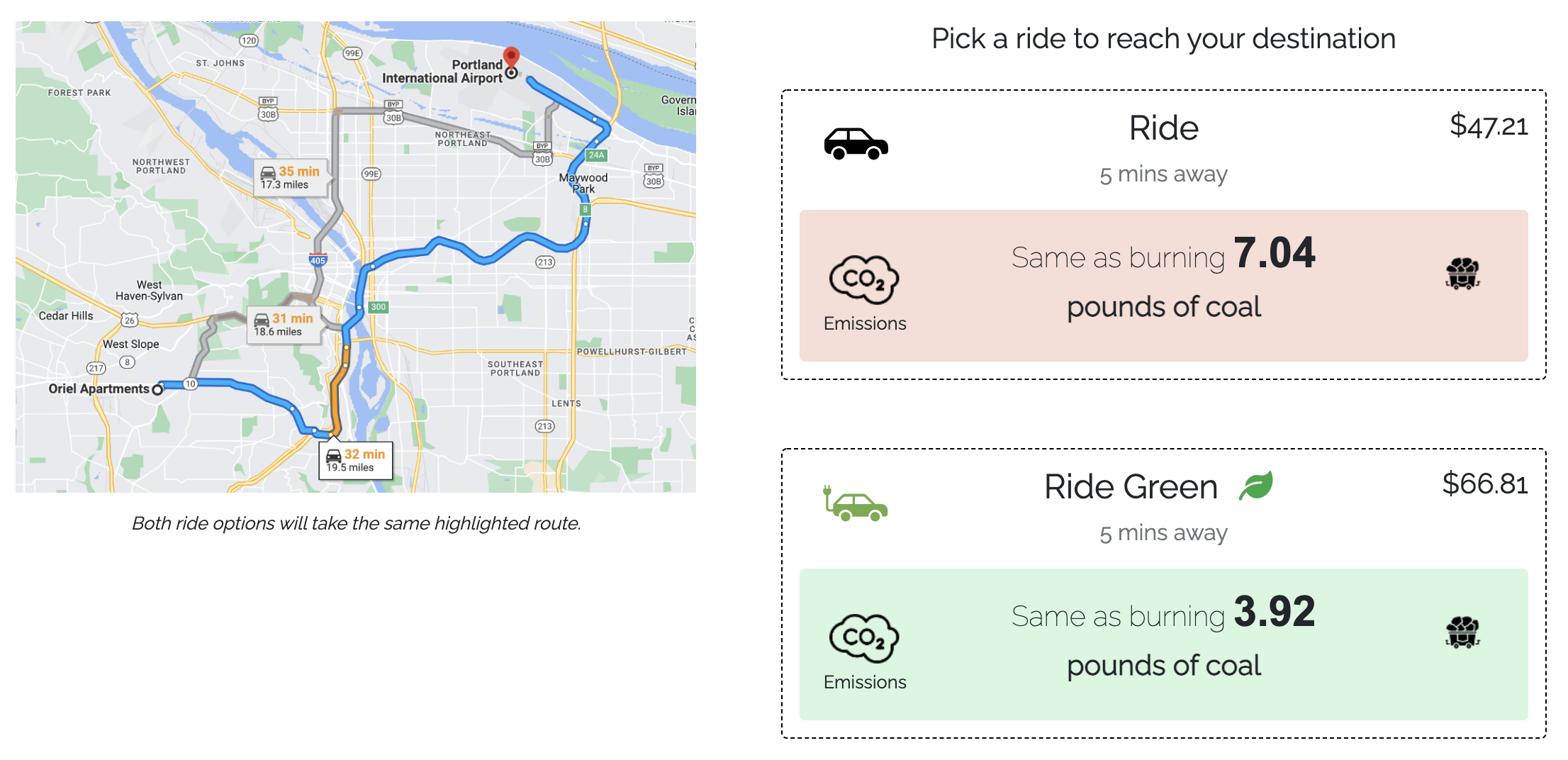}
  \caption{An example ``Pick your Rideshare'' survey question showing information about \COTwo\ emissions in equivalent terms (\emph{coal} here) for two ridesharing options---a standard (``Ride'') and an eco-friendly alternative (``Ride Green'').}
  \Description{A map of a 30 mile route in Austin on the left and two ride sharing options on the right. One ride green for \$42.99 burning 6.26 lbs of coal.  The other standard ride for \$35.93 burning 12.13 lbs of coal.}
\label{fig:study1}
\end{figure*}

\subsection{Experiment Setup}
To address these questions, we designed an online survey experiment to simulate choices made in a rideshare app (Figure~\ref{fig:study1}).
On each trial, participants saw a map with a hypothetical trip and chose between two rideshare options: \emph{Ride} and \emph{Ride Green}.
The prepared study dataset consisted of 15 trips based in 15 major US cities that were distributed evenly across short ($<15$ miles), medium ($<25$ miles), and long ranges ($>25$ miles), overall varying between 5--44 miles.
The \emph{Ride} option was always presented with a black car whereas the \emph{Ride Green} option was presented as a green car with a green leaf icon (Figure~\ref{fig:study1}).
The price for both options was either identical or higher for the \emph{Ride Green} option (between \$0.50--\$24.00 more), and ride option choice positions were randomized (i.e., whether \emph{Ride Green} appeared on top/bottom). 
Each participant completed 30 trials, each consisting of a randomly picked trip from the dataset and a randomly picked intervention from the following list: 

\begin{description}
    \item[Baseline:] No additional information provided about the ride other than a green leaf icon on the \emph{Ride Green} option.
    
    \item[Raw Emissions:] \COTwo\ emissions (in pounds) provided for both ride options. We used the US Fuel Economy calculator~\cite{epa_2022} to calculate the emissions of a gas vehicle (i.e., \emph{Ride}) and an electric vehicle (i.e., \emph{Ride Green}) for a given trip distance.  
    
    \item[Equivalencies:] Information about the \COTwo\ emissions for both ride options was provided in terms of \emph{measurable actions} that either emitted (e.g., charging smartphones, daily energy usage of houses, and burning pounds of coal or gallons of gasoline or barrels of oil) or sequestered (e.g., recycling bags of waste, switching incandescent lamps to LEDs, growing trees or forests) equivalent amounts of \COTwo\ (calculated using the EPA calculator~\cite{epa_calc_2022}).
    
    \item[Social:] Information about hypothetical social trends was provided, including a ``popular'' condition in which participants were told that 75--99\% of other riders chose the \emph{Ride Green} option and an ``unpopular'' condition in which 1--25\% of riders chose the green option. We also displayed a ``collective action'' condition, where participants saw both how many other riders had picked the green option as well as the collective environmental impact across all riders (e.g., \textit{16 trees were saved by 318 passengers who rode this today}). 
    
    \item[Points:] Participants were told they would ``receive 2x points'' for choosing the \emph{Ride Green} option (versus ``1x points'' for \emph{Ride}). Participants were deliberately not given any additional information about the points.
\end{description}

After completing 30 trials, participants answered a survey questionnaire that included questions related to their demographics and attitudes about climate change, renewable energy, and electric vehicles.
They also answered a MaxDiff questionnaire~\cite{qualtrics_2021} about the usefulness and relatability of the different equivalencies presented during the main experiment, and open-ended questions about their choice process.

\subsection{Analysis}

We measured the effect of each intervention using a logistic mixed-effects model (LME) implemented with the \texttt{lme4} library in \texttt{R}~\cite{bates2014fitting}.
This model predicted the likelihood of making \emph{Ride Green} choices using the difference in price between the \emph{Ride Green} and \emph{Ride} options and intervention type as fixed effects, and random intercepts for participants and city.

\subsection{Participants}
We recruited 1002 US-based participants from Prolific~\cite{palan2018prolific} (responses from all 50 US states;\@ 47\% women, 50\% men, 2\% genderqueer or non-binary, and 1\% preferring not to answer; mean age = 37 years, SD = 13 years).
Participants were all at least 18 years of age, fluent in English, and had used either Uber or Lyft (the two most popular US ride-sharing options).
All participants provided informed consent before beginning the task.

\subsection{Data and Code}
All of the collected data from this paper is available as supplemental material.
Additionally, all the \texttt{R} code used in the analysis is bundled with the data. 
The data and code is located as supplement in article's DOI in the
digital library and is also hosted in the Open Science
Framework\footnote{\label{f:osf} ACM Digital Library: \url{https://doi.org/10.1145/3544548.3580675}, OSF: \url{https://osf.io/79txw/} (Accessed February 2023)} (OSF).

\subsection{Findings}
\subsubsection{Interventions increase \emph{Ride Green} choices}
\begin{figure*}
  \centering
  \begin{subfigure}[b]{\columnwidth}
    \centering
    \includegraphics[width=\columnwidth]{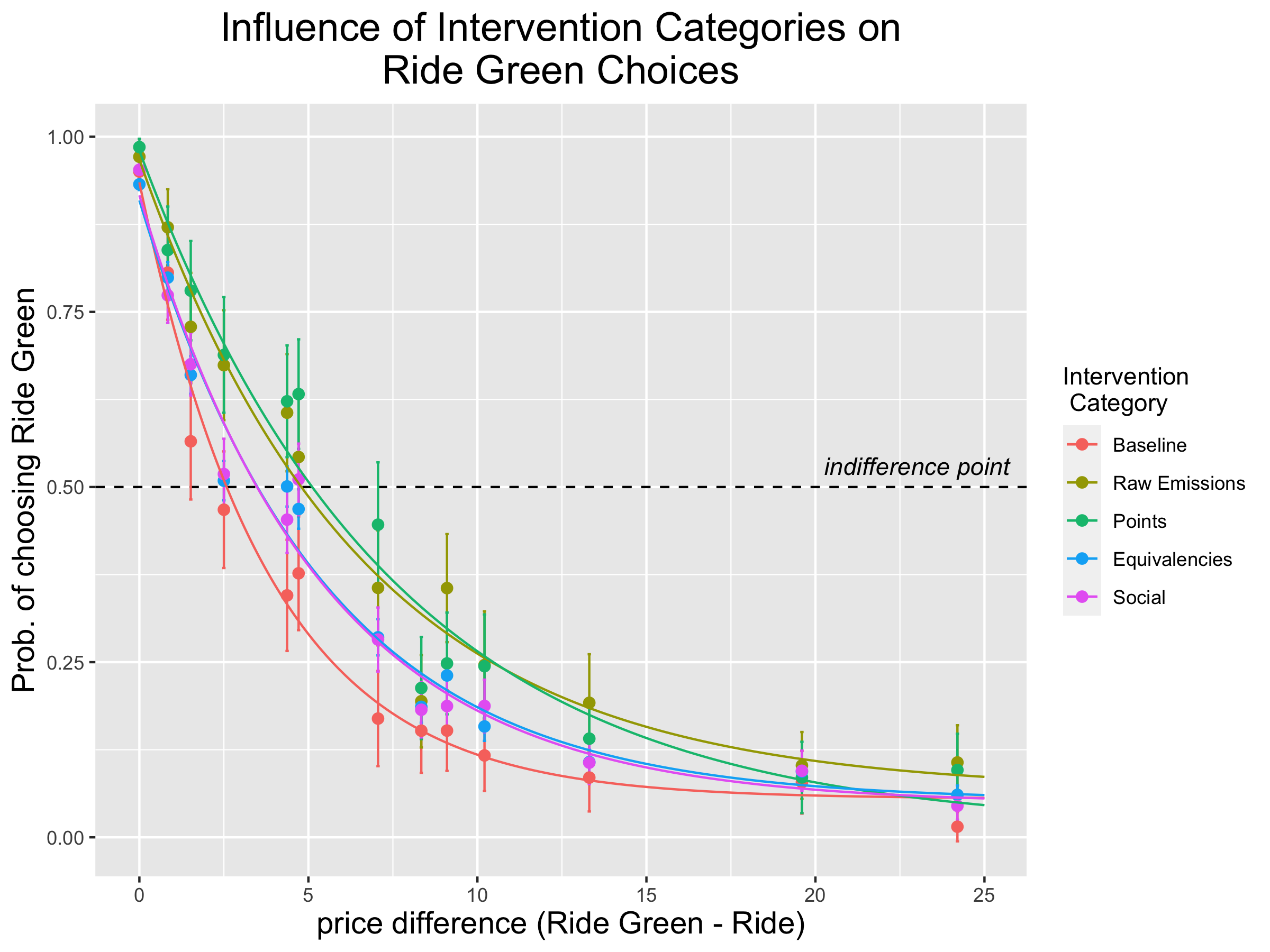} % ChkTeX 25
    \caption{\label{fig:rcurves}}
    \Description{An exponential decay curve that shows the probability of choosing ride green on Y-axis (0--1) and the price difference between Ride Green and Ride on the X-axis (\$0--\$25) for five different intervention types (baseline, social, equivalencies, raw emissions, and points). The curves for each of the interventions intersect the indifference point at 0.50 on the Y-axis; the intersection point of the baseline curve is the left-most point, while all the other curves intersect further right on the indifference line, with the points and raw emissions intervention curves being the right-most ones. This indicates people are willing to pay more for Ride Green even with the baseline intervention, and they are willing to pay even more with the other interventions.}
  \end{subfigure}  
  \hspace{1pc}
  \begin{subfigure}[b]{\columnwidth}
    \centering
    \includegraphics[width=\columnwidth]{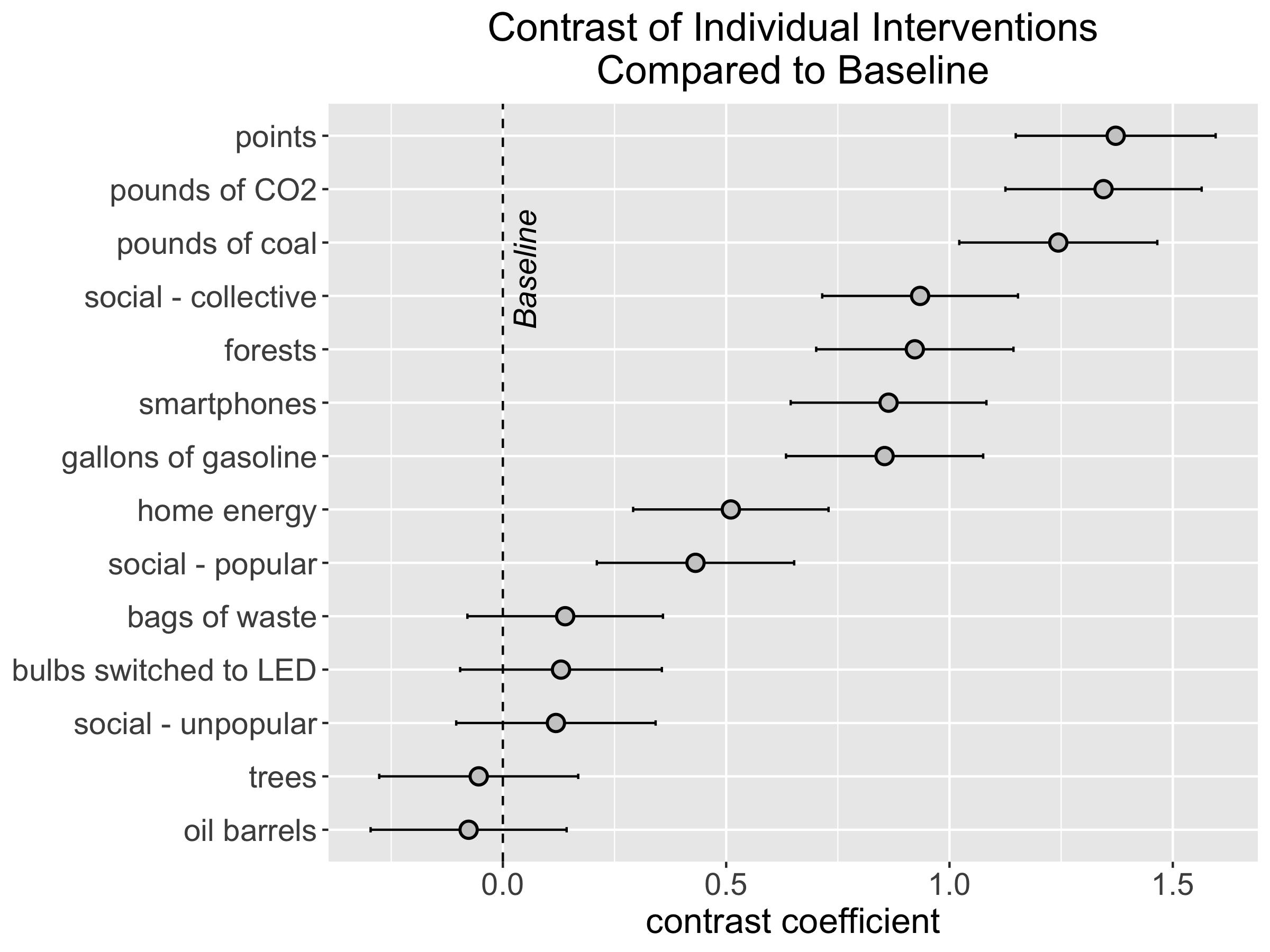}
    \caption{\label{fig:effect}}
    \Description{A chart displaying the probability of participants picking a green ride in log odd scale. The X-axis has beta coefficients for the different interventions. The Y-axis has all the different interventions sorted according to the beta coefficients in descending order. The beta coefficients are compared to the baseline condition, denoted by a dashed vertical line. Points, lbs. of CO2, and lbs. of coal are the top-performing interventions and are shown at the top of Y-axis, whereas trees, social-unpopular, bulbs switched to LED are at the bottom of Y-axis.} % ChkTex 12
  \end{subfigure}
  \caption{\label{fig:curves} Study~\ref{study:1} (a) Influence of different intervention types on probability that participants chose the \emph{Ride Green} option for different price differences. Colored dots indicate average probabilities across subjects, error bars indicate 95\% confidence intervals, and trend lines indicates the best fitting exponential decay curve.  (b) Logistic mixed-effects contrast coefficients comparing the influence of individual intervention types compared to the Baseline condition (dashed line). Error bars indicate 95\% confidence intervals.}
\end{figure*}

Overall, interventions with some form of information increased the likelihood of participants picking the \emph{Ride Green} option compared to the Baseline condition (Figure~\ref{fig:rcurves}). Interventions using Raw Emissions were highly effective, increasing the likelihood of choosing the green option by 3.7x, followed by Equivalencies and Social interventions (each increasing the likelihood by 1.6x; LME contrasting intervention types with Baseline condition: $\beta_\mathrm{Raw Emissions}=1.31$, $\beta_\mathrm{Equivalencies}=0.49$, $\beta_{\mathrm{Social}}=0.49$, all $z>5.39$, all $p<0.001$). P126 emphasized the usefulness of these interventions:
\begin{quote}
  ``Metrics that would be useful for me need to be both relatable/tangible and have a direct correlation to environmental impact that I understand. For example, I find pounds of \COTwo\ (which directly indicates carbon emissions) and number of trees needed to counteract the impact particularly helpful in understanding the environmental impact of daily activities.'' 
\end{quote}

As anticipated, participants became less likely to select \emph{Ride Green} as it got more expensive (LME estimating the influence of price on probability of choosing \emph{Ride Green}: $\beta_\mathrm{Price Difference}=-0.41$, $z=-7.62$, $p<0.001$). However, they were willing to pay more for \emph{Ride Green} as long as it was no more than \$4.37 over the \emph{Ride} option (Figure~\ref{fig:rcurves}). In the words of P755:
\begin{quote}
``I looked at the information on how much more efficient it was from an energy/fuel perspective and as long as the price wasn't too different I would choose the green option. The only time I would pick the non-green option was if there was a significant price difference.'' 
\end{quote}

\subsubsection{Raw \COTwo\ is effective}
Contrary to our predictions, providing raw \COTwo\ emission number was more effective than presenting emissions in terms of carbon equivalencies
(LME contrasting Equivalencies with Raw Emissions: $\beta_\mathrm{Raw Emissions}=0.82$, $z=10.06$, $p<0.001$).
P482 described what \COTwo\ emission numbers meant to them: \emph{``I would want to know how much \COTwo\ (carbon dioxide) I use up each day since this has a pretty direct correlation to the harm one does to the atmosphere.''}

\subsubsection{People like points}
Surprisingly, even though points were not tethered to any rewards, assigning 2x points for \emph{Ride Green} was as effective as providing raw \COTwo\ emission information, increasing the likelihood of choosing the green option by 3.9x over the Baseline condition (LME contrasting interventions with Baseline condition: $\beta_\mathrm{Points}=1.35$, $z=11.92$, $p<0.001$).
P371 highlighted the appeal of accumulating points: \emph{``\ldots getting more points for doing so is great, so I liked that as well''}. When asked about what information was used to make choices, P830 added: \emph{``\ldots points\ldots might lead to future discounts''}.

\subsubsection{Not all equivalencies were effective}
\begin{figure*}
  \centering
  \begin{subfigure}[b]{\columnwidth}
    \centering
    \includegraphics[width=\columnwidth]{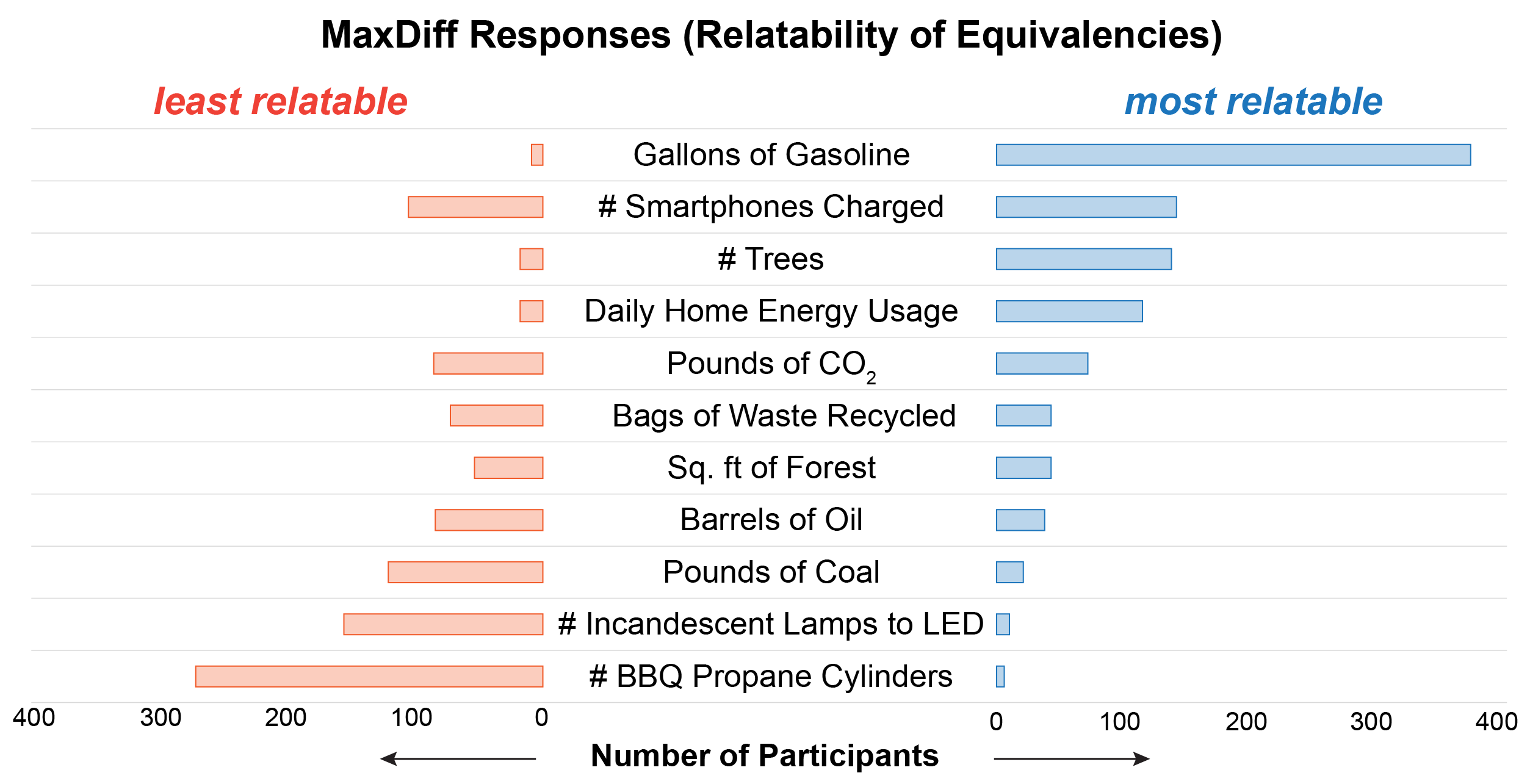}
    \caption{\label{fig:relatability}}
    \Description{A double-ended bar chart displaying aggregated participant responses for how relatable they found each of the interventions. For each intervention, the chart displays the total number of participants who found it to be the least relatable and most relatable. The chart shows the interventions in decreasing order of how many participants found it to be the most relatable intervention. Here are all the interventions listed from top to bottom: gallons of gasoline, smartphones charged, trees, daily home energy usage, pounds of CO2, sq. ft of forest, bags of waste recycled, barrels of oil, pounds of coal, incandescent lamps to LED, and BBQ propane cylinders.} % ChkTeX 12
  \end{subfigure}  
  \begin{subfigure}[b]{\columnwidth}
    \centering
    \includegraphics[width=\columnwidth]{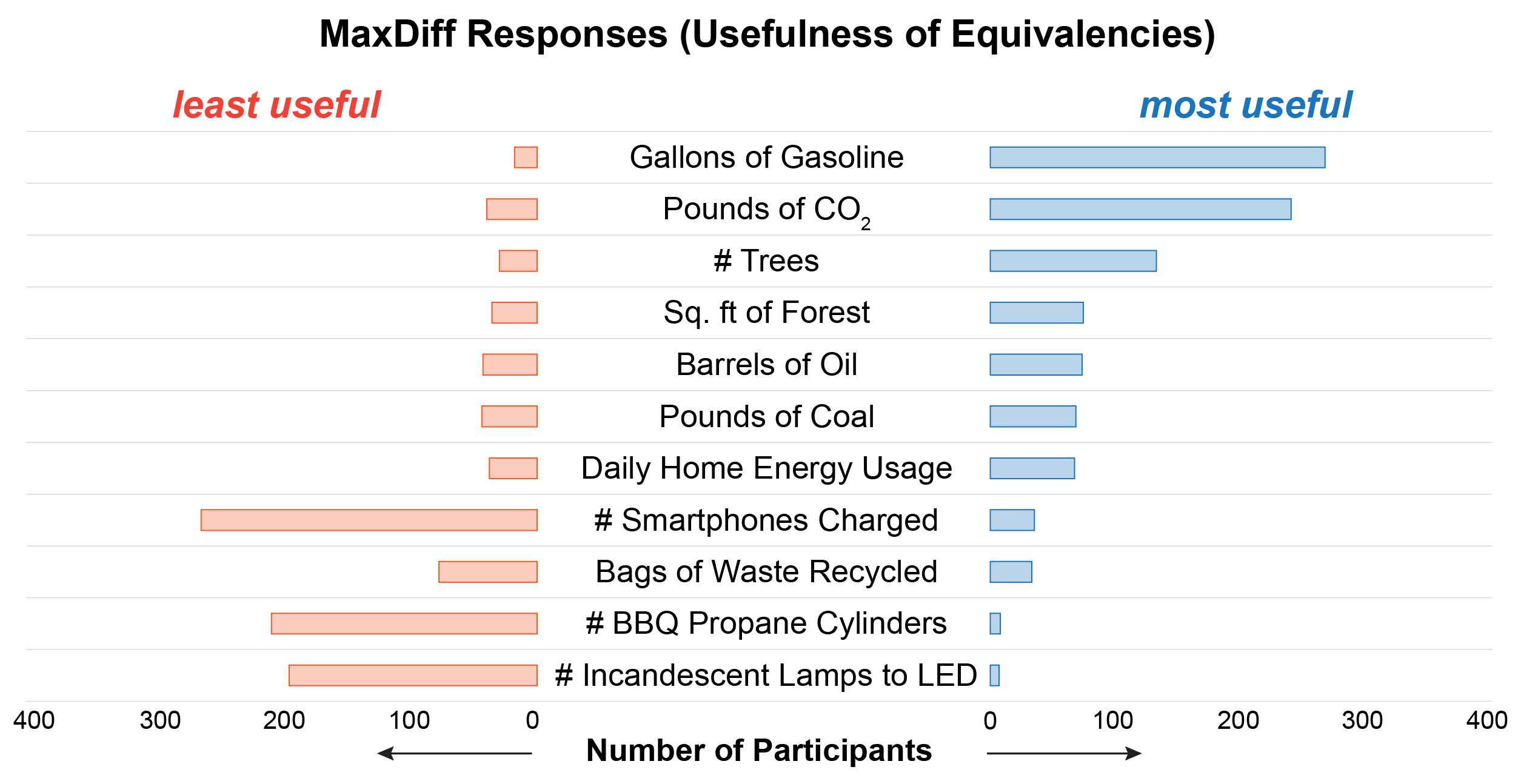}
    \caption{\label{fig:usefulness}}
    \Description{A double-ended bar chart displaying aggregated participant responses for how useful they found each of the interventions. For each intervention, the chart displays the total number of participants who found it to be the least useful and most useful. The chart shows the interventions in decreasing order of how many participants found it to be the most useful intervention. Here are all the interventions listed from top to bottom: pounds of CO2, gallons of gasoline, trees, square feet of forest, daily home energy usage, barrels of oil, pounds of coal,  bags of waste recycled, smartphones charged, BBQ propane cylinders, and incandescent lamps to LED.}
  \end{subfigure}
  \caption{\label{fig:maxdiff} MaxDiff responses for the (a) relatability and (b) usefulness of different equivalencies.}
\end{figure*}

Not all equivalencies increased the likelihood of participants picking green options (see Figure~\ref{fig:effect}).
Among the equivalencies, ``burning pounds of coal'', ``growing forest cover'', ``charging smartphones'', ``burning gallons of gasoline'', and ``daily energy usage of homes'' all increased the likelihood that participants chose the green option above the Baseline condition.
All equivalencies except ``pounds of coal'' aligned with participants' self-reports about how useful and relatable they found equivalencies (Figure~\ref{fig:maxdiff}). Conversely, sequestering equivalencies---``recycling bags of waste'', ``switching incandescent bulbs to LEDs'', and ``growing number of trees''---were less effective than other interventions.
Apart from ``number of trees'', participants also rated these
equivalencies as the least relatable and useful.

%None of the interventions \emph{lowered} the probability of making green choices compared to Baseline.

\subsubsection{Not all social interventions are equal}
Displaying social interventions yielded mixed results.
Although better than Baseline, social interventions performed worse than Points or Raw Emissions (LME contrasting interventions with Social condition: $\beta_\mathrm{Points}=0.85$, $\beta_\mathrm{Raw Emissions}=0.82$, all $z>9.18$, all $p<0.001$), but were similar to Equivalencies ($\beta_\mathrm{Equivalencies}=0.002$, $z=0.05$, $p=0.958$; Figure~\ref{fig:rcurves}).
Among the individual social interventions, ``
Social-Collective'' and ``Social-Popular'' both increased the likelihood of green choices over baseline whereas ``Social-Unpopular'' did not result in any change (Figure~\ref{fig:effect}).

\subsection{Discussion}

Our findings show that information interventions were overall effective in nudging riders towards making green ridesharing choices. However, we found that some interventions performed better than others. 

Offering points, even without any information about how they can be redeemed, turned out to be an effective intervention, supporting prior work on extrinsic rewards being used as a motivator for sustainable decision-making~\cite{froehlich2010design}. This finding is certainly interesting, and merits additional future investigation, but is beyond this article's focus on investigating carbon information.  

% to understand why they were highly effective in the current context, where ridesharing companies are ending rewards programs~\cite{leigh_2022}. 

Consistent with~\aptLtoX[graphic=no,type=env]{\textbf{H1.1}}{\ref{hypo:1:core}} interventions that provided information about carbon emissions or equivalences increased the likelihood that participants chose eco-friendly options.
We additionally observed that emission equivalencies (i.e., burning pounds of coal, charging smartphones, burning gallons of gasoline, and home energy usage) were more effective than the sequestering equivalencies (i.e., recycling bags of waste, switching incandescent lamps to LEDs, and planting trees) with the exception of forests.
There might be multiple factors at play here; one possibility is that emission equivalencies are closer to the source of pollution than sequestering equivalencies and may therefore register a stronger effect in the people's minds.
This difference in effectiveness between emissions and sequestering also draws attention to how these intervention statements were framed, whether \emph{positive valence} (i.e., saving something) or \emph{negative valence} (i.e., burning/wasting something) was employed~\cite{schwartz1977normative}.
Coal and forests also happen to be common topics generally discussed in the context of climate change across news media and advertising, which might have contributed to their effectiveness amongst participants.

The inconsistency between the participants' observed behavior and self-reported MaxDiff responses (see Figure~\ref{fig:maxdiff}) suggests there might be gaps in conveying how these equivalencies relate to emissions. For example, it might not be directly intuitive how charging smartphones or recycling bags of waste contribute towards more or less carbon emissions. Some equivalencies may also be outdated and no longer be relevant in people's minds. For example, switching incandescent lamps to LEDs was one of the least effective and least relatable interventions, which could potentially be due to the incandescent lamps being largely phased out around the globe~\cite{enwiki:1105075892}. These results suggests the need to explore whether there was a general lack of understanding about these equivalencies, and if that played a role in nudging participants towards making eco-friendly choices. 

Contrary to~\aptLtoX[graphic=no,type=env]{\textbf{H1.2}}{\ref{hypo:1:equiv}}, \COTwo\ was among the most effective interventions at pushing people towards greener ride-sharing choices.
This results was surprising as we expected that raw \COTwo\ units would be considered too scientific and not actionable for consumers compared to more relatable equivalencies~\cite{fischer2008feedback}.
One possibility is that \COTwo\ was mainly effective when displayed as a relative difference between options (i.e., in our current study, \COTwo\ information could be compared between the \emph{Ride} and \emph{Ride Green} options) and not because people have an inherent understanding of \COTwo\ as a metric for describing environmental impact.
% (See Study~\ref{study:3} for further detail).

In addition to interventions related to carbon emissions, we found that social interventions also influenced eco-friendly choices. Participants were more likely to choose more ``popular'' options, consistent with trends observed in psychology and marketing~\cite{cialdini1990focus, duan2008online}. Additionally, interventions that described the collective impact of certain equivalencies (``Social-Collective'') also proved to be effective, thus supporting prior work on collective eco-feedback~\cite{gabrielli2014design}.
Our next studies extend these findings by exploring the role of \emph{dynamic} social norms~\cite{sparkman2017dynamic} and whether collective impact influences certain equivalencies more than others.
% However, further analysis is needed to understand whether the effectiveness of certain equivalencies is influenced by how they are addressed---individual impact or collective impact.
Last, while this study allowed us to gather insights about different carbon and non-carbon related interventions, ride-sharing choices are made as single choices with limited temporal durations. Our next studies extend these findings to decisions with longer time horizons. 
%The ride-sharing context therefore provides a limited framework with which to measure the influence of temporal granularity on vehicle-based choices~\cite{van2001prototype,sanguinetti2018information}.

% repetitive with next section
% may not be the most appropriate scenario for understanding longer-term impact of eco-friendly decisions as participants may not care about the emissions from a single, short ride. 
%Study~\ref{study:4:rental} examines the role of emission information in the context of vehicle rentals, providing the opportunity to examine whether the result found in ride-sharing contexts extend to other vehicular domains and to explore the role temporal granularity plays in the way \COTwo\ information is processed.

%%%% 3.study1.tex ends here %%%%

% \section{STUDY 2: INFLUENCE OF CONTEXTUAL FRAMING ON CARBON EQUIVALENCIES FOR RIDE-SHARING CHOICES}

% % Brief introduction with hypotheses
% The results from study 1 provide insights into the effectiveness of different carbon equivalencies when communicating emissions information in the context of ride-sharing. In addition to equivalencies, the way information is framed can also impact how it is processed [REF ON FRAMING EFFECTS LIKE KAHNMAN ETC].

% These fr

% \subsection{Participants and Methods}

% A total of XXX participants completed the task. We excluded XX participants leaving a final sample of 201 participants (DEMOGRAPHIC BREAKDOWN - GENDER AND AGE).

%%%% 4.second.tex starts here %%%%

\section{Phase 2: Examining context, ambiguity, and temporality}
Study~\ref{study:1} provided a number of insights into the effectiveness of different forms of carbon messaging that we interrogate in the next set of studies. One key finding from Study~\ref{study:1} was that many equivalencies were effective at influencing people's choice and that contextual factors could also play a role. Study~\ref{study:2} expands on these findings by exploring the ways in which social context, information valence, and additional information influence eco-friendly choices.
Another surprising finding from Study~\ref{study:1} was that \COTwo\ information had a strong influence on participant choices. Study~\ref{study:3} examines this effect more thoroughly, exploring the extent to which absolute \COTwo\ values are used and/or understood by users when making decisions.
Finally, the ride-sharing context of Study~\ref{study:1} provided only a limited framework with which to test effects of temporal granularity on people's decisions. Study~\ref{study:4:rental} examines the role of emission information in the context of vehicle rentals, providing the opportunity to examine whether the results found in ride-sharing contexts extend to other vehicular domains and to explore the role temporal granularity plays in the way \COTwo\ information is processed~\cite{van2001prototype,sanguinetti2018information}.
Similar to Phase 1, all the collected data and analysis code from
Phase 2 is also included as supplemental material and in OSF\textsuperscript{\ref{f:osf}}.

\subsection{Contextual Framing on \COTwo\  Equivalencies for Ride-Sharing Choices}

% Brief introduction with hypotheses
The results from Study~\ref{study:1} provide insights into the effectiveness of different carbon equivalencies when communicating emissions information in the context of ride-sharing.
In addition to equivalencies, the way information is framed can also impact how it is processed~\cite{tversky1985framing,plous1993psychology,sanguinetti2018information}.  Study~\ref{study:2} examines the influence of three specific framings: social, valence, and detailed explanation.

\begin{study}\label{study:2}
  Examines the role of three potential framing effects of carbon equivalency information on ride-share choices:
\aptLtoX[graphic=no,type=env]{% Coding for XML/HTML generation
\begin{enumerate*}%[label=\ref{study:2}\alph*]
\item[2a]\label{study:2:social} social framing in popularity or community,
\item[2b]\label{study:2:valence} information valence as positive or negative impact,
\item[2c]\label{study:2:explain} adding detailed explanations.
\end{enumerate*}}{
\begin{enumerate*}[label=\ref{study:2}\alph*]
\item\label{study:2:social} social framing in popularity or community,
\item\label{study:2:valence} information valence as positive or negative impact,
\item\label{study:2:explain} adding detailed explanations.
\end{enumerate*}}
\end{study} 

% \missingfigure{Show some screenshots of Study~\ref{study:2}.}

\begin{figure}
    \centering
        \begin{subfigure}[b]{\columnwidth}
        \centering
        \includegraphics[trim=0 475 600 100,clip,width=.75\columnwidth]{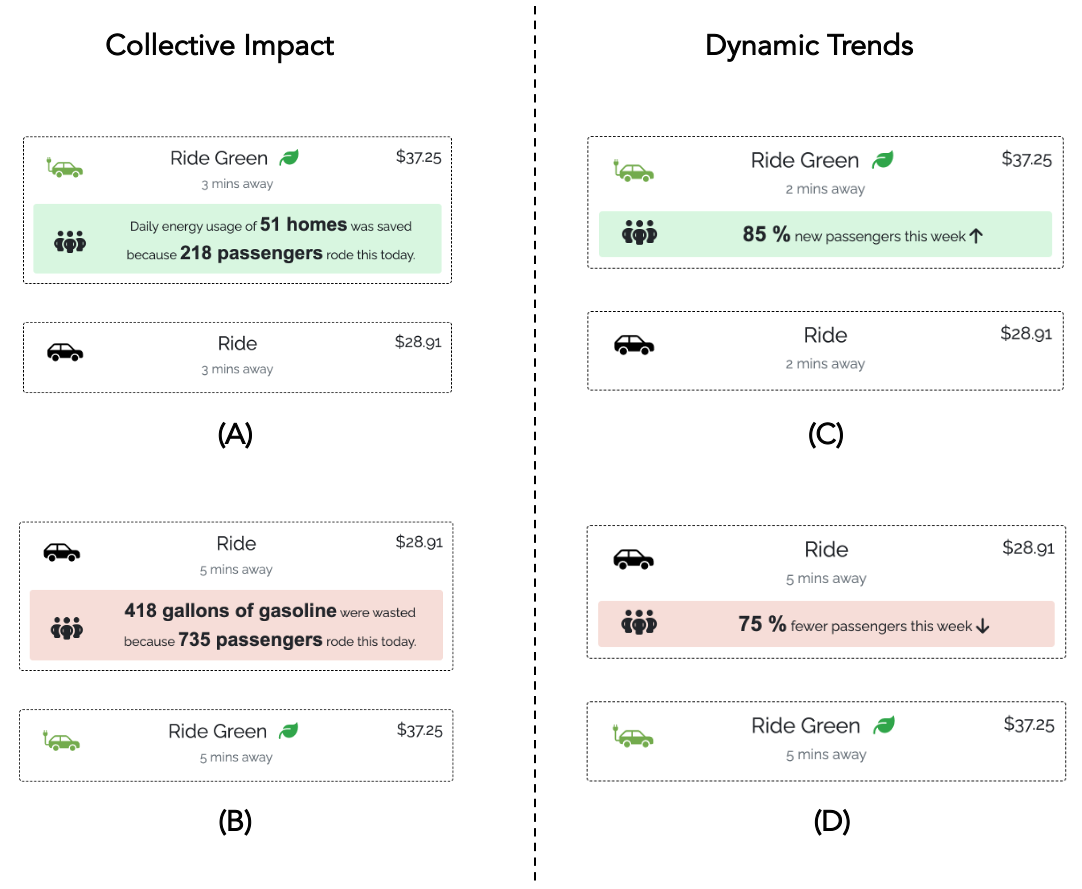}
        \caption{Collective Impact (Positive)\label{fig:studysocialcip}}
        \Description{Example of collective--positive framing. Two ride sharing options, with a Ride Green option on top that costs $\$37.25$ and Ride option on bottom that costs $\$28.91$. The Ride Green option indicates ``Daily energy usage of 51 homes was saved because 218 passengers rode this today''. The words ``51 homes'' and ``218 passengers'' are written in bold.}
        \vspace{1pc}
        \includegraphics[trim=0 100 600 500,clip,width=.75\columnwidth]{figures/studysocial.png}
        \caption{Collective Impact (Negative)\label{fig:studysocialcin}}
        \Description{Example of collective--negative framing. Two ride sharing options, with a Ride option on top that costs $\$28.91$ and Ride Green option on bottom that costs $\$37.25$. The Ride option indicates ``418 gallons of gasoline were wasted because 735 passengers rode this today''. The words ``418 gallons of gasoline'' and ``735 passengers'' are written in bold.}

    \end{subfigure}
    \begin{subfigure}[b]{\columnwidth}
        \centering
        \includegraphics[trim=575 475 0 100,clip,width=.75\columnwidth]{figures/studysocial.png}
        \caption{Dynamic Social Norms (Positive)\label{fig:studysocialdynp}}
        \Description{Example of positive dynamic social norms framing. Two ride sharing options, with a Ride Green option on top that costs $\$37.25$ and Ride option on bottom that costs $\$28.91$. The Ride Green option indicates ``85\% new passengers this week'' with an arrow pointing upwards beside the text. The percentage ``85\%'' is written in bold.}
        \vspace{1pc}
        \includegraphics[trim=575 100 0 500,clip,width=.75\columnwidth]{figures/studysocial.png}
        \caption{Dynamic Social Norms (Negative)\label{fig:studysocialdynn}}
        \Description{Example of negative dynamic social norms framing. Two ride sharing options, with a Ride option on top that costs $\$28.91$ and Ride Green option on bottom that costs $\$37.25$. The Ride option indicates ``75\% fewer passengers this week'' with an arrow pointing downwards beside the text. The percentage ``75\%'' is written in bold.}
    \end{subfigure}
    \caption{Example screenshots of the interface tested in Study~{2a}. %%%{\ref{study:2:social}}. 
Participants chose between two rides (a green ride and a regular ride) similar to Study~\ref{study:1}, but were now shown different social interventions. In Figures~\ref{fig:studysocialcip} and~\ref{fig:studysocialcin}, the interventions described the collective impact of people picking the different rides in terms of different equivalencies. In Figures~\ref{fig:studysocialdynp} and~\ref{fig:studysocialdynn}, the interventions showed trends of newer or fewer people picking the green or the non-green ride.}
\label{fig:studysocial}
\end{figure}

First, people's decisions can be strongly influenced by societal norms, with people tending to prefer behaviors that conform to social norms and avoiding those that do not~\cite{ajzen1991theory, bosnjak2020theory, cialdini1990focus, schultz2007constructive}.
Consistent with these tendencies, psychology and marketing research consistently finds that framing options as ``popular'' will increase the likelihood that these will be chosen~\cite{cialdini1990focus, schultz2007constructive, duan2008online}.
%This is what was observed in Study~\ref{study:1} and we sought to replicate the result here.
Moreover, recent work has demonstrated that highlighting \emph{dynamic changes} to social norms can also have a strong influence on participants behavior~\cite{sparkman2017dynamic}.
Consistent with this prior work, we observed in Study~\ref{study:1} that people were more likely to choose ``Ride Green'' if it was a framed as a popular option or if information about collective carbon savings were communicated alongside it (Figure~\ref{fig:effect}).
Study~\aptLtoX[graphic=no,type=env]{2a}{\ref{study:2:social}} examines the specific role of different types of social framing on people's ride-sharing choices.
We specifically hypothesized that
\aptLtoX[graphic=no,type=env]{% Coding for XML/HTML generation
\begin{enumerate*}%[label=\textbf{H\ref{study:2}.\arabic*}]
\item[\textbf{H2.1}]\label{hypo:2:pop} \textit{people are more likely to choose green options if carbon equivalencies are framed as being popular rather than unpopular},
\item[\textbf{H2.2}]\label{hypo:2:norms} \textit{framed as as \emph{starting} to be chosen by more people (dynamic social norms)}, and
\item[\textbf{H2.3}]\label{hypo:2:collective} \textit{framed around collective rather than individual impact}.
\end{enumerate*}}{
\begin{enumerate*}[label=\textbf{H\ref{study:2}.\arabic*}]
\item\label{hypo:2:pop} \textit{people are more likely to choose green options if carbon equivalencies are framed as being popular rather than unpopular},
\item\label{hypo:2:norms} \textit{framed as as \emph{starting} to be chosen by more people (dynamic social norms)}, and
\item\label{hypo:2:collective} \textit{framed around collective rather than individual impact}.
\end{enumerate*}}

\begin{figure}
    \centering    
    \begin{subfigure}[b]{0.75\columnwidth}
        \centering
        \includegraphics[trim=0 100 800 0,clip,width=\columnwidth]{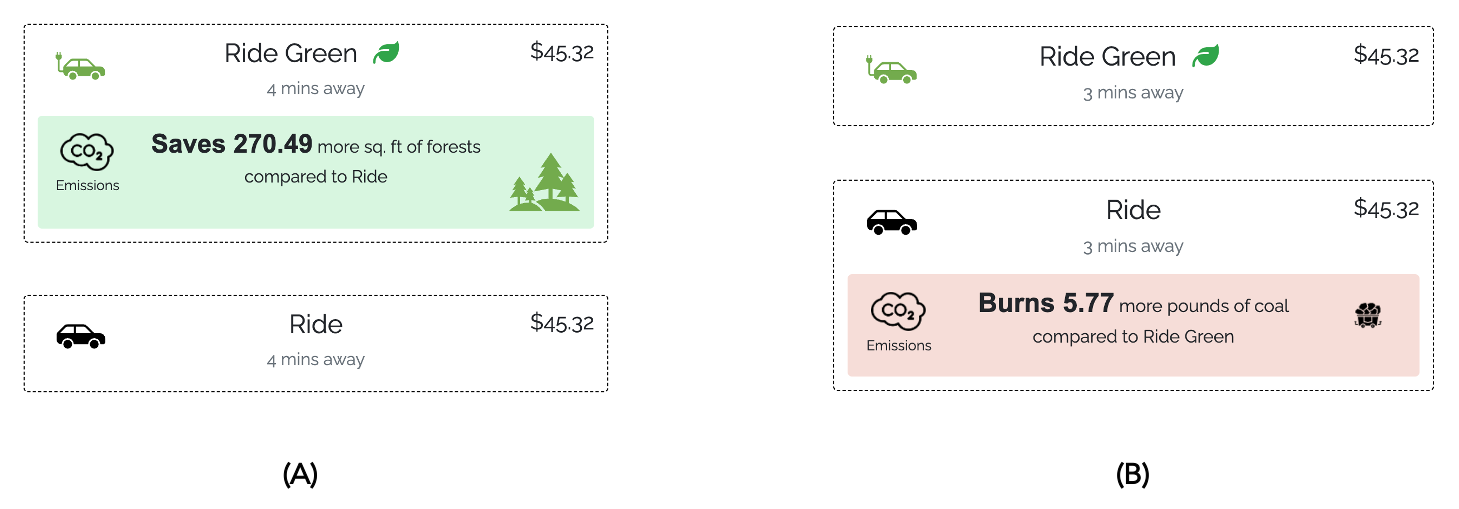}
        \caption{\label{fig:val:plus} Positive Valence}
        \Description{Example of positive valence framing. Two ride
          sharing options, with a Ride Green option on top that costs
          $\$45.32$ and Ride option on bottom that costs
          $\$45.32$. The Ride Green option indicates ``Saves 270.49
          more square feet of forests compared to Ride''. The text ``Saves 270.49'' is written in bold.}
    \end{subfigure}
    \begin{subfigure}[b]{0.75\columnwidth}
        \centering
        \includegraphics[trim=800 100 0 0,clip,width=\columnwidth]{figures/studyvalence.png}
        \caption{\label{fig:val:minus} Negative Valence}
        \Description{Example of negative valence framing. Two ride sharing options, with a Ride Green option on top that costs $\$45.32$ and Ride option on bottom that costs $\$45.32$. The Ride option indicates ``Burns 5.77 more pounds of coal compared to ride green''. The text ``Burns 5.77'' is written in bold.}
    \end{subfigure}
    \caption{Example screenshots of the interface tested in Study~{2b}. %%{\ref{study:2:valence}}. 
Participants chose between two rides (a green ride and a regular ride) similar to Study~\ref{study:1}, but were now shown interventions that were either framed positively (Figure~\ref{fig:val:plus}) or negatively (Figure~\ref{fig:val:minus}).}
\label{fig:studyvalence}
\end{figure}

Second, people's choices can be influenced by the valence of the information they are presented.
Research into human decision making consistently finds that people are more averse to losses then they are motivated by potential gains~\cite{schwartz1977normative, litovsky2022loss}.
This tendency has a strong implication for the display of carbon information since messaging is often framed around either gains (e.g., amount of carbon saved by not taking a trip) or losses (e.g., equivalent square feet of forest lost by a trip).
Study~\aptLtoX[graphic=no,type=env]{2b}{\ref{study:2:valence}} further examines whether framing equivalencies around losses or gains has a differential influence on ride-sharing choices.
Here, we hypothesize that
\aptLtoX[graphic=no,type=env]{% Coding for XML/HTML generation
\begin{enumerate*}%[label=\textbf{H\ref{study:2}.\arabic*},resume]
\item[\textbf{H2.4}]\label{hypo:2:valence}\textit{people are more likely to choose green options for negatively framed carbon equivalencies.}
\end{enumerate*}}{
\begin{enumerate*}[label=\textbf{H\ref{study:2}.\arabic*},resume]
\item\label{hypo:2:valence}\textit{people are more likely to choose green options for negatively framed carbon equivalencies.}
\end{enumerate*}}

\begin{figure}
    \centering
    \includegraphics[width=.75\columnwidth]{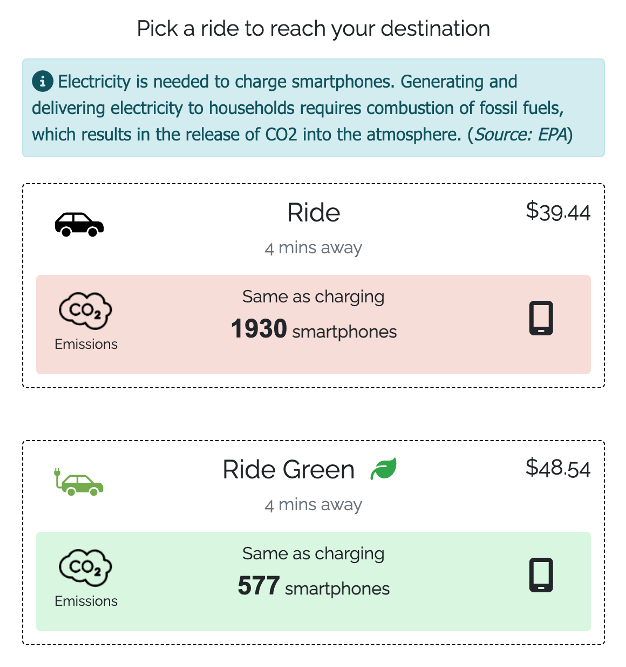}
    \caption{Example screenshots of the interface tested in Study~{2c}. %%%{\ref{study:2:explain}}. 
Participants chose between two rides (a green ride and a regular ride) similar to Study~\ref{study:1}, but now each interventions had a text blurb explaining how the equivalency (here, smartphones charging) relate to the carbon emissions from the two rides.}
    \Description{Example of additional information framing. Two ride sharing options, with a Ride option on top that costs $\$39.44$ and Ride Green option on bottom that costs $\$48.54$. The Ride option indicates ``Same as charging 1930 smartphones'' with the text ``1930 smartphones'' written in bold. The Ride Green option indicates ``Same as charging 577 smartphones'' with the text ``577 smartphones'' written in bold. Above the two options a blue banner show a small lower case ``i'' information icon with the text ``Electricity is needed to charge smartphones. Generating and delivering electricity to household requires combustion of fossil fuels, which results in the release of CO2 into the atmosphere (Source: EPA)''.}
\label{fig:studyinfo}
\end{figure}

Finally, people are less likely to choose options with information that is ambiguous, a phenomenon known as an \emph{ambiguity aversion}~\cite{fox1995ambiguity, ellsberg1961risk}.
In the context of carbon information, it is possible that equivalencies are not familiar to people and thus reduce the likelihood that they will take the information into consideration when making their decisions.
Study~\aptLtoX[graphic=no,type=env]{2c}{\ref{study:2:explain}} examines whether providing a clear explanation about how each equivalency is calculated increases it's effectiveness.
Our last hypothesis is that
\aptLtoX[graphic=no,type=env]{% Coding for XML/HTML generation
\begin{enumerate*}%[label=\textbf{H\ref{study:2}.\arabic*},resume]
\item[\textbf{H2.5}]\label{hypo:2:detail} \textit{people are more likely to choose green options with more detailed explanations about how the equivalency relates to the emissions from the ride}.
\end{enumerate*}}{
\begin{enumerate*}[label=\textbf{H\ref{study:2}.\arabic*},resume]
\item\label{hypo:2:detail} \textit{people are more likely to choose green options with more detailed explanations about how the equivalency relates to the emissions from the ride}.
\end{enumerate*}}

\subsubsection{Experiment Setup}

We ran Study~\ref{study:2} to examine the influence of (Study \aptLtoX[graphic=no,type=env]{2a}{\ref{study:2:social}}) social framing, (Study \aptLtoX[graphic=no,type=env]{2b}{\ref{study:2:valence}}) information valence, and (Study~\aptLtoX[graphic=no,type=env]{2c}{\ref{study:2:explain}}) additional information on ride green choices. 

The experimental set up used in Study~\ref{study:2} was identical to the set up in Study~\ref{study:1}; participants were asked to make 30 choices between two ride-sharing options randomly drawn from the the same maps, distances, and price differences.

Study~\aptLtoX[graphic=no,type=env]{2a}{\ref{study:2:social}} explored the influence of social
information on ride-sharing choices.
The wording added to each trial was randomly drawn from the following conditions: 
\begin{description}
\item[Popular:] participants were told that between \emph{75--99\%} of users chose the green option.
\item[Unpopular:] participants were told that between \emph{1--25\%} of users chose the green option.
\item[Positive collective impact:] information about the carbon equivalent saved by the \emph{Ride-Green} option (see Figure~\ref{fig:studysocialcip}).
  This condition compared the impact of collective framing on information about pounds of coal, gallons of gasoline, daily home energy usage, trees, square feet of forest, and bags of waste.
\item[Negative collective impact:] information about the aggregate carbon equivalent emitted by the \emph{Ride} option using the same units as in the positive collective impact condition (see Figure~\ref{fig:studysocialcin}).
\item[Individual:] This condition replicated Study~\ref{study:1}: i.e., interventions showed the individual impact in terms of the following equivalencies: pounds of coal, gallons of gasoline, daily home energy usage, trees, square feet of forest, and bags of waste. 
\item[Positive dynamic social norms:] information about the percentage of new passengers who had chosen the \emph{Ride-Green} option (see Figure~\ref{fig:studysocialdynp}).  The percentage was either >75\% (popular) or <25\%(unpopular).
\item[Negative dynamic social norms:] information about the percentage of passengers who had recently stopped choosing \emph{Ride} option (see Figure~\ref{fig:studysocialdynn}). The percentage was either >75\% (popular) or <25\%(unpopular).
\end{description}

Study~\aptLtoX[graphic=no,type=env]{2b}{\ref{study:2:valence}} specifically examined the role of
valence on green ride-sharing choices (see Figure~\ref{fig:studyvalence}).
On each trial, participants were given information about two emission equivalencies (gasoline, coal) and two sequestering equivalencies (forests, trees) framed in one of two ways:
\begin{description}
\item[Positive valence:] information was provided about the emission equivalencies saved by the \emph{Ride-Green} choice.
\item[Negative valence:] information was provided about the additional equivalencies burnt (for gasoline and coal) or destroyed (for forests and trees) by the \emph{Ride} choice.
\end{description}

Last, Study~\aptLtoX[graphic=no,type=env]{2c}{\ref{study:2:explain}} examined whether adding additional
information about each equivalency would increase their
effectiveness (see Figure~\ref{fig:studyinfo}). The equivalencies used in this study were:
% (1) pounds of \COTwo{}, (2) pounds of coal, (3) gallons of gasoline, (4) number of smartphones charged, (5) home energy usage, (6) trees saved, (7) square feet of forests saved, and (8) bags of waste.
pounds of \COTwo{}, pounds of coal, gallons of gasoline, number of smartphones charged, home energy usage, trees saved, square feet of forests saved, and bags of waste.

\changemarker{Participants, after completing 30 trials, answered a survey questionnaire that included questions about their demographics, attitudes towards climate change, renewable energy, and electric vehicles (we used the same survey questionnaire for Studies~\ref{study:3} and~\ref{study:4:rental} as well). This was followed by a few supplemental, open-ended questions about their choices (see Appendix~\ref{sec:appendix_study_2}).}

\subsubsection{Analysis}

Similar to Study~\ref{study:1}, we measured the effect of each intervention using a logistic mixed-effects model (LME) that predicted the likelihood of making \emph{Ride Green} choices. We used the difference in price between the \emph{Ride Green} and \emph{Ride} options as well as intervention types as fixed effects; we used random intercepts for participants and city.

\changemarker{
  \paragraph{Limitations:} For Study~\aptLtoX[graphic=no,type=env]{2b}{\ref{study:2:valence}}, the focus was on understanding the role of valence in the messaging of the equivalencies, which, at times, was not straightforward.
  We acknowledge that the wording, even though intended to communicate relative benefits or drawbacks, can be misinterpreted for absolute effects.
  For example, ``Ride Green saves 270.49 more sq.\ ft.\ of forests than Ride'' does not mean that \emph{Ride} saves any sq.\ ft.\ of forests, nor does it mean that \emph{Ride Green} saves any forest cover; this essentially meant that sequestering the emissions from both the rides would require growing new forest cover, where \emph{Ride Green} would need fewer sq.\ ft.\ of forest cover than \emph{Ride}.
  However, for the purpose of this study, we sacrificed accuracy to test whether words like \emph{save} or \emph{destroys} impact the effectiveness of these equivalencies. We did not explicitly gauge whether participants found the messaging to be confusing---a potential limitation of the study.}

\subsubsection{Participants}

For each study, we recruited 500 unique US-based participants from Prolific (Study~\aptLtoX[graphic=no,type=env]{2a}{\ref{study:2:social}}: 48.8\% men, 49.0\% women, 2\% genderqueer, agender, or non-binary, 0.2\% preferred not to answer, mean age = 38, SD = 13; 
Study~\aptLtoX[graphic=no,type=env]{2b}{\ref{study:2:valence}}: 44.5\% men, 51.9\% women, 3.4\% genderqueer, agender, or non-binary, 1\% preferred not to answer, mean age = 38, SD = 14; 
Study~\aptLtoX[graphic=no,type=env]{2c}{\ref{study:2:explain}}: 41.0\% men, 56.4\% women, 2.2\% genderqueer, agender, or non-binary, 0.4\% preferred not to answer, mean age = 38, SD = 13).

\subsubsection{Findings}
\paragraph{Popularity influenced participant choices but dynamic social norms had mixed effects.}

\begin{figure*}
  \centering
  \begin{subfigure}[b]{\columnwidth}
    \centering
    \includegraphics[width=\columnwidth]{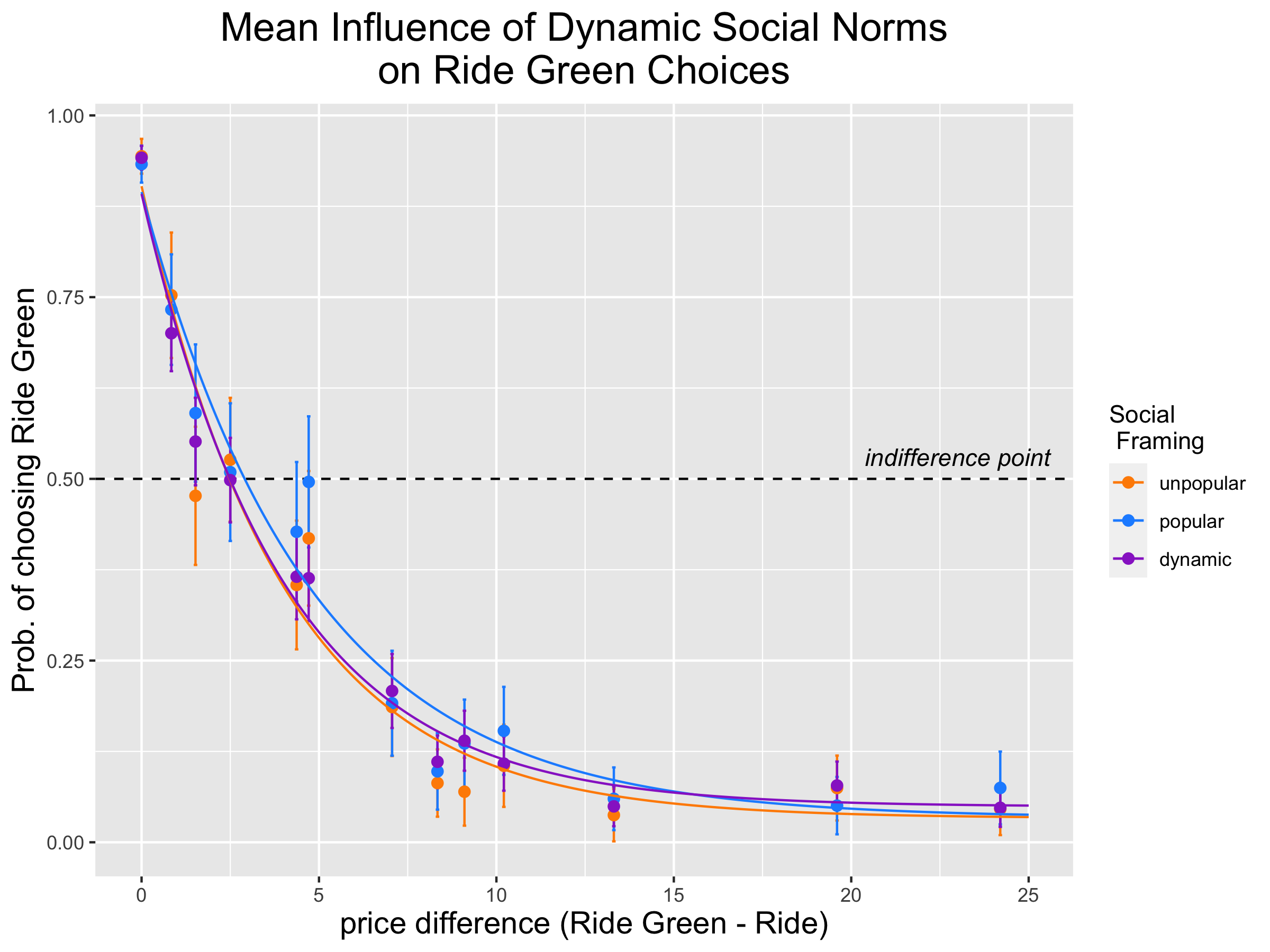} % ChkTeX 25
    \caption{\label{fig:social_curve}}
    \Description{An exponential decay curve that shows the probability of choosing ride green on Y-axis (0--1) and the price difference between Ride Green and Ride on the X-axis (\$0--\$25) for three different intervention types (popular, unpopular, and average of all dynamic social norms). The curves for each of the interventions intersect the indifference point at 0.50 on the Y-axis; the intersection point of the unpopular and dynamic social norm curve are the leftmost, whereas the popular curve intersects further right on the indifference line. This indicates people are willing to pay more for Ride Green in all conditions, but more so when the option is popular.}
  \end{subfigure}  
  \hspace{1pc}
  \begin{subfigure}[b]{\columnwidth}
    \centering
    \includegraphics[width=\columnwidth]{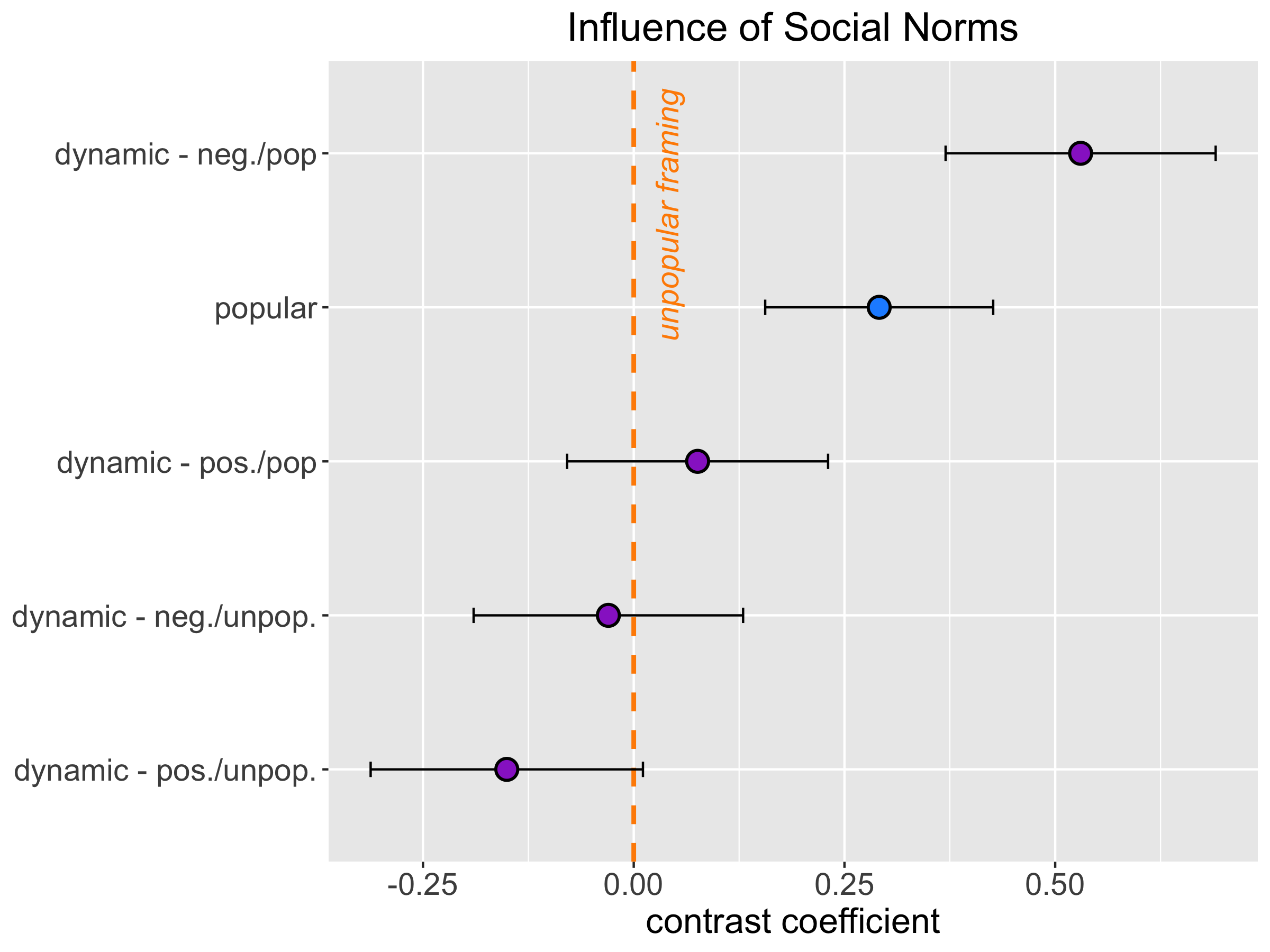}
    \caption{\label{fig:social_beta}}
    \Description{A chart displaying the probability of participants picking a green ride in log odd scale. The X-axis has beta coefficients for the different interventions. The Y-axis has all the different interventions sorted according to the beta coefficients in descending order. The beta coefficients are compared to the unpopular condition, denoted by a dashed vertical line. Dynamic social norms presented as many people leaving the Ride option was the most effective social intervention followed by the popular framing, which were both significantly better than the unpopular interventions. Dynamic social norms that highlighted either low number of people leaving the Ride option, a high number of people choosing the Ride Green options, or a low number of people choosing the Ride Green option did not perform better than the unpopular framing.}
  \end{subfigure}
  \caption{Influence of social framing on
    ride-sharing choices in Study~{2a}.  %%%{\ref{study:2:social}}. 
(a) Choice curves indicating the probability that participants chose the \emph{Ride-Green} option as a function of the price difference between \emph{Ride-Green} and \emph{Ride}. Participants were more likely to choose the \emph{Ride-Green} option if they saw that 75--99\% of other riders chose this option (popular; blue) compared to if they saw that 1--25\% of users chose the option (unpopular; orange) or where told about dynamic changes in rider preferences (dynamic; purple). Points indicate mean choice probabilities across participants, error-bars indicate 95\% confidence intervals, and lines indicate best fitting exponential decay curves. (b) Logistic mixed-effects contrast coefficients comparing the influence of the ``popular'' i dynamic social norm interventions compared to the ``unpopular'' condition (dashed orange line). For the Dynamic social norm y-axis labels, ``neg.'' and ``pos.'' correspond to negative and positive framings respectively, and ``pop.'' and ``unpop.'' correspond to popular and unpopular framings. Points indicate contrast beta weights and error bars indicate 95\% confidence intervals.}
\label{fig:social}
\end{figure*}

Overall in Study~\aptLtoX[graphic=no,type=env]{2a}{\ref{study:2:social}}, participants were slightly more likely to choose the \emph{Ride-Green} option when they were told that 75--99\% of riders chose that option (popular) than if 1--25\% of riders chose the option (unpopular; LME contrasting the popular vs unpopular condition: $\beta_{\mathrm{popular}}=0.29, z=2.15, p=0.031$; Figure~\ref{fig:social}).
This effect provides support for~\aptLtoX[graphic=no,type=env]{\textbf{H2.1}}{\ref{hypo:2:pop}} and is consistent with the small effect of popularity observed in Study~\ref{study:1}.
However, interventions that highlighted dynamic changes in social norms had a less consistent influence on participant choices.
The best performing dynamic social norm intervention was when participants were told that 75--99\% fewer riders chose the \emph{Ride} option (LME contrasting the dynamic condition with a high proportion fewer \emph{Ride} riders to the unpopular condition: $\beta_{\mathrm{dynamic}}=0.53, z=3.30, p=0.001$; Figure~\ref{fig:social_beta}).
All other dynamic social norm conditions did not differ from the
unpopular condition (LME contrasting the remaining dynamic social norm
conditions with the unpopular condition; all $\beta$ between $-0.15$
and $0.08$, all $|z|<0.93$, all $p>0.351$; Figure~\ref{fig:social_beta}).
We therefore find partial support for~\aptLtoX[graphic=no,type=env]{\textbf{H2.2}}{\ref{hypo:2:norms}}, but only under conditions where a high percentage of people are seen to be \emph{switching away} from a particular choice. The self-reported open-ended responses from participants showed that many of them did not find the interventions that showed ``\% of participants'' (i.e., both popularity or dynamic trends) appealing.

\begin{figure*}
  \centering
  \begin{subfigure}[b]{\columnwidth}
    \centering
    \includegraphics[width=\columnwidth]{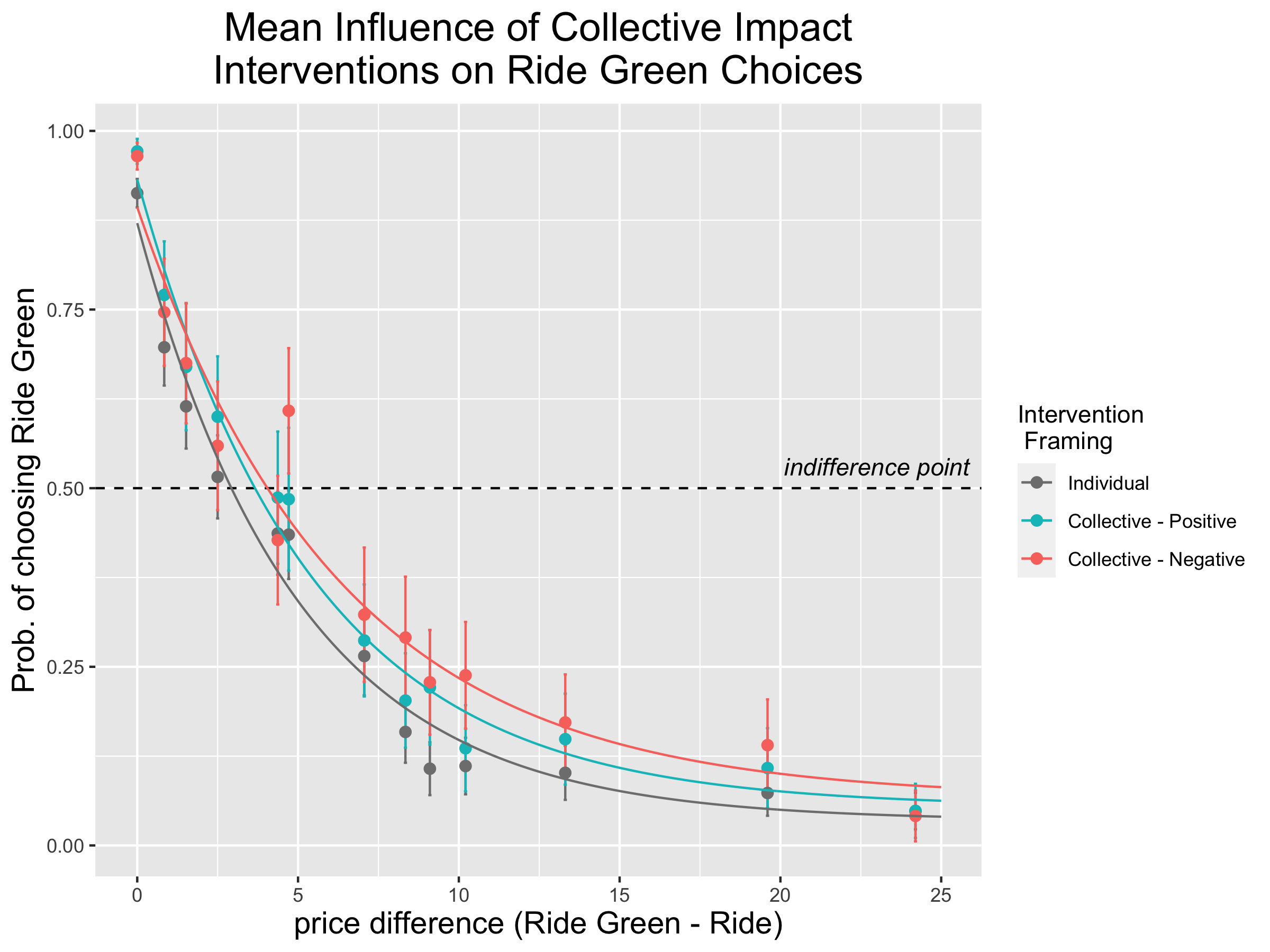} % ChkTeX 25
    \caption{\label{fig:collective_curve}}
    \Description{An exponential decay curve that shows the probability of choosing ride green on Y-axis (0--1) and the price difference between Ride Green and Ride on the X-axis (\$0--\$25) for three different intervention types (individual, collective--positive, and collective--negative). The curves for each of the interventions intersect the indifference point at 0.50 on the Y-axis; the intersection point of the individual intervention is the leftmost, whereas the collective--positive and collective--negative interventions intersect further right on the indifference line. This indicates people are willing to pay more for Ride Green in all conditions, but more so under the collective framing.}
  \end{subfigure}  
  \hspace{1pc}
  \begin{subfigure}[b]{\columnwidth}
    \centering
    \includegraphics[width=\columnwidth]{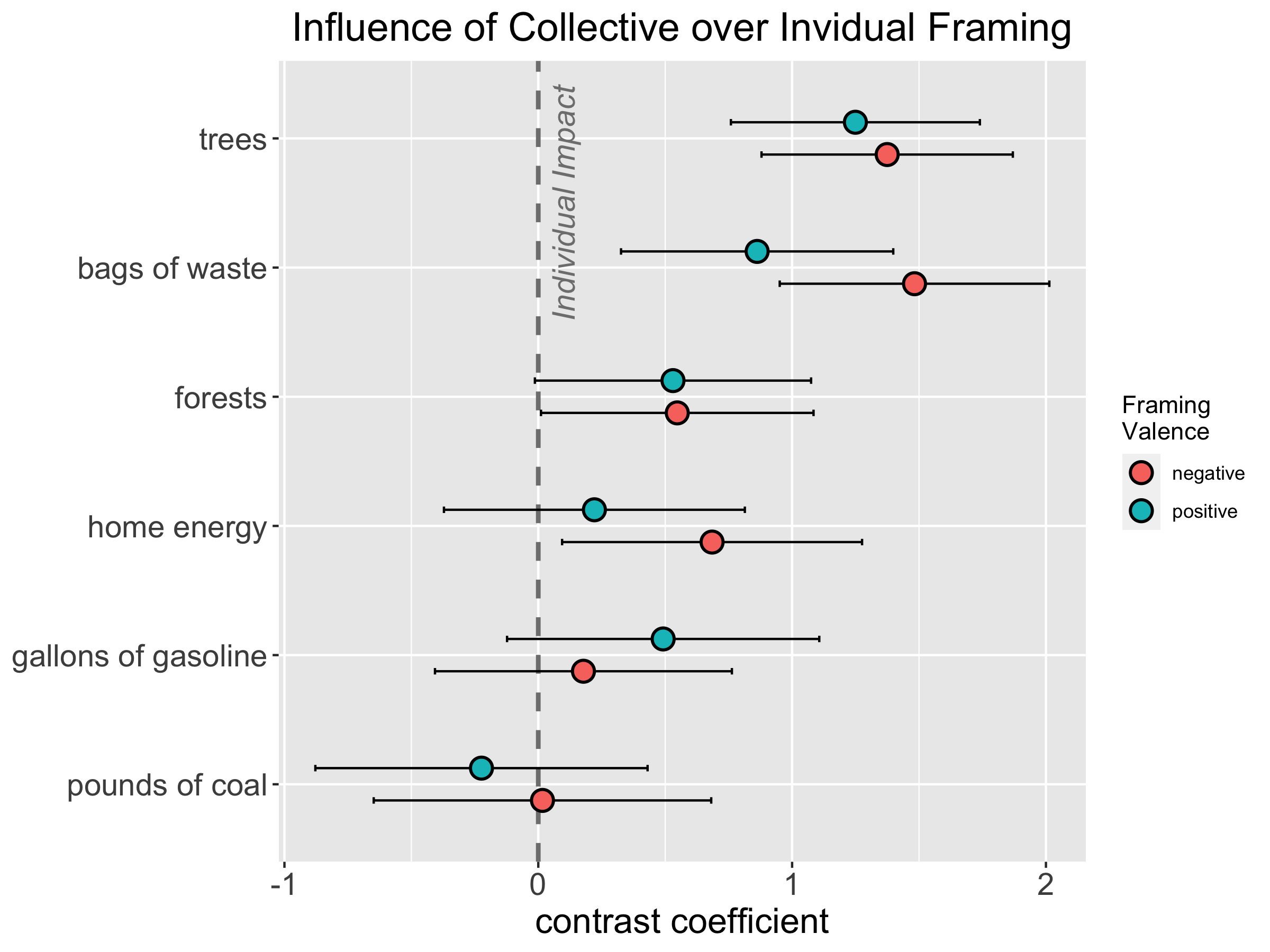}
    \caption{\label{fig:collective_beta}}
    \Description{A chart displaying the probability of participants picking a green ride in log odd scale. The X-axis has beta coefficients for the different collective interventions (both negative and positive). The Y-axis has all the different interventions sorted according to the beta coefficients in descending order. The beta coefficients are compared to the individual intervention, denoted by a dashed vertical line. Both the negative and positive collective interventions were significantly effective for trees, bags of waste, and forests in descending order along the Y-axis. Collective interventions were not effective for home energy, gallons of gasoline, and pounds of coal in descending order along the Y-axis.}
  \end{subfigure}
  \caption{Influence of collective social framing on ride-sharing choices. (a) Choice curves indicating the probability that participants chose the \emph{Ride-Green} option as a function of the price difference between \emph{Ride-Green} and \emph{Ride}. Participants were more likely to choose the \emph{Ride-Green} option if they were given information about the collective emissions generated by people who chose the \emph{Ride} option (collective--negative; red) or the collective emission reductions that resulted from people choosing the \emph{Ride-Green} option (collective--positive; teal) than if they were only given emission information resulting from their choices (individual; grey). (b) Logistic mixed-effects contrast coefficients comparing the change in likelihood that participants chose the \emph{Ride-Green} in the collective--negative (red) and collective--positive (teal) conditions compared to the individual condition (grey dashed line). Points indicate contrast beta weights and error bars indicate 95\% confidence intervals.}
\label{fig:collective}
\end{figure*}

\paragraph{Collective impact increases the likelihood of choosing green.}
Information about collective impact in Study~\aptLtoX[graphic=no,type=env]{2a}{\ref{study:2:social}} had a consistent influence on participant ride-sharing choices.
Interventions that highlighted either the negative or positive collective impact on \COTwo{} equivalencies made people overall more likely to choose \emph{Ride-Green} options (LME contrasting positive and negative collective impact interventions compared to interventions centered around individual action: $\beta_{\mathrm{collective-positive}}=0.62, z=6.20, p<0.001$; $\beta_{\mathrm{collective-negative}}=0.75, z=7.53, p<0.001$.
See Figure~\ref{fig:social}).
The collective framing primarily influenced responses by increasing the effectiveness of equivalencies that were not effective under the individual framing. Specifically, both positive and negative collective framing increased the effectiveness of the trees and waste equivalencies, two equivalencies that were among the least effective equivalencies when given under an individual framing in Study~\ref{study:1} (LME contrasting the influence of positive and negative collective impact interventions compared to interventions framed around individual action: all $\beta$s for the waste intervention $>0.86$, all $z>3.15$, all $p<0.002$; all $\beta$s for the tree intervention $>1.24$, all $z>4.99$, all $p<0.001$).
Additionally, the negative collective framing provided marginal increases to the effectiveness of the home energy and forest interventions ($\beta_{\mathrm{collective-negative-home}}=0.68, z=2.27, p=0.023$; $\beta_{\mathrm{collective-positive-forest}}=0.55, z=1.99, p=0.046$). The collective framing did not increase the effectiveness of any other interventions (all $\beta<0.53$, all $z<1.92$, all $p>0.055$.
See Figure~\ref{fig:collective_beta}).
Together these results provide support for~\aptLtoX[graphic=no,type=env]{\textbf{H2.3}}{\ref{hypo:2:collective}} in that framing equivalencies around the collective action of others increases their efficacy as part of certain interventions.
More specifically, collective framing provided a boost to equivalency interventions that were ineffective when framed around individual action (e.g., tree and waste equivalencies).

The self-reported responses suggested that participants had mixed feelings towards collective impact information. While some participants did not see any value in learning about the collective impact, others saw its value: 
\begin{quote}
``It definitely made me more likely to pick the green option. I think once the impact is put in terms of group behavior and not just individual, its easier to see how those decisions lead to better environmental outcomes. It feels like working within a collective that can make more of a difference together than just I would individually.'' (P1337)
\end{quote}

Most participants felt their actions were insignificant compared to wealthy individuals, big corporations, and countries: ``\emph{I do what I can, but I'm also aware that a small number of corporations are responsible for the vast majority of pollution, and my singular efforts can't combat that.} (P1199)''

\begin{figure*}
  \centering
  \begin{subfigure}[b]{\columnwidth}
    \centering
    \includegraphics[width=\columnwidth]{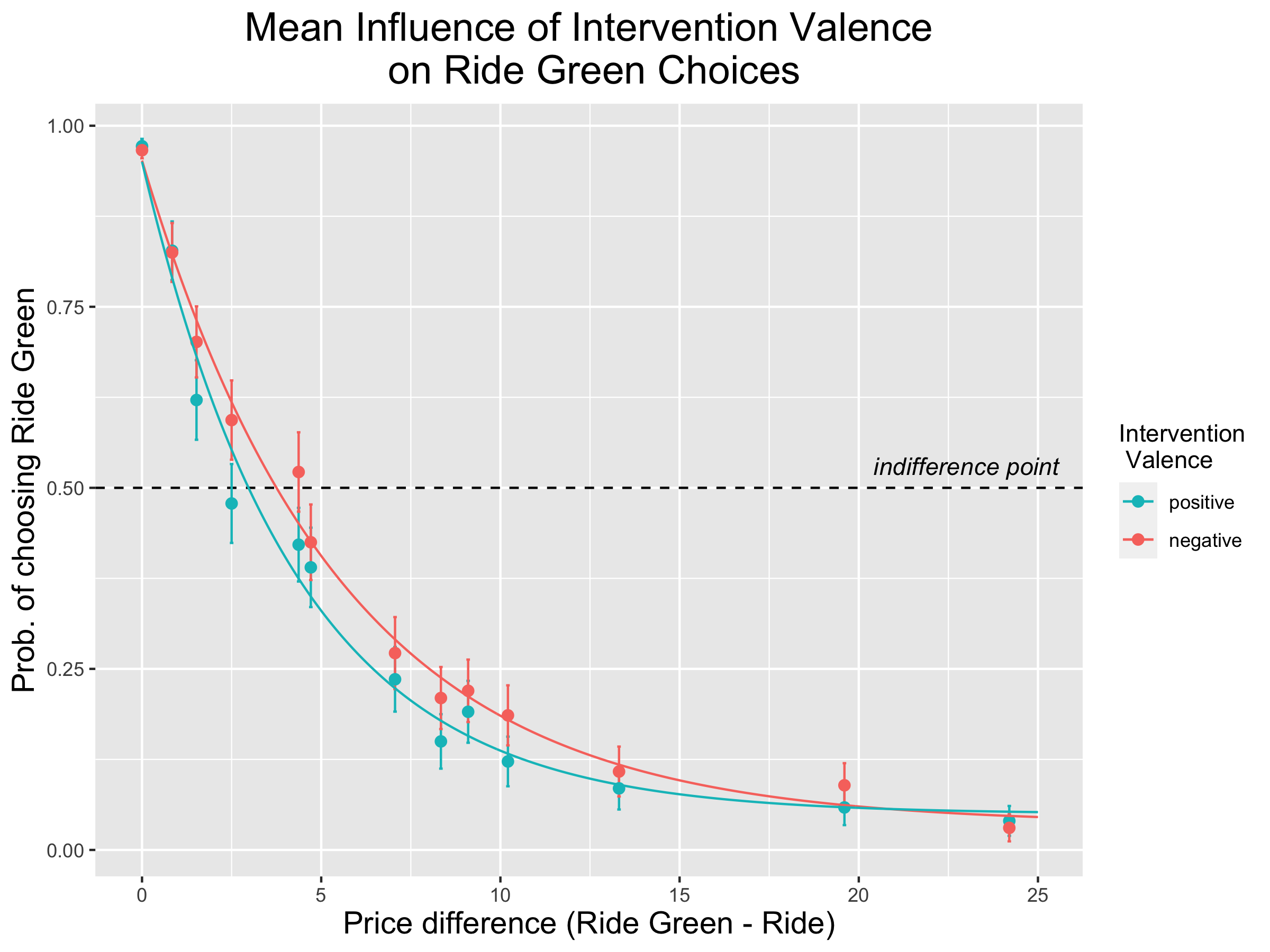} % ChkTeX 25
    \caption{\label{fig:valence_curve}}
    \Description{An exponential decay curve that shows the probability of choosing ride green on Y-axis (0--1) and the price difference between Ride Green and Ride on the X-axis (\$0--\$25) for two different intervention types (positive and negative valence). The curves for each of the interventions intersect the indifference point at 0.50 on the Y-axis; the intersection point of the positive valence intervention is the leftmost, whereas the negative valence intervention curve intersects further right on the indifference line. This indicates people are willing to pay more for Ride Green options under negatively rather than positively valenced conditions.}
  \end{subfigure}  
  \hspace{1pc}
  \begin{subfigure}[b]{\columnwidth}
    \centering
    \includegraphics[width=\columnwidth]{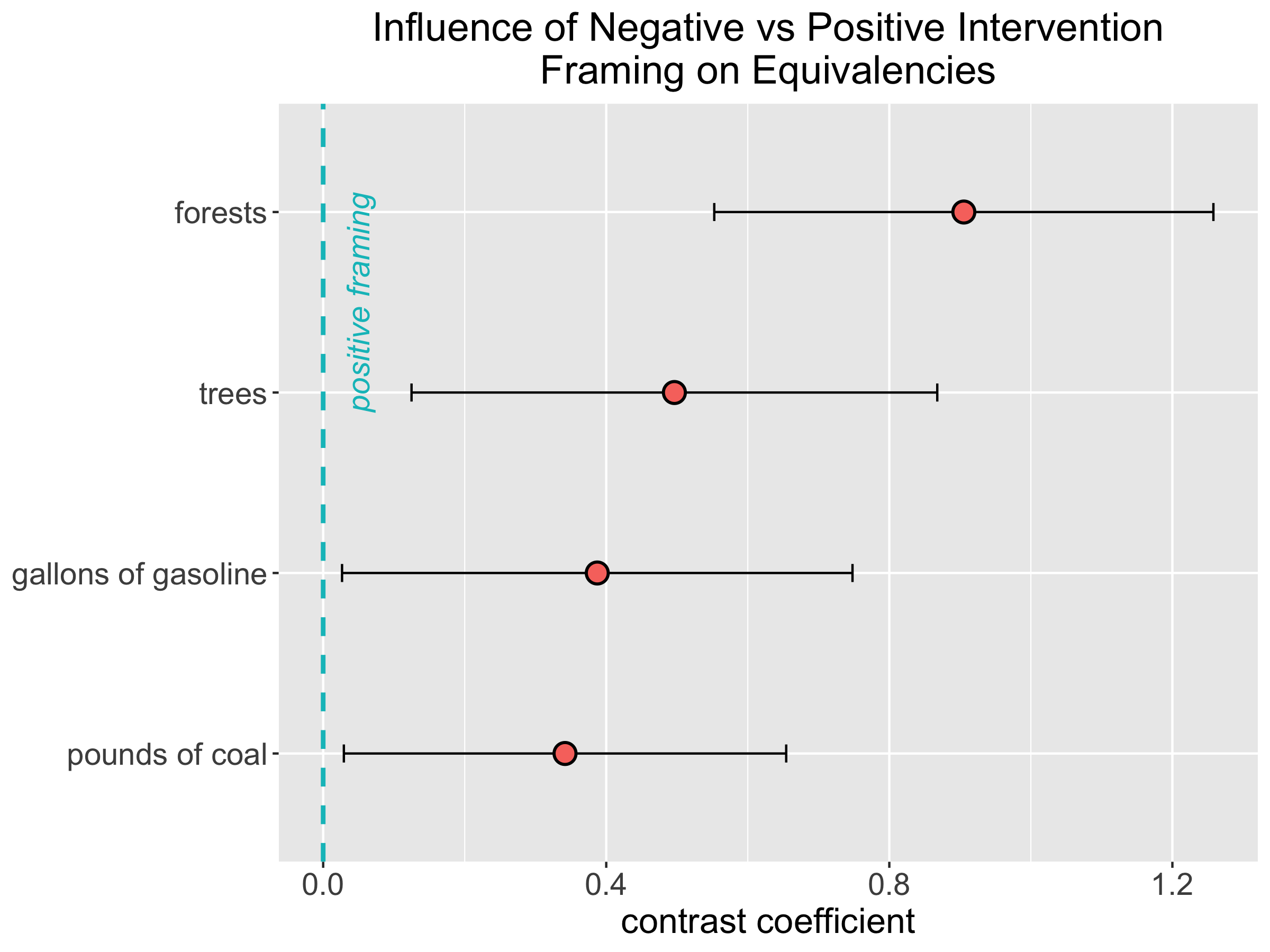}
    \caption{\label{fig:valence_beta}}
    \Description{A chart displaying the probability of participants picking a green ride in log odd scale. The X-axis has beta coefficients for the different interventions. The Y-axis has all the different interventions sorted according to the beta coefficients in descending order. The beta coefficients are compared to the popular valence condition, denoted by a dashed vertical line. Forests, trees, gallons of gasoline, and pounds of coal were all significantly higher in the negatively valenced framing compared with the positively valenced framing (in descending order along the Y-axis)}
  \end{subfigure}
  \caption{Influence of valence on ride-sharing choices. (a) Choice curves indicating the probability that participants chose the \emph{Ride-Green} option as a function of the price difference between \emph{Ride-Green} and \emph{Ride}. Participants were more likely to choose the \emph{Ride-Green} option if they were informed about the emissions they would produce by choosing the \emph{Ride} option (negative; red) compared to information about the emissions they would save by choosing the \emph{Ride-Green} option (positive; teal). (b) Logistic mixed-effects contrast coefficients comparing the change in likelihood that participants chose the \emph{Ride-Green} in the negative valence condition (red) compared with the positive valence condition (teal dashed line). Points indicate contrast beta weights and error bars indicate 95\% confidence intervals.}
\label{fig:valence}
\end{figure*}

% \subsubsection{Study 2b: Negatively valenced wording influences choices more than positively valenced wording}
\paragraph{Negative valence wording influences choices more than
  positive valence.}
Study~\aptLtoX[graphic=no,type=env]{2b}{\ref{study:2:valence}} examined the influence of wording valence
on the the effectiveness of different carbon equivalency
interventions.
Interventions with negatively valenced wording (i.e., that highlighted
the additional emission equivalents produced by the \emph{Ride} option)
increased the likelihood that people chose the \emph{Ride-Green}
choice over positively valenced wording (i.e., that highlighted the
emission equivalents saved by the \emph{Ride-Green} option; LME
contrasting the influence of negatively valenced interventions over
positive interventions: $\beta_{\mathrm{negative}}=0.53, z=7.00,
p<0.001$; Figure~\ref{fig:valence_curve}).
Moreover, negatively valenced wording increased the likelihood of \emph{Ride-Green} choices in all tested equivalency conditions (all $\beta>0.34$, all $z>2.10$, all $p<0.033$; Figure~\ref{fig:valence_beta}).
These results support~\aptLtoX[graphic=no,type=env]{\textbf{H2.4}}{\ref{hypo:2:valence}}, indicating that highlighting the emissions produced by a less eco-friendly option can have a stronger repelling effect than the attractive effect of highlighting the carbon savings provided by a greener option.

This finding was also supported by self-reported responses: 
\begin{quote}
   ``I think seeing negative impacts made me feel more guilty- like I should choose the environmentally-responsible option. It was great seeing the benefits of my actions, but I think most people wouldn't care about this as much. Guilt tripping is a successful tactic here.'' (P0686)
\end{quote}

At the same time, participants also complained about the shaming effect of these statements: ``\emph{The “negative” wording kind of annoyed me. Especially when the much cheaper one was described so negatively. I resented feeling guilted.}'' (P0870)

\begin{figure*}
  \centering
  \begin{subfigure}[b]{\columnwidth}
    \centering
    \includegraphics[width=\columnwidth]{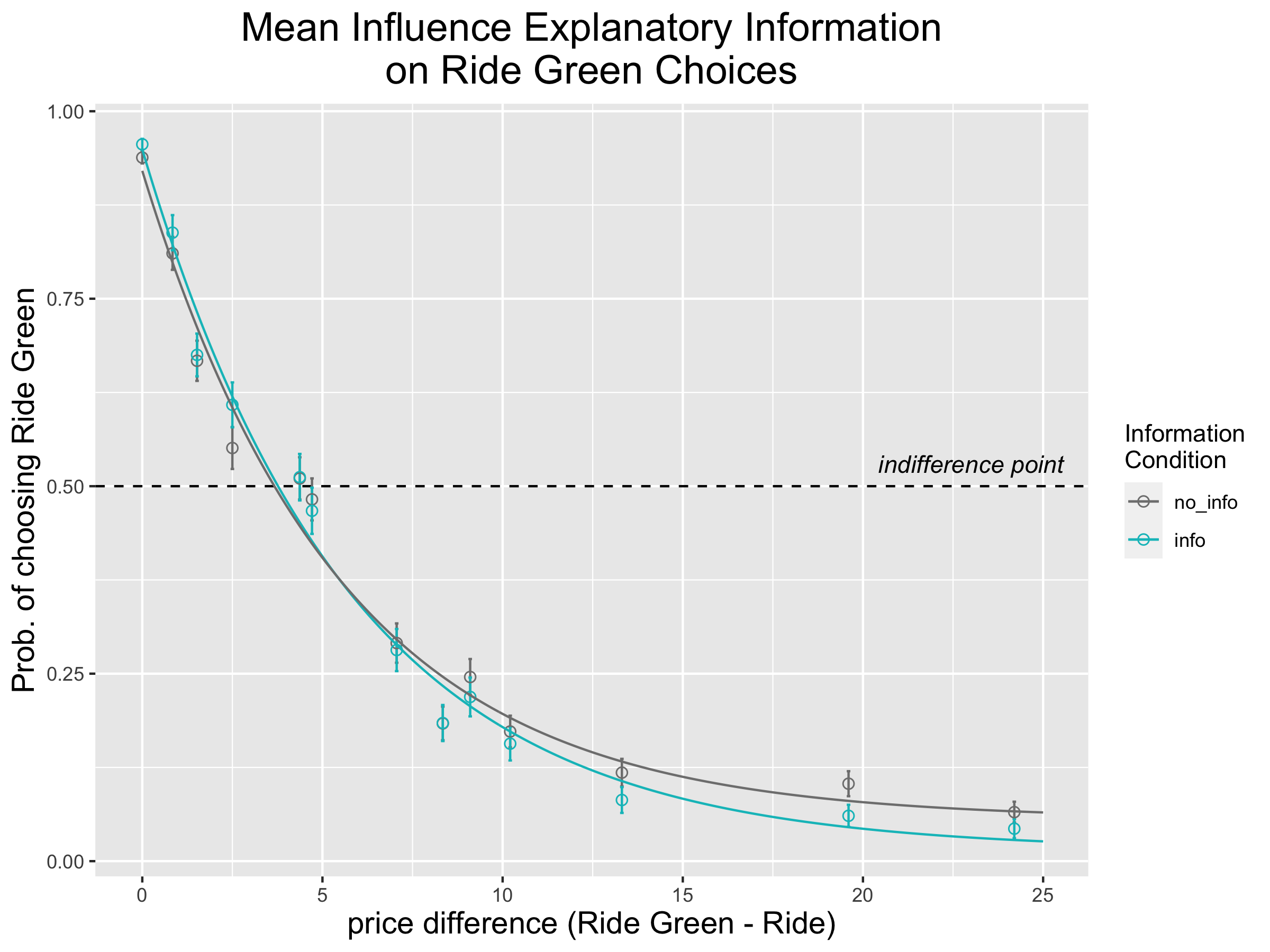} % ChkTeX 25
    \caption{\label{fig:info_curve}}
    \Description{An exponential decay curve that shows the probability of choosing ride green on Y-axis (0--1) and the price difference between Ride Green and Ride on the X-axis (\$0--\$25) for three different intervention types (info and no info). The curves for each of the interventions intersect the indifference point at 0.50 on the Y-axis; the intersection point of the info and no info curves are approximately identical, indicating no differences between intervention conditions.}
  \end{subfigure}  
  \hspace{1pc}
  \begin{subfigure}[b]{\columnwidth}
    \centering
    \includegraphics[width=\columnwidth]{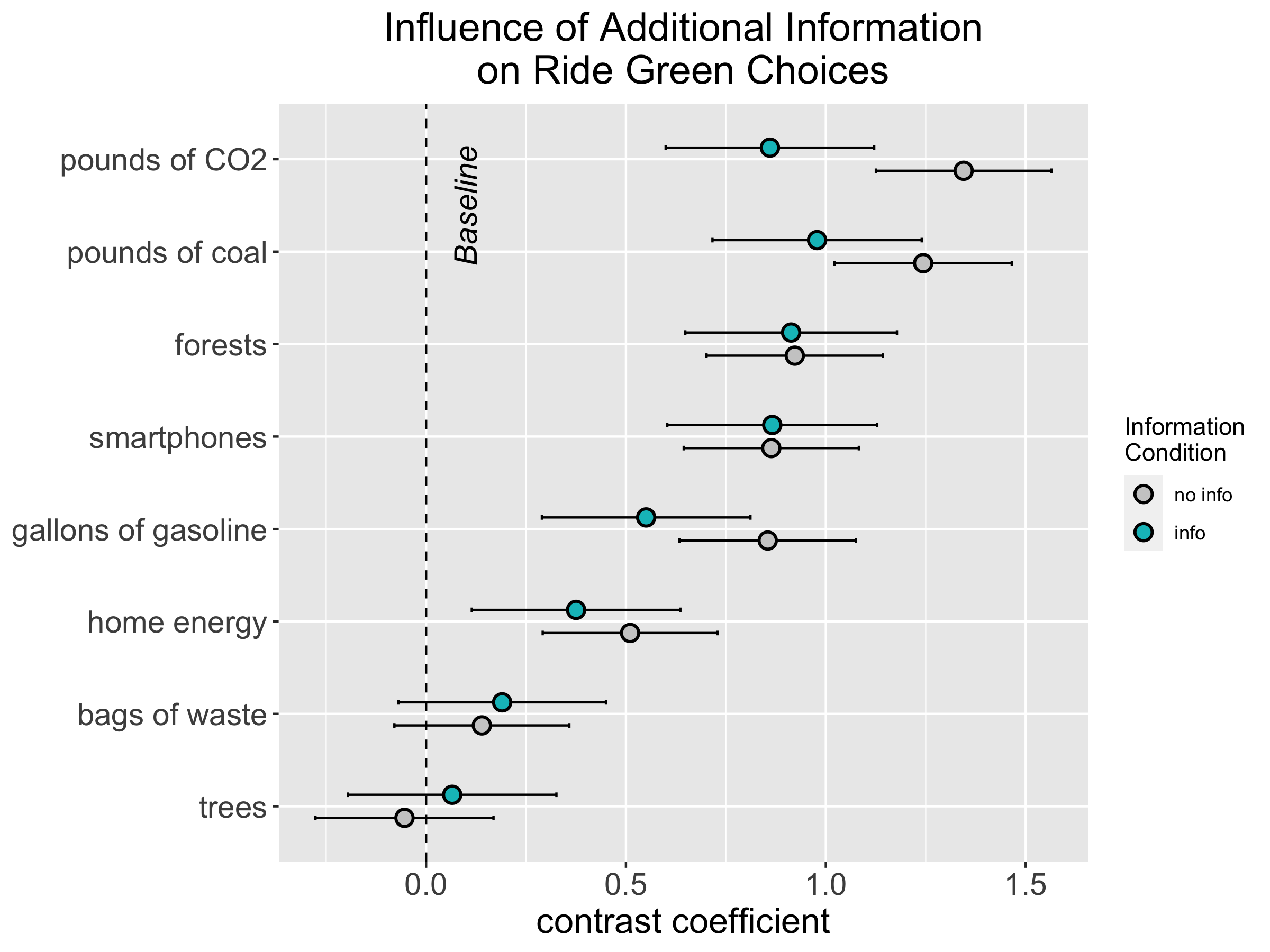}
    \caption{\label{fig:info_beta}}
    \Description{A chart displaying the probability of participants picking a green ride in log odd scale. The X-axis has beta coefficients for the different interventions. The Y-axis has all the different interventions sorted according to the beta coefficients in descending order. The beta coefficients are compared to the baseline condition, denoted by a dashed vertical line. Pounds of CO2, pounds of coal, forests, smartphones, gallons of gasoline, and home energy all performed significantly better than baseline in descending order on the Y-axis. Bags of waste and trees were not significantly different from baseline, which are at the bottom of the Y-axis. None of the conditions differed between the info and no info conditions.}
  \end{subfigure}
  \caption{Influence of additional information on ride-sharing choices. (a) Choice curves indicating the probability that participants chose the \emph{Ride-Green} option as a function of the price difference between \emph{Ride-Green} and \emph{Ride}. Participants were no more likely to choose the \emph{Ride-Green} option when provided additional information about each equivalency (info; teal) than they were in Study 1 where no additional information was provided (no info; grey). (b) Logistic mixed-effects contrast coefficients comparing the change in likelihood that participants chose the \emph{Ride-Green} option compared to the baseline green leaf intervention (black dashed line) in both the info (teal) and no-info (grey) conditions. Points indicate contrast beta weights and error bars indicate 95\% confidence intervals.}
\label{fig:info}
\end{figure*}

% \subsubsection{Study 2c: Additional equivalency explanations do not influence the likelihood of choosing green ride-sharing options }
\paragraph{Equivalency explanations do not influence  choosing green ride-sharing options.}
Finally, Study~\aptLtoX[graphic=no,type=env]{2c}{\ref{study:2:explain}} examined whether providing additional information about carbon and carbon equivalencies improved their influence on participant choices.
Contrary to~\aptLtoX[graphic=no,type=env]{\textbf{H2.5}}{\ref{hypo:2:detail}}, adding additional information about equivalencies did not have an overall influence on the likelihood of \emph{Ride-Green} choices (LME contrasting the influence of additional information over conditions from Study~\ref{study:1} with no additional information: $\beta_{\mathrm{information}}=-0.02, z=-0.13, p=0.895$; Figure~\ref{fig:info_curve}).
When examining individual equivalencies, additional information slightly \emph{decreased} the influence of direct \COTwo{} emissions (LME contrasting the influence of additional information compared to no information for trials with \COTwo{} emissions: $\beta_{\mathrm{CO_2}}=-0.52$, $z=-2.22$, $p=0.026$).
However, additional information did not influence any of the other interventions (LME contrasting the influence of additional information compared to no information for each intervention: all $|\beta|<0.30$, all $|z|<1.81$, all $p>0.069$; Figure~\ref{fig:info_beta}). As such, we find little support for~\aptLtoX[graphic=no,type=env]{\textbf{H2.5}}{\ref{hypo:2:detail}} as additional information has little to no influence on participant choices. Some participants expressed feelings of guilt when presented with these explanations: 

\begin{quote}
  ``As a student, cost is my number one priority, so the explanations provided often just made me feel guilty even though I knew I would pick the cheapest option every time. If the prices were the same, then I'd pick the green option because I'd like to be kinder to the environment where it's feasible for me, so that's where the explanations were helpful.'' (P1729)
\end{quote}

%%%% sections/4.study2.tex ends here %%%%

%%%% sections/4.study3.tex starts here %%%%

\subsection{How People Think About \COTwo\ Emissions}
In other domains, target numbers have been shown to provide a touchstone that contextualizes information and helps people achieve their goals. For example, in US-based nutrition, the 1990 Nutrition Labeling and Education act standardized the target ``2,000 calories a day'' to contextualize numbers about food calories.
Likewise, the Yamasa Corporation introduced a benchmark of 10,000 steps a day to help contextualize healthy activity habits and promote their company's step counter. 
Although there is long-standing controversy around both of these examples (e.g., 2,000 calories a day represents the average US-based caloric intake rather than the calories in a healthy diet; 10,000 steps a day was created as part of a marketing campaign rather than from scientific studies; a more accurate target may be lower~\cite{lee2019association}), target numbers nevertheless provide reference points that people can use to determine how healthy a particular food might be, or how active they really are. 
Study~\aptLtoX[graphic=no,type=env]{3b}{\ref{study:3:target}} explores the influence of providing people with target emission numbers on ride-sharing choices. We additionally explore interventions aimed at familiarizing people with carbon emission values. 

\begin{study}\label{study:3}
Examines the role of relative and absolute carbon values on participant ride-share choices.
\aptLtoX[graphic=no,type=env]{% Coding for XML/HTML generation
\begin{enumerate}%[label=\ref{study:3}\alph*]
\item[3a]\label{study:3:rel_ab} Measures the influence of relative versus absolute \COTwo\ emissions values.
\item[3b]\label{study:3:target} Examines whether providing carbon targets nudges people to consider absolute carbon values.
\end{enumerate}}{\begin{enumerate}[label=\ref{study:3}\alph*]
\item\label{study:3:rel_ab} Measures the influence of relative versus absolute \COTwo\ emissions values.
\item\label{study:3:target} Examines whether providing carbon targets nudges people to consider absolute carbon values.
\end{enumerate}}
\end{study}

In this study, we hypothesized that
\aptLtoX[graphic=no,type=env]{% Coding for XML/HTML generation
\begin{enumerate*}%[label=\textbf{H\ref{study:3}.\arabic*}]
\item[\textbf{H3.1}]\label{hypo:3:rel} \textit{people use relative but not absolute carbon emission values when making choices},
\end{enumerate*} and that 
\begin{enumerate*}%[label=\textbf{H\ref{study:3}.\arabic*},resume]
\item[\textbf{H3.2}]\label{hypo:3:abs} \textit{targets increase people's use of absolute carbon emission values}.
\end{enumerate*}}{
\begin{enumerate*}[label=\textbf{H\ref{study:3}.\arabic*}]
\item\label{hypo:3:rel} \textit{people use relative but not absolute carbon emission values when making choices},
\end{enumerate*} and that 
\begin{enumerate*}[label=\textbf{H\ref{study:3}.\arabic*},resume]
\item\label{hypo:3:abs} \textit{targets increase people's use of absolute carbon emission values}.
\end{enumerate*}}

\subsubsection{Experiment Design}

\begin{figure}
    \centering
    \begin{subfigure}[b]{0.45\columnwidth}
        \includegraphics[width=\columnwidth]{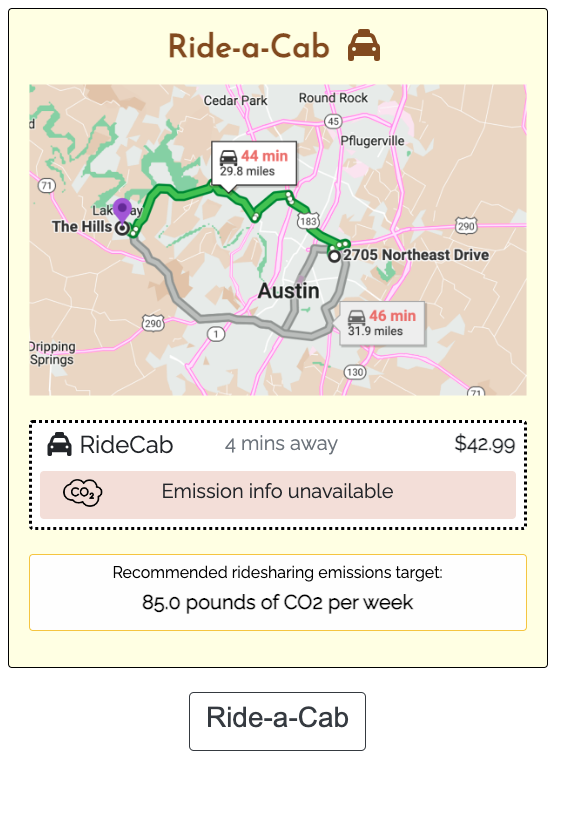}
        \caption{\label{fig:choose:a}}
    \end{subfigure}
    \begin{subfigure}[b]{0.45\columnwidth}
        \includegraphics[width=\columnwidth]{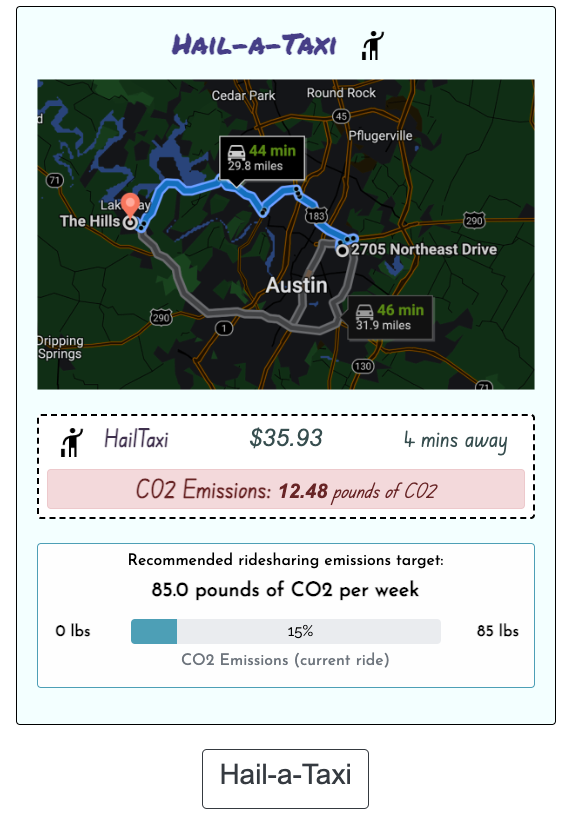}    
        \caption{\label{fig:choose:b}}
    \end{subfigure}
    \caption{Example screen shot of the ride-sharing choice study used for Study~\ref{study:3} asking ``Which company's ride would you pick to reach your destination? Click on an app to select your preference''. Participants chose between rides from two hypothetical ride-sharing apps:  \emph{Ride-a-Cab}, shown in Option (b), %(\subref{fig:choose:b}), 
or \emph{Hail-a-Taxi}, shown in Option 
(b). %(\subref{fig:choose:b}). 
Carbon information was provided either for both options, only one option (e.g., in this example \COTwo{}\ information is only provided for \emph{Ride-a-Cab}), or neither option. In Study~{3b},  %%{\ref{study:3:target}}, 
information about emission targets were provided below the \COTwo{}\ information (in this example, participants are shown a recommended weekly \COTwo{}\ emission target and the percentage of that target taken up by the ride).}
    \Description{Example screenshot of the ride-sharing choice study used for Study~\ref{study:3} asking ``Which company's ride would you pick to reach your destination? Click on an app to select your preference''. It shows two hypothetical ride-sharing apps --- Hail-a-Taxi (on the left side) and Ride-a-Cab (on the right side). Both apps show a map with the same route but in different colors. Below the map, Hail-a-Taxi shows a ride called ``HailTaxi'' which costs \$39.44 and is 4 mins away. For that ride, \COTwo{} emission information is unavailable. Below the ride information, a recommended emissions target of 85 pounds of \COTwo{} per week is given. On the right side, Ride-a-Cab displays a ride below the map called ``RideCab'' which costs \$48.54 and is 4 mins away. The interface also shows that ``RideCab'' emits 10.45 pounds of \COTwo{}. Below the ride information, Ride-a-Cab displays a recommended target of 85 pounds of \COTwo{} per week as well. On top of the page, there is a banner saying ``Which company's ride would you pick to reach your destination?''. Below it, there is an instruction asking the participant to click on an app to select their preference.}
\label{fig:study3}
\end{figure}

In this study, participants chose between rides from two hypothetical ride-sharing services (\emph{Ride-a-Cab} and \emph{Hail-a-Taxi}) that were shown side-by-side (see Figure~\ref{fig:study3}). The order in which the services were presented (e.g., whether \emph{Ride-a-Cab} was on the left or right of the screen) was counterbalanced between participants.  On each trial, information about the mapped route, time until service arrival, and price were provided for each service. All possible emission values, trip characteristics, and price differences were identical to those used in Study~\ref{study:1}.

In Study~\ref{study:3}, \COTwo\ emission values were presented in one of three ways:
\begin{enumerate}
    \item \textbf{Price only:} No \COTwo\ emission information provided for either ride-sharing option (only price).
    \item \textbf{Carbon for one option:} \COTwo\ emission information provided for one but not the other option.
    \item \textbf{Carbon for both options:}\COTwo\ emission information provided for both options.
\end{enumerate}

% \missingfigure{Show some screenshots of Study~\ref{study:3}.}

In Study~\aptLtoX[graphic=no,type=env]{3a}{\ref{study:3:rel_ab}}, \COTwo\ emission information was presented on each trial without any additional context or carbon targets. In Study~\aptLtoX[graphic=no,type=env]{3b}{\ref{study:3:target}}, participants were given one of three targets to help contextualize \COTwo\ emission numbers:

\begin{enumerate}
    \item \textbf{Weekly emission target:} Participants were told on each trial that their \COTwo\ emissions should not exceed 85 pounds of \COTwo\ per week. This number was based on an annual recommendations of 2 metric tons of \COTwo\ made after the Paris Accord~\cite{hall_2016}. We chose to provide a weekly number as the total emission numbers for some trips exceeded the recommended daily total.
    \item \textbf{Average weekly emissions in participant zip code:} Participants were given an arbitrary \COTwo\ emission number meant to represent the average weekly emissions in their community. We set the number to be 85 pounds of \COTwo\ in this case.
    \item \textbf{Average weekly emissions of eco-conscious community members:} Participants were given an arbitrary \COTwo\ emission number meant to represent the average weekly emissions from the most eco-conscious community members. We used the same 85 pounds of \COTwo\ in this case also. 
\end{enumerate}

To test~\aptLtoX[graphic=no,type=env]{\textbf{H3.1}}{\ref{hypo:3:rel}}, we examined the influence of numerical emission values on people's ride-sharing choices.
We hypothesized that numerical values will only influence choices when both options are presented with emission values, but not when \COTwo\ emissions are presented for one option because people lack a frame of reference for numerical \COTwo{}\ values.
To test~\aptLtoX[graphic=no,type=env]{\textbf{H3.2}}{\ref{hypo:3:abs}}, we additionally ran experiments where we provide different frames of reference for \COTwo{}\ emissions to examine whether these increase the importance of numerical \COTwo{}\ values on people's ride-sharing choices.

\changemarker{These 30 trials were followed by the survey questionnaire and supplemental, open-ended questions about their choices, understanding of \COTwo{}, and how they perceive targets (see Appendix~\ref{sec:appendix_study_3}).} 

\subsubsection{Analysis}

We used a LME model to predict the likelihood of choosing the \emph{Ride-a-Cab} option.  Each model included the price difference between both ride sharing options as a fixed effect and random intercepts for participants and city. All additional fixed effects are outlined in the results.

\changemarker{
\paragraph{Limitations:} We assumed that an ``overall'' weekly emissions target might not be relatable for everyone as it assumes that the emissions from all other daily activities are getting measured. Further, participants would have to mentally allot their weekly quota of ``ridesharing'' emissions as a proportion of ``overall'' emissions, which introduces additional complexities. In order to maintain simplicity and study the impact of targets, we used ridesharing, not overall, emissions for the three targets to allow participants to make a direct comparison (since ridesharing emissions are what is being measured in the displayed interface). While the interface (see Figure~\ref{fig:carbon_targets}) clearly mentions the ``recommended emissions target'', we did not conduct any usability study to understand how participants interpreted these targets. 
}

\subsubsection{Participants}

For Study~\aptLtoX[graphic=no,type=env]{3a}{\ref{study:3:rel_ab}}, we recruited 201 participants from Prolific (44.5\% men, 54.5\% women, 1\% genderqueer or non-binary, mean age = 36, SD = 13). For Study~\aptLtoX[graphic=no,type=env]{3b}{\ref{study:3:target}}, a total of 601 participants were randomly assigned to one of three possible conditions (49.25\% men, 48.25\% women, 2.3\% genderqueer, agender, or non-binary, 0.2\% preferred not to answer, mean age = 38, SD = 13). 

\subsubsection{Findings}
\paragraph{People respond to relative \COTwo{}\ values more than absolute values.}

\begin{figure*}
  \centering
  \begin{subfigure}[b]{\columnwidth}
    \centering
    \includegraphics[width=\columnwidth]{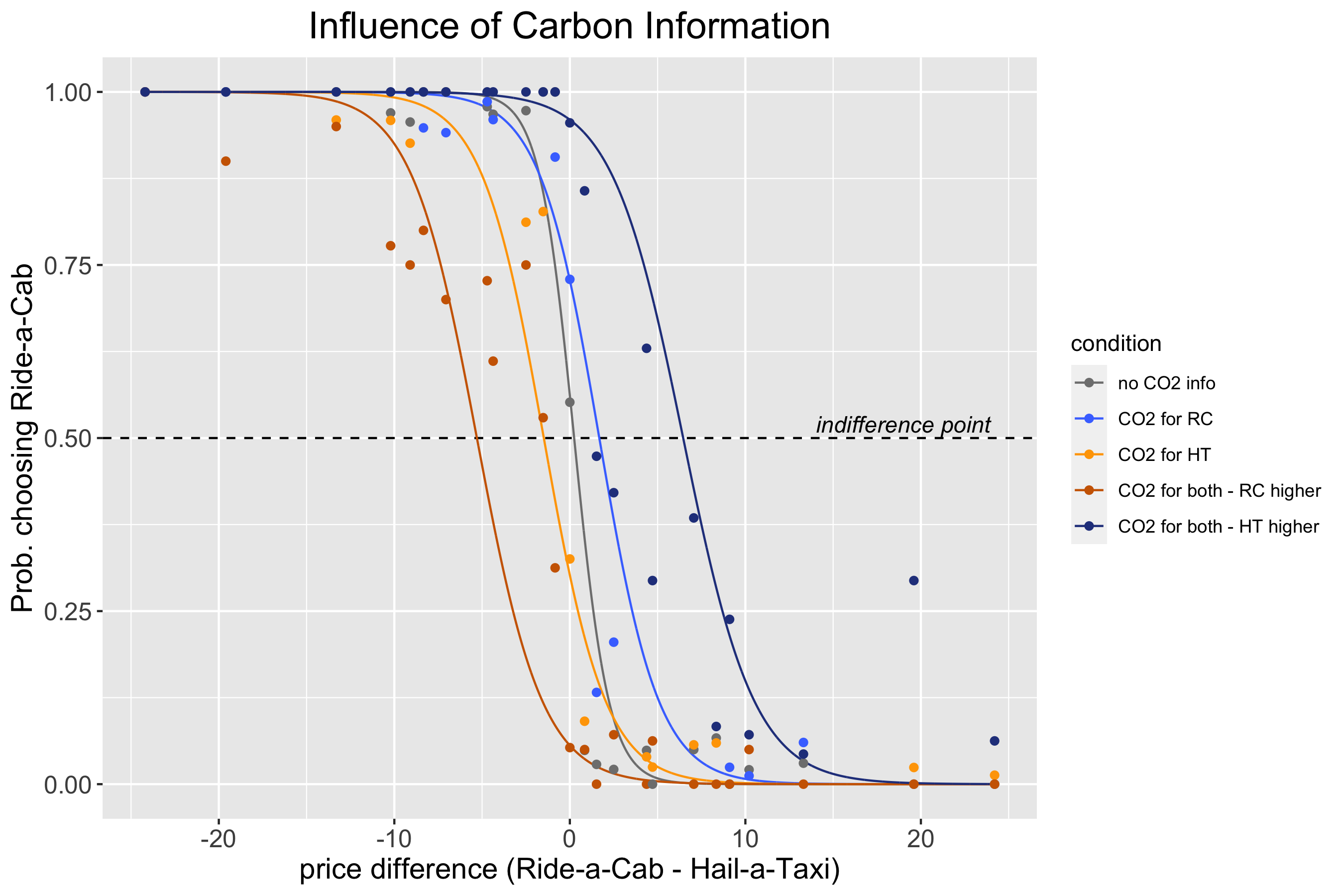} % ChkTeX 25
    \caption{\label{fig:carbon_curve}}
    \Description{An logistic choice curve that shows the probability of choosing Ride-a-Cab on Y-axis (0--1) and the price difference between Ride-a-Cab and Hail-a-Taxi on the X-axis (\$0--\$25) for five different contrasts (no carbon info, carbon info for ride-a-cab only, carbon info for hail-a-taxi only, carbon info for both: higher ride-a-cab carbon values, and both: higher hail-a-taxi carbon value). The curves for each of the interventions intersect the indifference point at 0.50 on the Y-axis; the intersection point of the both: higher ride-a-cab carbon values is leftmost, followed by carbon info for hail-a-taxi only, no carbon info, carbon info for ride-a-cab only, and both: higher hail-a-taxi carbon values furthest right on the indifference line. This indicates that people are willing to pay more for lower carbon options when carbon information is available for both options, and more for options that provide carbon information when info is only available for one option.}
  \end{subfigure}  
  \hspace{1pc}
  \begin{subfigure}[b]{\columnwidth}
    \centering \includegraphics[width=\columnwidth]{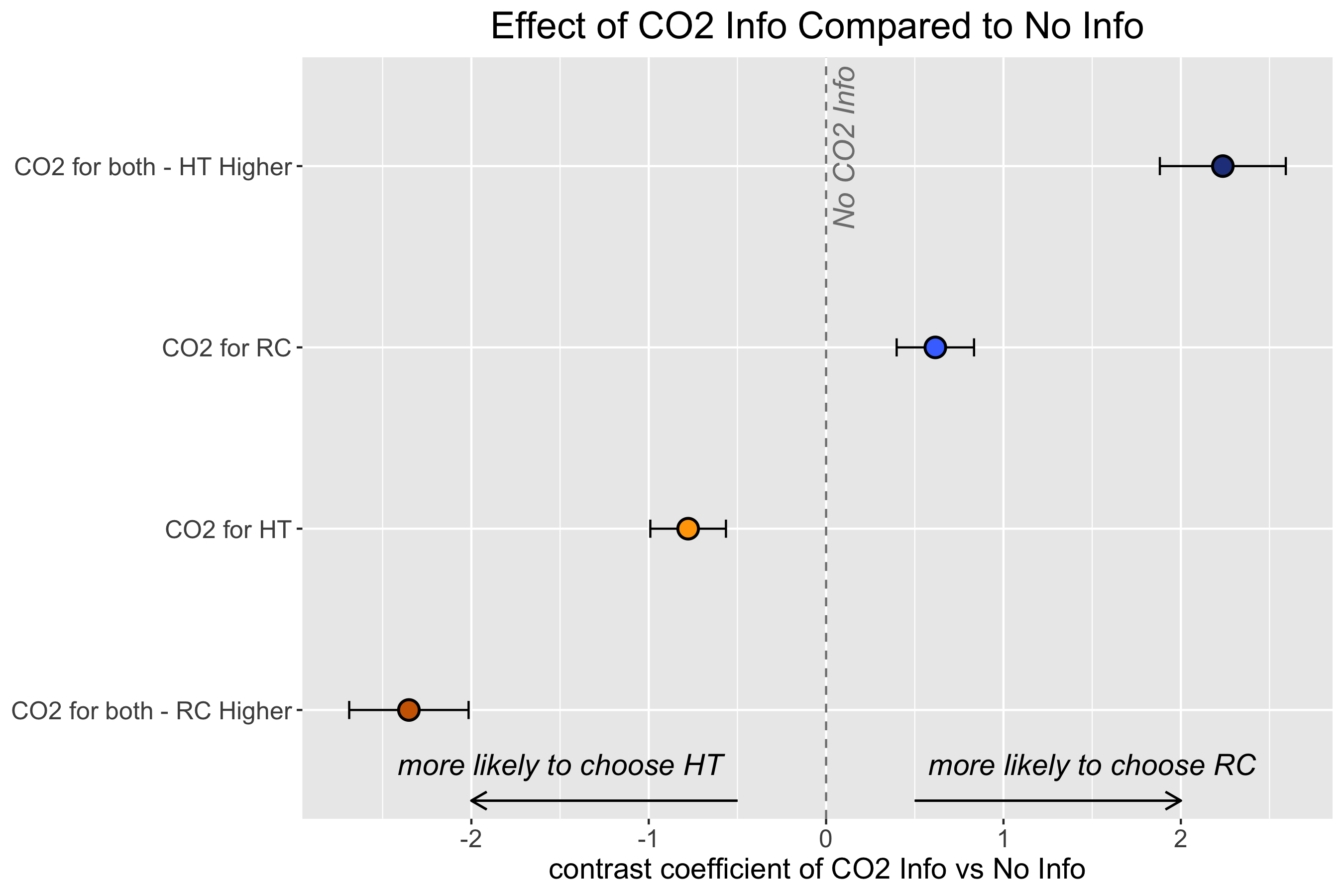}
    \caption{\label{fig:carbon_beta}}
    \Description{A chart displaying the probability of participants picking ride-a-cab in log odd scale. The X-axis has beta coefficients for the different conditions. The Y-axis has all the different conditions sorted according to the beta coefficients in descending order. The beta coefficients are compared to the condition with no carbon information, denoted by a dashed vertical line. The beta coefficient for both: hail-a-taxi higher is the farthest the the right on the X-axis, followed by carbon info for ride-a-cab only, which is also to the right of the vertical dashed line. The beta coefficient for info for hail-a-taxi only is negative and falls to the left of the dashed line and both: ride-a-cab higher is the farthest the the left on the X-axis.}
  \end{subfigure}
  \caption{Influence of \COTwo\ emission information on ride-sharing choices. (a) Choice curves indicating the probability that participants chose the \emph{Ride-a-Cab} (RC) option as a function of the price difference between \emph{Ride-a-Cab} and \emph{Hail-a-Taxi} (HT). Compared to the condition where no \COTwo\ information was presented (grey), participants where more likely to choose \emph{Ride-a-Cab} when \emph{Hail-a-Taxi} emissions were higher (dark blue) and more likely to choose \emph{Hail-a-Taxi} when \emph{Ride-a-Cab} emissions were higher (dark orange). When carbon information was only provided for one option, participants were more likely to choose \emph{Ride-a-Cab} when emissions were only presented for \emph{Ride-a-Cab} (light orange) and \emph{Hail-a-Taxi} when emissions were only presented for \emph{Hail-a-Taxi} (light blue). Lines indicate best fitting logistic regression lines and points indicate average choice probabilities across all participants. (b) Logistic mixed-effects contrast coefficients comparing the change in likelihood that participants chose \emph{Ride-a-Cab} when carbon was presented for both (dark blue and dark orange) or only one ride option (light blue and light orange) compared to when no \COTwo\ was presented (grey dashed line). Points indicate contrast beta weights and error bars indicate 95\% confidence intervals.}
\label{fig:carbon0}
\end{figure*}

\proofreadtrue%
\changemarker{
When both \emph{Ride-a-Cab} and \emph{Hail-a-Taxi} displayed numerical values for \COTwo{} emissions (in lbs.), participants primarily used the emissions information to make their decision.}
As is evident from Figure~\ref{fig:carbon0}, when both options were presented with \COTwo{}\ emission values, participants were much more likely to choose the ride-sharing option that had lower \COTwo{}\ values, even when accounting for price differences (LME contrasting the likelihood of choosing \emph{Ride-a-Cab} over \emph{Hail-a-Taxi} when \COTwo{}\ emissions were higher for \emph{Hail-a-Taxi}: $\beta=4.17, z=19.81, p<0.001$).

When \COTwo\ emission values were present for only one option we found two distinct results. On one hand, the displayed \COTwo\ \emph{level} had a small effect. Specifically, when \COTwo\ values were shown exclusively for \emph{Ride-a-Cab}, participants were only slightly more likely to choose \emph{Ride-a-Cab} if emissions were relatively low (LME measuring the influence of \COTwo\ emission values while controlling for price: $\beta_{\mathrm{CO_2}}=0.19, z=2.28, p=0.023$). Furthermore, \COTwo\ values did not have an influence on participant choices in the case of \emph{Hail-a-Taxi} ($\beta_{\mathrm{CO_2}}=0.12, z=1.00, p=0.317$). This result supports~\aptLtoX[graphic=no,type=env]{\textbf{H3.1}}{\ref{hypo:3:rel}}, that participants better understand \COTwo\ emission values when they are being compared between two options rather than presented as separate, absolute numbers.

On the other hand, we found that \changemarker{the \emph{presence}} of \COTwo\ information, when analyzed irrespective of \COTwo\ \emph{levels}, had a significant effect on participant choice in the one-option conditions (LME predicting the probability of choosing the option with \COTwo\ numbers provided in the one option condition: $\beta_{\mathrm{Ride-a-Cab}}=0.62, z=5.53, p<0.001$; $\beta_{\mathrm{Hail-a-Taxi}}=0.78, z=7.15, p<0.001$; Figure~\ref{fig:carbon0}). Indeed, qualitative reports suggest that participants may have preferred options that they deemed as being more transparent about carbon emissions. In the words of P4307: ``\emph{I found the prices listed and the transparency of info provided on CO emissions to be very useful.}'' Further, participants also found ``\emph{emissions data to be useful if the data was available for both.}  (P4312)'' At the same time, when asked what ``100 pounds of \COTwo{}'' or ``1 pound of CO2'' means, most participants were unsure, or gave vague answers like ``pollution'' or ``greenhouse gas''; some even compared them to ``garbage'' or ``cigarettes'', suggesting major gaps in carbon literacy: \emph{``I don't know what this means. Is that a lot? Is that a little? How meaningful is it? No idea.''} (P4252)

\changemarker{
To summarize, this finding suggests that participants, when presented with absolute values, based their decisions largely on the availability of \COTwo{} information, and not necessarily on what that information represents (since we did not observe any significant effect for \COTwo{} level thresholds). However, when a comparison is possible, they lean towards the ride-sharing option that shows the lower \COTwo{} values.}

\begin{figure*}
    \centering
    \includegraphics[width=0.85\textwidth]{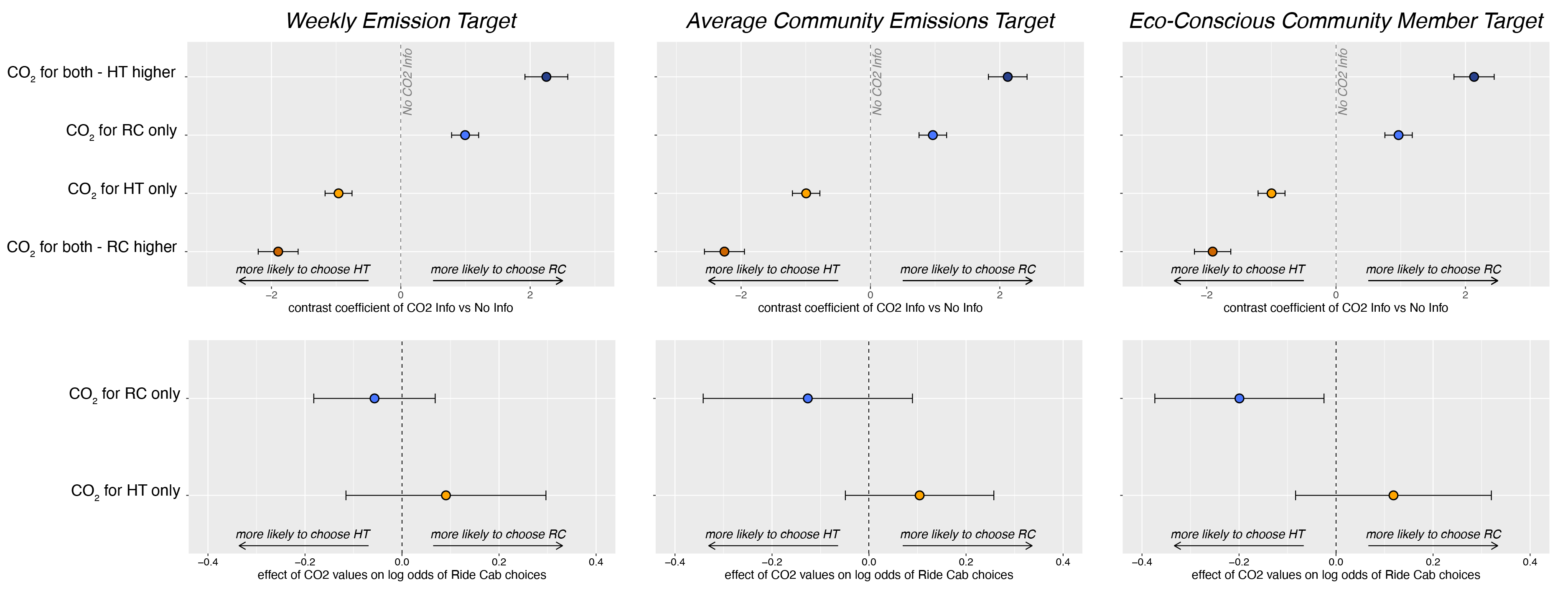}
    \caption{\label{fig:carbon_targets}
    Influence of adding different emission targets on ride-sharing choices. 
    Participants made ride sharing choices while being provided with emission targets about their recommended weekly emissions (left column), average emissions in their zipcode (middle column), and average emissions from the most eco-conscious members of their zipcode (right column). 
    The three panels in the top row show the results of logistic mixed-effects contrasts comparing the change in likelihood of choosing the \emph{Ride-a-Cab} (RC) as a function of option when different levels of carbon information are shown compared to when no carbon information shown (grey dashed line). 
    In all three target conditions, when \COTwo\ information was provided for both options, participants were more likely to choose \emph{Ride-a-Cab} if this option had the lowest emissions (dark blue) and choose \emph{Hail-a-Taxi} if this option had the lowest emissions (dark orange). 
    When information was only provided for one option, participants were more likely to choose \emph{Ride-a-Cab} when \COTwo\ was only presented for this option (light orange), and \emph{Hail-a-Taxi} when \COTwo\ was only presented for this option. 
    The three panels on the lower row show the influence of absolute \COTwo\ emission values on participant choices in conditions where emission values were only provided for one option. In the eco-conscious community members target intervention, participants were slightly less likely to choose the \emph{Ride-a-Cab} option when it had higher absolute \COTwo\ values (blue point in lower right panel). However, absolute \COTwo\ values had no influence on any other choice in any other emission target context. All points indicate beta coefficients from logistic mixed effects models and error-bars indicate 95\% confidence intervals.
    }
    \Description{A chart with three columns and two rows of panels. Each column corresponds to results from a different emission target intervention (titles above each column from left to right: Weekly Emission Target, Average Community Emission Target, and Eco-Conscious Community Member Target). The top row of three panels shows charts displaying the probability of participants picked ride-a-cab in log odd scale in all three target conditions. The X-axis in each chart has beta coefficients for the different conditions. The Y-axis in each chart has all the different conditions (carbon info for ride-a-cab only, carbon info for hail-a-taxi only, carbon info for both: higher ride-a-cab carbon values, and both: higher hail-a-taxi carbon value) sorted according to the beta coefficients in descending order. The beta coefficients are compared to the condition with no carbon information, denoted by a dashed vertical line in all panels. In all three panels, the beta coefficient for both: hail-a-taxi higher is the farthest the the right on the X-axis, followed by carbon info for ride-a-cab only, both right of the vertical dashed line. The beta coefficient for info for hail-a-taxi only is negative and falls to the left of the dashed line and both: ride-a-cab higher is the farthest the the left on the X-axis. The three panels in the bottom row show the influence of emission values on the probability that participant chose ride-a-cab in log odds scale. The Y-axis in each chart has all the different conditions where only one option was presented with carbon information (carbon info for ride-a-cab only, carbon info for hail-a-taxi only). Each chart has a dashed black line at the zero point on the X-axis which denotes no influence of emissions on participant choices. In all three panels, the points indicating ride-a-cab only are all slightly left of the black dashed line and the points indicating hail-a-taxi only are slightly right of the black dashed line. None of the points differ significantly from zero except the ride-a-cab only condition within the Eco-Conscious Community Member Target intervention.}
\vspace*{-6pt}
\end{figure*}

\paragraph{Providing emission targets does not push people think about absolute \COTwo\ values}
As is evident from Figure~\ref{fig:carbon_targets}, none of our attempts to provide targets to contextualize carbon emissions succeeded in increasing people's use of absolute \COTwo\ values when choosing between options.
In all three Study~\aptLtoX[graphic=no,type=env]{3b}{\ref{study:3:target}} target conditions, we consistently found that participants used \COTwo\ values when emission information was provided for both options, similar to Study~\aptLtoX[graphic=no,type=env]{3a}{\ref{study:3:rel_ab}} (LME contrasting the likelihood of choosing \emph{Ride-a-Cab} over \emph{Hail-a-Taxi} when \COTwo\ emissions were higher for \emph{Hail-a-Taxi} in all three target conditions: all $\beta>3.73$, all $z>16.13$, all $ p<0.001$; Figure~\ref{fig:carbon_targets}).
We additionally consistently found the same bias observed in Study~\aptLtoX[graphic=no,type=env]{3a}{\ref{study:3:rel_ab}}, where participants were more likely to choose the the option with \COTwo\ values provided in the conditions where only one option contained emission information (LME predicting the probability of choosing the option with \COTwo\ numbers provided in the one option condition: all $\beta>0.96$, all $z>8.8$, all $p<0.001$; Figure~\ref{fig:carbon_targets}).

However, \COTwo\ levels had little to no impact on participant choices when only one option was presented alongside targets.
We found a small effect of \COTwo\ levels on \emph{Ride-a-Cab} choices when targets related to eco-conscious community members were presented (LME measuring the influence of \COTwo\ emission values while controlling for price: $\beta_{\mathrm{CO_2}}=0.20, z=2.23, p=0.025$) but not for any other choices under any of the other target conditions (all $\beta<0.13$, all $z<1.14$, all $p>0.251$; Figure~\ref{fig:carbon_targets}).
This result fails to support~\aptLtoX[graphic=no,type=env]{\textbf{H3.2}}{\ref{hypo:3:abs}}: emission targets did not increase people's reliance on \COTwo\ emission values in the context of ride-sharing choices.

Overall the results from Study~\ref{study:3} provide compelling information that although people do consider \COTwo\ emission values for ride-sharing choices, they primarily do so when they are able to compare these values between options.
Our results also suggest that one-time targets may not provide enough context to increase people's carbon literacy in a way similar to targets used in healthy eating and exercise domains.

Self-reported responses showed that participants were on the fence about trusting these target emission numbers. In the words of P0096:``\emph{I did, but I had to. I'm not educated enough about those numbers now, but I would make sure that I did educate myself soon after seeing them.  I don't like to think I'm being fooled by companies.''}. Some suggested that ``\emph{the EPA or a public environmental organization that is independent of industry}'' (P0150) should provide these targets. When asked who they \emph{would} trust to provide carbon footprint information, participants suggested friends and family.

\subsection{Green Interventions in a Longer Term Rental Context}
% \section{STUDY 4: INFLUENCE OF TEMPORAL GRANULARITY ON RENTAL CHOICES}

The design framework proposed by Sanguinetti and colleagues highlights ``temporal granularity'' as an important component of eco-feedback information. Studies 1--3 tested the the influence of different equivalencies on ride-sharing choices, a context in which choices are limited to more immediate time-frames. In Study~\ref{study:4:rental}, we use a simulated rental vehicle paradigm to test the the influence of temporal granularity on carbon friendly vehicle choices. Rental choices are very often completed on online platforms, where consumers are presented with a number of vehicle options that vary on a number of features (e.g., price, size, fuel efficiency). Unlike ride-sharing choices, vehicle rentals can span various temporal horizons (e.g., days, weeks, months) and provide a framework to study the role of temporal granularity plays on people's perception carbon equivalency information.

\begin{study}\label{study:4:rental}
Examines the influence of presenting \COTwo{} equivalency information over different temporal horizons on people's perception of \COTwo{} equivalencies in the context of rental cars. 
\end{study}

\begin{figure}
    \centering
    \includegraphics[width=0.9\columnwidth]{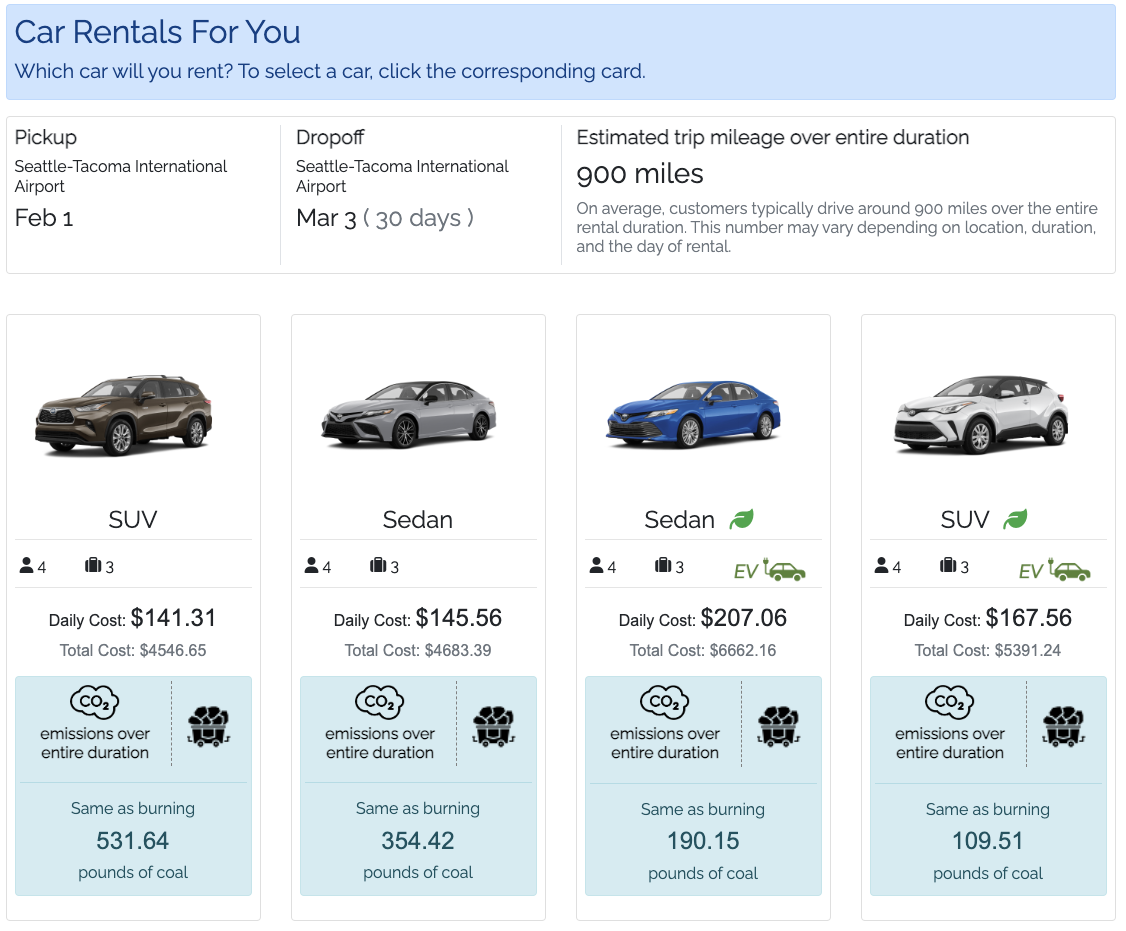}
    \caption{\label{fig:studyrentals} Example screen shot from the rental car choice task. Participants chose between four rental options that always contained two sedans and two SUVs and two options marked as ``EVs''. On each trial, total mileage for all choices was displayed in the top right  and specific emission values were shown below vehicle: either for the full duration of the trip or per day in the trip.}
    \Description{An example screenshot of a trial from Study~\ref{study:4:rental}. There is a blue banner at the top of the screen shot with the title ``Car Rentals For You'' and a subheader reading ``Which car will you rent? To select a car, click the corresponding card''. Below the banner are three rectangles layed out side by side}
\end{figure}

One effect of longer temporal horizons is that they are likely associated with higher overall \COTwo{} emissions (e.g., a person's emissions from a car rented for one month are likely higher than those rented for one day). Thus, estimates provided about the potential carbon emitted by different vehicle rentals will depend in part on the temporal duration of the anticipated trip. 

Research into numerical cognition proposes a specific way in which larger emission values associated with longer trips could bias the way people process emission information. The \emph{numerical distance effect} (related to Weber-Fechner laws) describes a process by which people's sensitivity to differences between numbers decreases as the size of the numbers increase~\cite{dehaene2003neural,gallistel2000non}. In the context of communicating emissions for vehicle rentals, this research proposes that people will be less sensitive to differences in higher \COTwo{} emission values, such as those associated with trips of longer temporal duration. We thus explore the possibility that people become less sensitive to \COTwo{} emission differences between rentals of longer duration. We hypothesize that 
\aptLtoX[graphic=no,type=env]{% Coding for XML/HTML generation
\begin{enumerate*}%[label=\textbf{H\ref{study:4:rental}.\arabic*}]
\item[\textbf{H4.1}]\label{hypo:4:rental} \textit{for longer trips, people are more sensitive to emission differences presented as smaller daily emission averages than larger total emission values}.
\end{enumerate*}}{\begin{enumerate*}[label=\textbf{H\ref{study:4:rental}.\arabic*}]
\item\label{hypo:4:rental} \textit{for longer trips, people are more sensitive to emission differences presented as smaller daily emission averages than larger total emission values}.
\end{enumerate*}}

\subsubsection{Experiment Setup}

In this study, participants were presented with four potential vehicle rental options for a hypothetical trip (see Figure~\ref{fig:studyrentals}). 
%% Not saying TOYOTA for submission: using anon command.
The options contained two mid-sized sedans %\anon[]{ 
(randomly selected from a Toyota Corolla, Prius, Camry, and Avalon) %} 
and two mid-sized SUVs %\anon[]{ 
(randomly selected from a Toyota Corolla Cross, RAV4, Highlander, and Land Cruiser), %}, 
and on each trial two randomly chosen vehicles were marked as battery electric vehicles, designated with a green ``EV'' symbol and a green car icon (see Figure~\ref{fig:studyrentals}). 
All cars were indicated to carry a maximum of 4 passengers and 3 pieces of luggage.

For each trial, information was provided about the location of the trip (chosen from 5 possible locations), the duration of the rental, estimated distance that will be traveled (in miles), price of the rental per travel day, and \COTwo{} emission information. 
The duration of the rental could either be 1, 7, 21, or 30 days. 
The distance in miles was chosen by first randomly selecting from a range of miles per day that was either short (range: \emph{20--40} miles/day), medium (range: \emph{50--70} miles/day), or long (range: \emph{90--110} miles/day) and multiplying this number by the trip duration (in days). 
Emission information on each trial was presented using one of the following equivalencies from Study~\ref{study:1}:

\begin{description}
    \item[Baseline:] No additional information provided about the ride other than a green ``EV'' icon for two vehicles on each trial.
    
    \item[Raw Emissions:] \COTwo{} emissions (in pounds) for all rental options calculated using the US Fuel Economy calculator~\cite{epa_2022} based on the car type and estimated trip distance.
    
    \item[Equivalencies:] Information about the \COTwo{} emissions for all rental options provided in terms of charging smartphones, daily energy usage of houses, burning pounds of coal, burning gallons of gasoline, recycling bags of waste, growing trees or square footage of forests. \changemarker{Unlike Studies~\aptLtoX[graphic=no,type=env]{2a}{\ref{study:2:social}} and~\aptLtoX[graphic=no,type=env]{2b}{\ref{study:2:valence}}, where equivalencies were presented in a comparative manner between the two ride-hailing options, the rental interface only displayed absolute emissions values in terms of the equivalencies (see Figure~\ref{fig:studyrentals})}. 
\end{description}

Critically, emission information was presented in one of two ways:
\begin{enumerate}
    \item \textbf{Full trip emissions:} Sum of the estimated emissions for the entire trip.
    \item \textbf{Emissions per day:} full trip emissions divided by the number of travel days
\end{enumerate}

As a test of~\aptLtoX[graphic=no,type=env]{\textbf{4.1}}{\ref{hypo:4:rental}}, we predicted that emission information would have a larger influence on people's choices when presented per day rather than for the full trip, and the influence of switching information per day would have a larger impact for trips of longer temporal duration.

\changemarker{Similar to the previous studies, participants also answered the survey questionnaire, followed by supplemental, open-ended questions about their preferred granularity for seeing their carbon footprints (see Appendix~\ref{sec:appendix_study_4}).}

\subsubsection{Analysis}

To measure the influence of different interventions on vehicle rental choices, we use a logistic mixed-effects model that predicted likelihood of choosing an ``EV'' rental. We included the average price difference between the two ``EV'' and two non-EV options and intervention type as fixed effects, and random intercepts for participants, city, and trip range category (short, medium, or long).

Because there were four possible rental options, we used multinomial logistic regression to estimate the influence of emission values on participant choices. To leverage the hierarchical nature of our data, we used the Begg and Gray approximation to multinomial logistic regression, which consists of logistic mixed-effects regressions between all pairwise rental choices~\cite{becg1984calculation} (see Figure~\ref{fig:trip_duration} in Appendix). This approximation offers more model flexibility but produces larger standard errors and is considered a more conservative approach to multinomial regression. 

Each pairwise logistic mixed-effects regression model included the pairwise price difference between rental options and pairwise emission difference as a fixed effects, and included participant identifiers, trip location, and trip range category (short, medium, or long) as random intercepts.

To compare the influence of emission values on choices between the ``per day'' and ``full trip'' conditions, we divided the emission values for the ``full trip'' condition by the trip duration. This ensured that emissions were on a similar scale for different duration conditions. 

\begin{figure}
    \centering
    \includegraphics[width=0.9\columnwidth,trim=0 0 15px 0, clip]{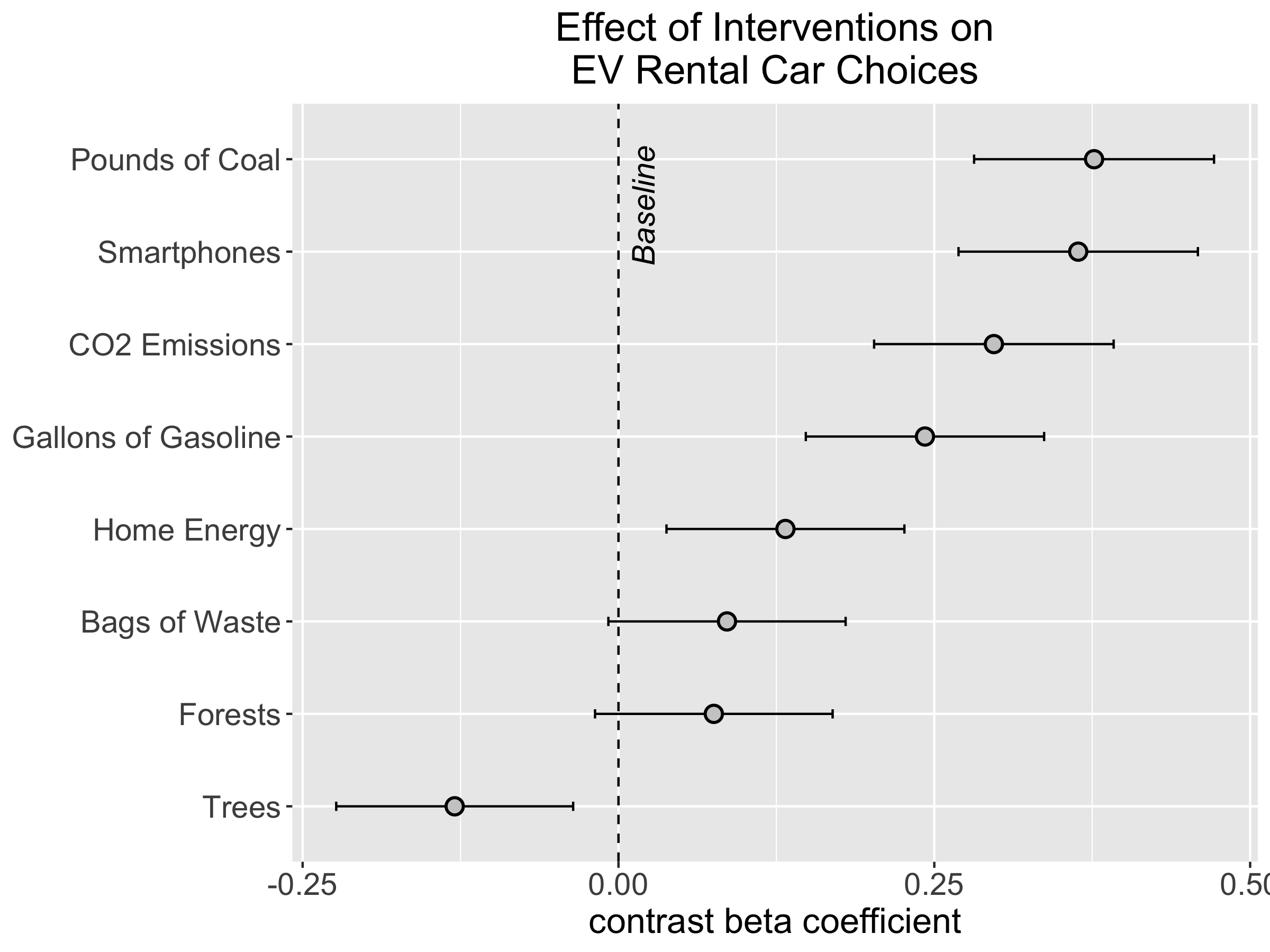}
    \caption{\label{fig:rental_int_betas} Influence of different emission equivalencies on rental choices. Points indicate logistic mixed-effects contrast coefficients estimating the change in likelihood that participant chose an ``EV'' rental option for different emission equivalencies compared to the baseline condition (black dashed line). Errorbars indicate 95\% confidence intervals.}
    \Description{A chart displaying the probability of participants picking an EV rental choice in log odd scale. The X-axis has beta coefficients for the different interventions. The Y-axis has all the different interventions sorted according to the beta coefficients in descending order. The beta coefficients are compared to the baseline condition, denoted by a dashed vertical line. Pounds of coal, smartphones, \COTwo\ emissions, gallons of gasoline, and home energy were all significantly better than the baseline condition. Bags of waste, forests, and trees are not significantly different from the baseline condition.}    
\end{figure}

% \begin{figure}
%     \centering
%     \begin{subfigure}[b]{0.4\columnwidth}
%         \centering
%         \includegraphics[width=\columnwidth]{figures/trip_druation_contrast_graphic.png}
%         \caption{\label{fig:tc_graphic}}
%         \Description{}
%     \end{subfigure}
%     \vspace{2pc}
%     \begin{subfigure}[b]{\columnwidth}
%         \centering
%         \includegraphics[width=\columnwidth]{figures/trip_druation_contrast_polished.png}
%         \caption{\label{fig:tc_betas}}
%         \Description{}
%     \end{subfigure}
%     \caption{\label{fig:trip_duration} Contrast of the influence of emission values between the ``Full Trip'' and ``Per Day'' conditions. (a) Schematic of all possible pairwise comparisons between the four possible rental options on each trial. (b) Contrast coefficients for pairwise comparisons between all possible rental options comparing the change in the influence of emission values when presented per day versus for the full trip. Negative values indicate that people are less willing to choose options with high emissions. Contrasts are shown for each emission equivalency and for all trip durations on the y-axis. Points indicate contrast coefficients and errorbars indicate 95\% confidence intervals.}
% \end{figure}

\subsubsection{Participants}

For Study~\ref{study:4:rental}, we recruited a total of 2020 participants from Prolific (43.75\% men, 54.30\% women, 1.36\% genderqueer or non-binary, 0.15\% agender, 0.42\% preferred not to answer, mean age = 42, SD = 15). 

\subsubsection{Findings}
\paragraph{\COTwo{} emission equivalencies influence rental choices in ways similar to ride-sharing choices} Overall, participants were more likely to choose rentals with an ``EV'' logo, choosing this option 60\% of the time. We then examined whether any of the tested carbon equivalencies influenced the probability of choosing an ``EV'' rental. Consistent with the results found in Studies~\ref{study:1} and~\ref{study:2}, pounds of coal, raw \COTwo{} emission values, smartphones, gallons of gasoline, and home energy usage all increased the likelihood that participants chose one of the EV rental options above the baseline condition (LME contrasting the influence of equivalencies on the likelihood of choosing an ``EV'' option over baseline: all $\beta>0.132$, all $z>2.75$, all $p<0.006$; Figure~\ref{fig:rental_int_betas}). Additionally, bags of waste did not differ significantly from baseline ($\beta_{\mathrm{bags}}=0.09, z=1.79, p=0.073$). Contrary to Studies 1 and 2, forests did not differ significantly from baseline ($\beta_{\mathrm{forests}}=0.08, z=1.57, p=0.116$) and trees slightly \emph{reduced} the likelihood that people chose the EV condition compared to baseline ($\beta_{\mathrm{trees}}=-0.13, z=2.75, p=0.007$). Nevertheless, these results help generalize the results we observed in our Ride-Sharing studies to a new vehicle choice context.

\paragraph{Emission values have the same influence on rental choices when presented per day or for the full trip} Contrary to~\aptLtoX[graphic=no,type=env]{\textbf{4.1}}{\ref{hypo:4:rental}}, we found no evidence that presenting emission information per day influences choices any more than presenting information for the full duration. 
We tested this by contrasting the influence of emission values on participant rental choices in the ``per day'' compared to the ``full trip'' condition for different trip durations (see analysis methods for contrast details). 
We found no clear differences in the influence of emission values between the ``per day'' and ``full trip'' conditions, and critically, no increasing benefit of presenting emission information ``per day'' as trips increase in duration and ``full trip'' emission values get larger (a total of 8 out of 192  contrasts (4\%) shown in Figure~\ref{fig:trip_duration} (Appendix) reached a threshold of $p<0.05$, with $\beta$s ranging from -0.55--0.50, and $p$s ranging from 0.005--0.043; however this is roughly the number expected by chance given the number of contrasts performed). 
We therefore find no evidence in our study that higher emission values associated with longer temporal trip durations influenced people's rental choices.  

Participants had a variety of self-reported responses related to how they wanted information about carbon presented over time. A few participants wanted their carbon aggregated over longer durations. P2244 said, \emph{``Aggregate would be more useful to show trends as shorter time periods, like day, wouldn't be very useful. How I do over a year is more important than if I have a good or bad carbon footprint day.''} However, most participants preferred a daily breakdown of their emissions as it felt more actionable: 
\begin{quote}
    ``I think breaking it down into shorter time periods would more useful because it's easier to visualize the impact. For example, I can know that I'll reduce emissions by X amount if I go a day without driving my car. If I was given the carbon footprint over a longer period of time, I wouldn't be able to as easily figure this out because I would have to do math to convert it to a daily figure.'' (P2212)
\end{quote}

%%%% sections/4.study4.tex ends here %%%%

%%%% 4.second.tex ends here %%%%

%%%% 5.discussion.tex starts here %%%%

\section{Discussion}
%\hl{Todo: Summarize the main take aways of all the studies 1--4 in a single paragraph. Mention raw \COTwo{} often wins regardless of how people may relate to it.}

% Please add the following required packages to your document preamble:
% \usepackage{multirow}

\begin{table*}
  \begin{tabular}{p{.45\textwidth}  p{.45\textwidth}}
    \multicolumn{2}{l}{\textbf{Communicating \COTwo{} equivalencies}} \\
    \midrule
    \begin{enumerate}%[label=1.\arabic*, leftmargin=*]
\leftskip-15pt
    \item[1.1] How does information about \COTwo{} emissions for available ride-hailing options influence a passenger's choice in terms of picking a green ride?
    \item[1.2] Does communicating \COTwo{} emissions in measurable equivalent actions influence a passenger's choice more than providing direct \COTwo{} emission numbers for picking a green ride?
    \item[1.3] How do carbon-based interventions compare with  other non-\COTwo{} related intervention types (e.g., extrinsic rewards, social motivators)?
    \end{enumerate}&
    \begin{itemize}[leftmargin=*]
    \item All interventions increase green ride choices.
    \item Raw \COTwo{} (as an intervention) is effective.
    \item People like extrinsic rewards (points).
    \item Not all \COTwo{} equivalencies were effective.
    \item Not all social interventions are equal.
    \end{itemize} \\
  \end{tabular}
  \caption{\label{tab:summary_table_1} Research questions (left) and
    findings (right) from Phase 1.}
\end{table*}

\begin{table*}
  \begin{tabular}{p{.45\textwidth}  p{.45\textwidth}}
    \multicolumn{2}{l}{\textbf{Contextual framing of \COTwo{} equivalencies}} \\
    \midrule
    \begin{enumerate}%[label=2.\arabic*, leftmargin=*]
\leftskip-15pt
    \item[2.1] How do popularity-based interventions (i.e., \% of people preferring a given ride type) influence a ride-hailing passenger's choice in terms of picking a green ride?
    \item[2.2] How do dynamic trends (i.e., increase in the adoption of a given ride type) influence a ride-hailing passenger's choice in terms of picking a green ride?
    \item[2.3] How do \COTwo{} equivalencies influence a a ride-hailing passenger's choice when framed in terms of collective impact rather than individual impact?
    \end{enumerate}&
    \begin{itemize}[leftmargin=*]
      \item[] Popularity influenced participant choices but dynamic social norms had mixed effects. Collective impact increases the likelihood of choosing green rides.
    \end{itemize}\\
%    \midrule
    \begin{enumerate}%[label=2.\arabic*, leftmargin=*, resume]
%      \stepcounter{enumi}
\leftskip-15pt
    \item[2.4] How do \COTwo{} equivalencies influence a ride-hailing
      passenger's choice  when framed negatively?
    \end{enumerate}&
    \begin{itemize}[leftmargin=*]
      \item[] Interventions with negatively valenced wording (i.e., that highlighted the  additional emission equivalents produced by the non eco-friendly ride) increased the likelihood of passengers picking a green ride over positively valenced wording  (i.e., that highlighted the emission equivalents saved by the eco-friendly ride).
    \end{itemize} \\
%    \midrule
    \begin{enumerate}%[label=2.\arabic*, leftmargin=*, resume]
\leftskip-15pt
    \item[2.3] How do \COTwo{} equivalencies influence a ride-hailing
      passenger's choice when detailed explanations about the
      equivalencies are provided?
    \end{enumerate}&
    \begin{itemize}[leftmargin=*]
    \item[] Equivalency explanations do not influence choosing green ride-sharing options.
    \end{itemize}\\
    %\midrule
    \multicolumn{2}{l}{\textbf{Relative and absolute \COTwo{} emissions values}} \\
    \midrule
    \begin{enumerate}%[label=3.\arabic*, leftmargin=*]
\leftskip-15pt
    \item[3.1] How effective are raw \COTwo{} emissions values for influencing
      people's choices when made available in absolute terms (i.e.,
      not used for comparison) compared to relative terms (i.e., used
      for comparison between choices)?
    \end{enumerate}&
    \begin{itemize}[leftmargin=*]
    \item[] People respond to relative \COTwo{} values more than absolute \COTwo{} values.
    \end{itemize}\\
%    \midrule
    \begin{enumerate}%[start=2, label=3.\arabic*, leftmargin=*]
\leftskip-15pt
    \item[3.2] How effective are \COTwo{} targets for nudging users to consider
      absolute carbon values?
    \end{enumerate}&
    \begin{itemize}[leftmargin=*]
    \item[] Providing emission targets does not push people to think about absolute \COTwo{} values.
    \end{itemize}\\
    % \midrule
    \multicolumn{2}{l}{\textbf{Temporal Effects of \COTwo{} equivalencies}} \\
    \midrule
    \begin{enumerate}%[label=4.\arabic*, leftmargin=*]
\leftskip-15pt
    \item[4.1] How sensitive are people towards emission difference s
      presented as smaller daily emission averages than larger total
      emission values in the context of longer trips (i.e., car
      rentals)?
    \end{enumerate}&
    \begin{itemize}[leftmargin=*]
    \item[] \COTwo{} emission equivalencies influence rental choices in ways similar to ride-sharing choices. Emission values have the same influence on rental choices when presented per day or for the full trip.
    \end{itemize} \\
  \end{tabular}
  \caption{\label{tab:summary_table} Research questions (left) and
    findings (right) from Phase 2.}
\end{table*}

The aim of this series of studies was to understand how people process emission information in the context of ride-sharing and rental vehicle user interfaces.
Our studies were designed in part to explore facets of Sanguinetti et al.'s \emph{design framework for eco-feedback information}~\cite{sanguinetti2018information}.
We begin by situating our main study findings \changemarker{(see Tables~\ref{tab:summary_table_1} and~\ref{tab:summary_table})} into this framework then discuss the broader design implications of our findings and how these results can inform emission feedback in other domains.

\subsection{Implications for the \emph{Design Framework for Eco-Feedback Information}}
Our results have implications for each dimension of Sanguinetti et al.'s information \emph{message} and \emph{granularity} categories:

\begin{description}
    \item[Metrics:] Across all four studies we find evidence that people respond well to different forms of messaging about carbon. Study 1 and 4 both find that people prefer ride-sharing options presented with a simple icon, be it a green leaf (Study~\ref{study:1}) or a green ``EV'' logo (Study~\ref{study:4:rental}). We additionally find that this tendency can be enhanced by additional forms of emission information, whether through raw \COTwo{} emissions (Studies 1--4), energy related equivalencies (e.g., pounds of coal, gallons of gasoline, home energy usage, smartphones; Studies~\ref{study:1},~\ref{study:2} and~\ref{study:4:rental}) or sequestering equivalencies (e.g., square feet of forests; Studies~\ref{study:1} and~\ref{study:2}). Indeed, our results suggest that people \emph{prefer} receiving emission-related messaging: the quantitative and qualitative results from Study~\ref{study:3} consistently show that people prefer ride-sharing options that are transparent about their emissions. These results suggest that there are numerous ways in which emission messaging can be effective, whether through more simple logo methods, more technical carbon information, or more relatable equivalencies.

    \item[Valence:] In addition to carbon-related messaging, we find that message valence can influence people's ride-sharing choices.
      Study~\aptLtoX[graphic=no,type=env]{2b}{\ref{study:2:valence}} finds that ride-sharing choices are
      more influenced by using negatively valenced emission framing,
      i.e., highlighting the extra carbon emitted by an option, than
      using positive framing, i.e., highlighting the emissions saved by an option (Figure~\ref{fig:valence}).
      These findings are consistent with research such as the Normative Activation Framework~\cite{schwartz1977normative} and research into loss aversion~\cite{novemsky2005boundaries, litovsky2022loss, tversky1992advances} that find asymmetric influences of negative versus positive framings.
      These results propose that framing valence plays an additional role in how people process information about carbon, and may act more readily when the harmful emission effects of an option are highlighted rather than the benefits.

    \item[Contextual Information:] Studies~\aptLtoX[graphic=no,type=env]{2c}{\ref{study:2:explain}} and~\ref{study:3} examined the influence of contextualizing information about emissions and equivalencies when people made their ride-sharing decisions.
      In both cases, adding contextual information did not influence people's choices: adding explanations for each equivalency in Study~\aptLtoX[graphic=no,type=env]{2c}{\ref{study:2:explain}} did not influence the effectiveness of any of the tested equivalencies (Figure~\ref{fig:info}), and adding reference emission targets in Study~\aptLtoX[graphic=no,type=env]{3b}{\ref{study:3:target}} did not push people to use absolute carbon values to inform their decisions (Figure~\ref{fig:carbon_targets}).
      These results suggest that lightweight contextual interventions may not be sufficient to influence user choices in ride-sharing settings and that longer-term interventions may be required for context to matter.
      Additional research will be required to examine the frequency and intensity required for contextual information to influence people's choices.

    \item[Behavioral Granularity:] We found a clear influence of social framings for ride-sharing choices. One of our most consistent findings was that highlighting the collective impact of other users' decisions increased the likelihood that participants chose green ride-sharing options (Figures~\ref{fig:curves},~\ref{fig:collective}). This finding was true for both positive and negative valence framings (Figure~\ref{fig:collective}). Social framings were especially effective at increasing the impact of equivalencies that were not effective when framed individually (e.g., trees, bags of waste, Figure~\ref{fig:collective_beta}). We additionally found that highlighting the popularity of an option could also influence people's choices if a high proportion of users were making the choice (Figure~\ref{fig:social}). Both of these examples are consistent with work in social norm theory, which consistently demonstrates the power of highlighting social norms and dynamics in shaping people's behavior~\cite{schultz2007constructive, schultz2018constructive, allcott2011social}. Highlighting dynamic changes to social norms had a less consistent effect in our study. Dynamic norms were primarily effective when a large proportion of users were reported to be switching \emph{away} from the \emph{Ride} option. This successful framing is consistent with previous work on dynamic social norms that highlights behaviors that people were \emph{reducing} (e.g., reducing meat consumption~\cite{sparkman2017dynamic}). However, our results may highlight situations in which dynamic norms are less effective (e.g., when framed around adopting new behaviors and/or when a lower percentage of people a switching).

    \item[Temporal Granularity:] Study~\ref{study:4:rental} examined whether the time horizon over which carbon information is provided influenced how people processed emission information. We found no evidence that people's use of emission information for longer rental trips changed as a function of whether the emission information was presented per day or for the full trip duration. We had originally proposed that \emph{numerical distance effects} (related to Weber-Fechner laws~\cite{dehaene2003neural,gallistel2000non}) may reduce the weight people gave to higher emission values associated with longer trips. However, we found no evidence of this in Study~\ref{study:4:rental}. It is possible that participants relied more on heuristic approaches, such as choosing options with lower emissions among all of the options, without processing fine-grained representations of emission values.

    \item[Data granularity:] Although our studies show that both coarse data (e.g., a green leaf logo) and detailed data (e.g., \COTwo\ information) influenced participant choices, we found many cases where more detailed information had a larger effect (Figures~\ref{fig:curves},~\ref{fig:carbon0},~\ref{fig:carbon_targets}). \changemarker{Studies~\ref{study:1} and~\aptLtoX[graphic=no,type=env]{2c}{\ref{study:2:explain}} showed that detailed emissions information (e.g., raw \COTwo{} emissions values or equivalency information) raised the likelihood of participants picking a green ride over abstract information (e.g., a green leaf logo). Further,} our findings from Study~\ref{study:3} show that people \emph{prefer} options with information, actively choosing options that provided emission information over those that did not. This want for information could be related to an \emph{ambiguity effect}, where people prefer choice options that are more certain and less ambiguous~\cite{fox1995ambiguity}. Interestingly, we found no evidence that people have a good understanding of absolute emission values (Figure~\ref{fig:carbon_targets}). This suggests that although people prefer more precise emission numbers, there is a potential limit to the benefit of data granularity in the context of emissions.
\end{description}

We next discuss the implications of our work more broadly, both in our understanding about how people think about carbon emission information, and the design implications for interfaces aimed at communicating emission information to users.

% \subsection{Cost of Going Green / Carbon Empathy}
% In a perfectly balanced market, people should make decisions about goods purely on price signals signals alone~\cite{simon1955behavioral}.
% When they do not, there is some other action at play, be it an inherent bias or a bias created by external forces that make people more aware of some aspect of a product (or the product itself in the case of advertising).
% In this work, we see that for most participants, transportation decisions are driven primarily by price.
% However, our studies also reveal that there are several negative externalities that, when people's attention is drawn to them, can impact their decisions.
% In particular, interventions that make people aware of the main
% negative externality of a transportation choice---primarily increased \COTwo\ output---cause people to reassess the social marginal benefit of the product, lowering its desirability with respect to other, less harmful options.
% Our work shows that the way this plays out across different products and interventions is not always perfectly rational.

% This should not be surprising since externalities occur ex post facto the rational pricing market.

\vspace*{-6pt}
\subsection{\COTwo\ Interventions Are Not ``One Size Fits All''}

Study~\ref{study:1} found that although people do not fully understand absolute \COTwo\ values,  \COTwo\ was nevertheless an effective intervention tool, consistently ranking among the top interventions in both our ride-sharing and rental studies. These results propose a separation between ``normative'' and ``revealed'' preferences, whereby people's reports are not always consistent with their behavior~\cite{beshears2008preferences}. This finding also demonstrates the importance of mixed-methods approaches, which measure both how people processes as well as understand different intervention approaches, while also measuring the impact of these interventions on behavior.

More broadly, this finding has a powerful implication: despite efforts to create more useful or relatable equivalencies to help carbon literacy, equivalencies may be no more effective than providing direct \COTwo\ emission values. Conversely, it is possible that \COTwo\ is primarily effective with respect to vehicles and may not translate as readily to other domains (e.g., food). Nevertheless, our results suggest that communicating equivalencies versus direct \COTwo\ emissions is a complicated and situation-dependent design challenge.

Furthermore, there is likely a degree of heterogeneity in the effectiveness of different interventions. Interventions that may be familiar and relevant to some participants may not resonate as much with others (e.g., incandescent bulbs may no longer be relevant to some participants). Our examination of participants' qualitative responses propose that different interventions had widely different, sometimes opposite, impacts on different people. Although a formal analysis of heterogeneous treatment effects is outside the scope of this particular study, we see this as a fruitful direction for efforts aimed at personalizing emission-based messaging. Indeed, recent efforts at more targeted climate-based messaging has shown promise, especially in populations that have traditionally been resistant to messaging about the environment~\cite{goldberg2021shifting, harinen2021machine}.  While further work is needed to better understand the impact of individual differences on carbon messaging, this possibility presents design opportunities for more tailored carbon-based intervention, including personalized messaging and recommendation systems.

\subsection{Embodying Carbon-based Interventions}
These studies tested how different information dimensions of the eco-feedback design framework proposed by Sanguinetti et al.~\cite{sanguinetti2018information} impact decision-making in different personal transportation contexts.
Indeed, \COTwo{} information was an effective \emph{metric} for communicating carbon emissions despite gaps in how people interpret this scientific unit.
\emph{Contextual information}, such as comparing \COTwo{} information between choices, allowed people to make well-informed, eco-friendly choices.
Even without any contextual information, people were drawn towards choices that were transparent in displaying \COTwo{} information.
These findings suggest a potential value in making carbon emissions information more commonplace for consumer products and services, similar to calorie posting~\cite{bollinger2011calorie}.

As an extension, these findings suggest that further study should be given to other embodiments for carbon-based interventions, such as automobile ownership or dashboard displays. Personal vehicle ownership, being a longer-term decision, adds another layer of temporality. Design interventions for owned vehicles can leverage dynamic counterfactuals to convey not only emissions information but also fuel and cost savings over longer periods of time~\cite{shamma2022ev}. \changemarker{Similarly, real-time feedback applications can incorporate similar design interventions to encourage sustainable driving~\cite{avolicino2022ecogo}.} While in-situ car dashboards currently employ different kinds of eco-feedback interventions, they have largely focused on communicating fuel efficiency using metrics such as \emph{miles/kWh} or abstract representations such as \emph{virtual leaves}~\cite{outlander_2014,dahlinger2018impact}.
Overall, there are many opportunities to display carbon-based interventions for encouraging eco-friendly driving, from in-situ driving interventions to trip summary displays. \changemarker{Further, our findings can also be extended for transportation modes beyond automobiles, be it for nudging users to choose low-emissions flights on Google Flights or FlyGreen~\cite{sanguinetti2017greenfly}, or for encouraging eco-friendly commute alternatives such as public transport, biking, carpooling, and walking using apps such as Ubigreen~\cite{froehlich2009ubigreen} or EcoTrips~\cite{park2017ecotrips}.}

\subsection{Fostering Carbon Literacy}
Continual exposure to \COTwo{} information in different aspects of our daily lives can help put the ``eco-friendliness'' of different items into perspective, bridging gaps in \emph{carbon literacy} over time.
Although people responded well to interventions with \COTwo\ information, we did not find evidence that people specifically understand or use absolute carbon values in the same way that people may understand number of calories or step counts. Although our findings did not show any significant impact when different targets were provided as contextual information (\aptLtoX[graphic=no,type=env]{\textbf{H3.1}}{\ref{hypo:3:rel}}), open-ended responses revealed the need for an independent organization such as the EPA to recommend targets for \COTwo{} emissions. In the context of our study, targets may have needed some refinement to be effective. For example, emission numbers for the rides never crossed 45\% of the targets we showed (85 lbs.\ of \COTwo{}). It is  possible that participants may have deprioritized such small proportions.
Further studies are needed to understand whether goal-setting is effective only beyond a certain threshold or not.
Finally, while our work aimed to encourage eco-friendly behavior amongst individual consumers, it does not de-emphasize the need for large corporations and policymakers to play their part in reducing carbon emissions---a concern raised by multiple participants in the open-ended responses and of growing concern from climate scientists around the world~\cite{mann2021new, chater2022frame}.

%%%% 5.discussion.tex ends here %%%%

%%%% 6.conclusion.tex starts here %%%%

\section{Conclusion and Future Work}
In this article, we ask: what \COTwo\ equivalencies will help people make eco-friendly choices when selecting a ride share or renting a car. We further ask: do consumers relate and understand direct feedback about \COTwo\ emissions in these transportation-related interfaces?
Or do they instead prefer simpler heuristic interventions (e.g., green logos) or more relatable \COTwo{}-equivalent activities?
Our work found that people responded well to interventions that communicated emission information, even if they did not perfectly understand the information units (e.g., absolute \COTwo\ emission values). Moreover, participants reported \emph{wanting} information about emissions and preferred options that provided emission information over ones that did not (Figure~\ref{fig:carbon0}). Beyond the information itself, we additionally find that framing can also influence intervention effectiveness, with a particular emphasis on social or valence-based framing contexts. These findings fit well with known behavioral science theories that highlight people's interest in information and aversion to ambiguity~\cite{tversky1992advances,ellsberg1961risk}, responsiveness to social norms~\cite{sparkman2017dynamic, schultz2018constructive, schultz2007constructive}, and the impact of valenced information framing~\cite{litovsky2022loss, schwartz1977normative, novemsky2005boundaries}. Taken together, these results provide an exciting starting point from which to further our understanding of how best to communicate carbon information in ways that are meaningful and impactful for users.

There is more work to be done to better understand how people integrate emissions information into their decisions.
For example, many interventions we tested worked well in controlled studies, but it is unclear how well these interventions will translate into \emph{everyday practices}---a concern that has previously been raised about eco-feedback systems in general~\cite{hansson2021decade}.
Can we design interventions that better activate people's empathetic response to climate change? How might we utilize ongoing interventions to strengthen people's empathetic responses? Finally, to what extent can these findings inform the design and development of sustainable practices by other stakeholders beyond individual consumers?
Climate change is one of, if not the, central challenge of our time.
It is imperative that we continue to develop our understanding of how
people interpret and utilize information about emissions so we can
increase carbon literacy and provide more effective eco-friendly options.

\begin{acks}
Many thanks to Monica Van, Matthew Lee, Candice Hogan, and Matthew Klenk for their feedback on this work.
\end{acks}

\balance%
\bibliographystyle{ACM-Reference-Format}
\bibliography{carbon}%% Commented by merge tool

%%%% appendix.tex starts here %%%%

\clearpage
\appendix

\begin{figure*}[p]
  \centering
  \begin{subfigure}[b]{0.4\textwidth}
    \centering
    \includegraphics[width=\textwidth]{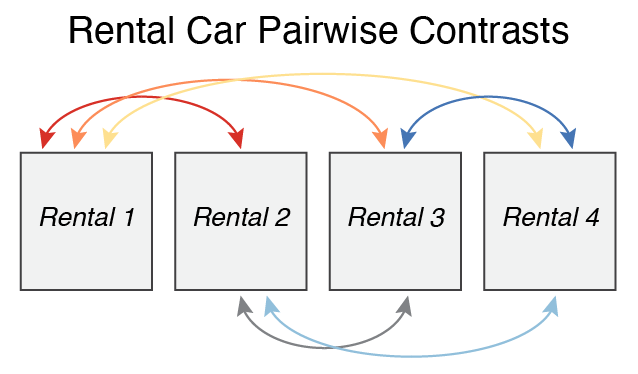}
    \caption{\label{fig:tc_graphic} Schematic of all possible pairwise comparisons between the four possible rental options on each trial.}
    \Description{Schematic of the different pairwise contrasts. Four squares aligned horizontally meant to represent the four different rental options on each trial. Each box has the words ``Rental'' written in the middle with a number ranging from 1--4 from left to right respectively. Arrows are drawn connecting the squares and used to illustrate all six possible pairwise comparisons between the four rental options.}
  \end{subfigure}
  \begin{subfigure}[b]{\textwidth}
    \centering
    \vspace{2pc}
  \includegraphics[width=\textwidth]{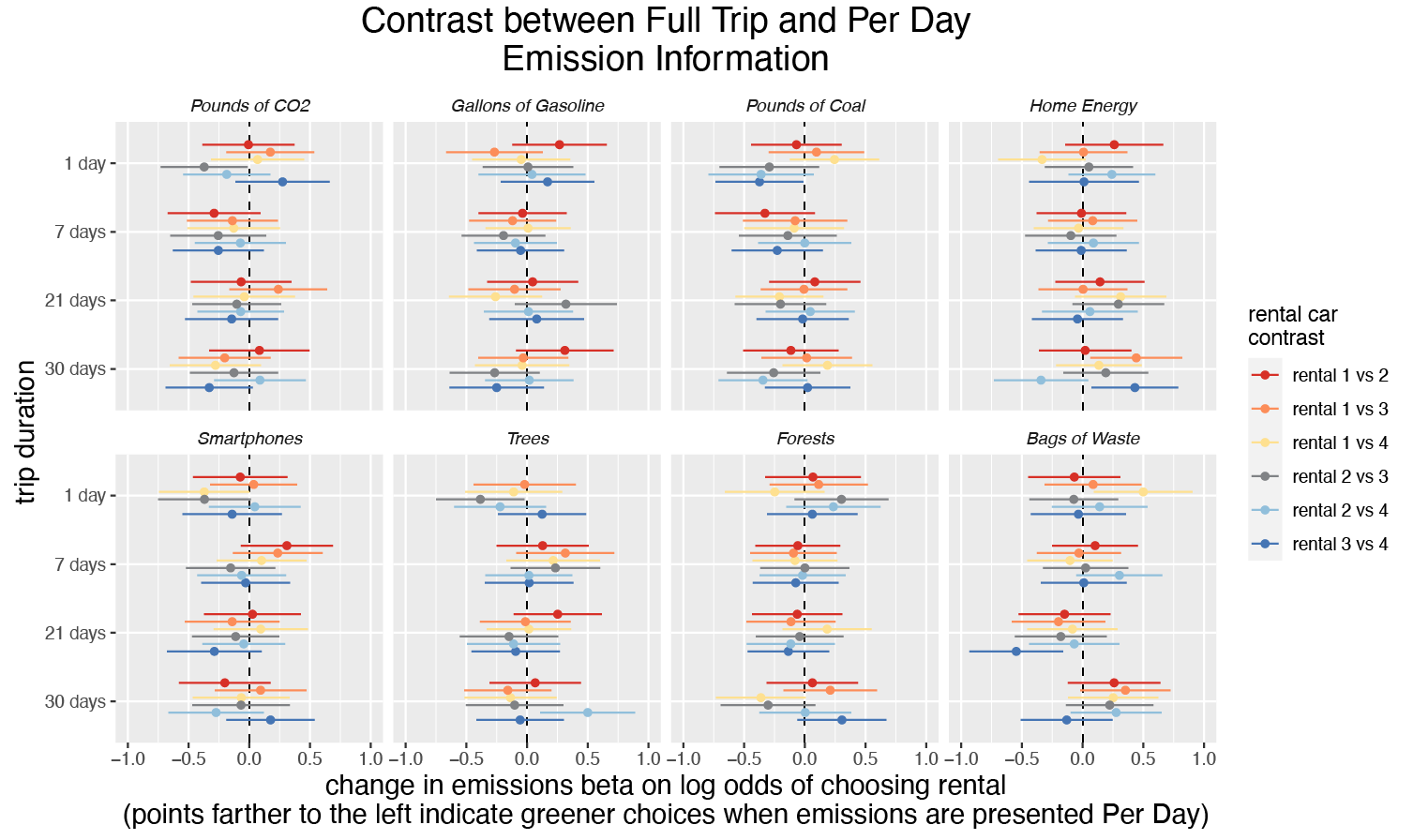}
    \caption{\label{fig:tc_betas} Contrast coefficients for pairwise comparisons between all possible rental options comparing the change in the influence of emission values when presented per day versus for the full trip. Negative values indicate that people are less willing to choose options with high emissions. Contrasts are shown for each emission equivalency and for all trip durations on the $y$-axis. Points indicate contrast coefficients and errorbars indicate 95\% confidence intervals.}
    \Description{A chart with eight panels arranged with four columns and two rows showing the change in the influence of \COTwo\ emission values between the per day and full trip conditions. The X-axis shows the influence of emission values on the probability that a rental choice will be made in log odds scale. The Y-axis shows different trip durations, from 1 day to 30 days in descending order. For each level of trip duration, points and error bars from the six pairwise contrasts are vertically aligned indicating the change in the influence of emissions between the per day and full trip intervention conditions. A black dashed line is added to all panels to indicate the zero point (indicating no difference between the per day and full trip conditions). Points that fall to the left of the dashed line indicate that people are less likely to choose options with higher emissions in the per day condition compared to the full trip, and points that fall to the right of the dashed line indicate that people are more likely to choose options with higher emissions in the per day condition compared to the trip duration condition. The top four panels show the analysis for (from left to right): Pounds of \COTwo\, Gallons of Gasoline, Pound of Coal, and Home Energy. The four bottom panels show the analysis for (from left to right): Smartphones, Trees, Forests, and Bags of Waste. None of the contrasts across all conditions were systematically to the left or to the right of the black dashed line, indicating that the influence of emission values did not differ between the per day and full trip conditions.}
  \end{subfigure}
  \caption{\label{fig:trip_duration} Contrast of the influence of
    emission values between the ``Full Trip'' and ``Per Day''
    conditions.}
\end{figure*}

\section{Self Reported Questions}
\subsection{Open-Ended Questions for Study 1} 
\begin{itemize}
\item Choose one item that was the most useful and least useful
  for making your decisions.
\item Choose one item that was the most relatable and least
  relatable from the list below.
\item If you had to measure the environmental impact of your daily activities, what information or metrics would be most meaningful to you? Please note that these metrics don't have to be scientific in nature and can be beyond the ones you have already seen in this study (e.g., gallons of gasoline).
\end{itemize}

\subsection{Questions for Study 2}\label{sec:appendix_study_2}
\begin{itemize}
\item What factors did you find most useful when choosing a ride in the study? Please list as many factors as you wish
\item What factors did you find least useful when choosing a ride in the study? Please list as many factors as you wish
\end{itemize} 
\subsubsection{Only for Study 2a participants} 
\begin{itemize}
\item Do you think your personal actions and choices have a significant impact on the environment? Why or why not?
\item How did information about the collective environmental impact of everyone’s actions affect your decision?
\item How does learning about new eco-friendly trends affect your decisions?
\end{itemize} 
\subsubsection{Only for Study 2b participants} 
\begin{itemize}
\item How did learning about the damaging impact your actions can have on the environment (e.g., more coal burnt, trees killed, etc) affect your decision?
\item How did learning about the positive impact your actions can have on the environment (e.g., more trees planted, less coal burnt, etc) affect your decision?
\end{itemize} 
\subsubsection{Only for Study 2c participants}
\begin{itemize}
\item In general, how well do you understand the impact of your actions on the environment?
\item How did you perceive the explanations provided for the environmental impact of your ride? How did it affect your decision?
\end{itemize} 

\subsection{Study 3a}\label{sec:appendix_study_3}
Each participant was asked only one out of these three questions:
\begin{itemize}
\item When you hear 100 pounds of CO2, what is the first thing that comes to mind?
\item When you hear 1 pound of CO2, what is the first thing that comes to mind?
\item 1 pound of CO2 is just as bad as \_\_\_\_\_
\end{itemize}
This question was shown to all participants:
\begin{itemize}
\item Do you use information about carbon emissions while making daily decisions (e.g., which appliance to use, which trips to take, etc.)?
\end{itemize} 

\textbf{Study 3b} 
\begin{itemize}
\item What factors did you find most useful when choosing a ride in the study? Please list as many factors as you wish
\item What factors did you find least useful when choosing a ride in the study? Please list as many factors as you wish
\end{itemize} 

\subsubsection{Only for Target 1 participants}
\begin{itemize}
\item What did you think about the recommended CO2 emission numbers?
  Do you trust these numbers?
\item In your opinion, who and/or what would be a reliable source
  for providing these numbers?
\item How did the recommendation affect your decision?
\end{itemize}
\subsubsection{Only for Target 2 participants}
\begin{itemize}
\item What do you think about the environmental impact of your
  actions on the community that you are a part of?
\item How does learning about the environmental impact of others in your community affect your decisions?
\end{itemize} 
\subsubsection{Only for Target 3 participants}

\begin{itemize}
\item How does information regarding the environmental impact of eco-friendly members in your community affect your decisions?
\item Whose carbon footprint information (e.g.,
community members, peer groups, family, celebrities, etc.) would
influence your decision and why?
\end{itemize} 

\subsection{Study 4}\label{sec:appendix_study_4}
\begin{itemize}
\item How would you prefer to see the information about the carbon footprint of your activities? (Options: Aggregated over longer time periods (e.g., per year), Broken into shorter time periods (e.g., per day), In some other way)
\item Can you elaborate your response to the previous question? How does the time period of carbon footprint information affect your decisions?
\end{itemize}

%%%% appendix.tex ends here %%%%

\end{document}